%% file: cds-format.tex
\documentclass[ALICE,manyauthors]{cernphprep}

\usepackage{eso-pic,picture}
\usepackage{amsmath}
\usepackage{amssymb}
\usepackage{wasysym}
\usepackage{units}
\usepackage{graphicx}
\usepackage{graphics, epsfig, subfigure}
\usepackage{fancyhdr}
\usepackage{multirow}
\usepackage{multicol}
\usepackage{placeins}
\usepackage{longtable}
\usepackage{color}
\usepackage{appendix}
\usepackage{afterpage}
\usepackage{ulem} 

\DeclareGraphicsExtensions{.pdf,.png,.jpg}

\definecolor{myblue}{rgb}{0.0,0.0,0.6}
\definecolor{darkred}{rgb}{0.7,0.0,0.0}
\definecolor{darkgreen}{rgb}{0,0.7,0.0}

\usepackage{ulem}

\newcommand{\pt}      {\ensuremath{p_{\mathrm{T}}}}
\newcommand{\pT}      {\pt}
\newcommand{\ptlead}      {\ensuremath{p_{\mathrm{T,LT}}}}
\newcommand{\pTlead}      {\ptlead}

\newcommand{\gmom}    {\mbox{${\rm GeV}/c$}}

\newcommand{\tev}     {\mbox{${\rm TeV}$}}

\newcommand{\bfigFullPage}{\begin{figure} \begin{center} \vspace{0pt}}
\newcommand{\bfig}[1][!]{\begin{figure}[#1] \begin{center}}
\newcommand{\efig}{\end{center} \rule{4cm}{0.4pt} \end{figure}}
\newcommand{\bfigonecol}[1][!]{\begin{figure*}[#1] \begin{center}}
\newcommand{\efigonecol}{\end{center} \end{figure*}}
\newcommand{\efigNoLine}{\end{center} \end{figure}}
\newcommand{\btab}[1][!]{\begin{table}[#1] \begin{center}}
\newcommand{\etab}{\end{center} \rule{4cm}{0.4pt} \end{table}}
\newcommand{\btabonecol}[1][!]{\begin{table*}[#1] \begin{center}}
\newcommand{\etabonecol}{\end{center} \end{table*}}
\newcommand{\etabNoLine}{\end{center} \end{table}}

\newcommand{\twoPlotsOneCol}[4][t]{\bfigonecol[#1] \begin{minipage}[t]{0.48\linewidth} \vspace{0pt} \centering #2 \end{minipage} \hfill \begin{minipage}[t]{0.48\linewidth} \vspace{0pt} \centering #3 \end{minipage} #4\efigonecol}
\newcommand{\twoPlotsNoLine}[4][hp!]{\bfig[#1] \begin{minipage}[t]{0.48\linewidth} \vspace{0pt} \centering #2 \end{minipage} \hfill \begin{minipage}[t]{0.48\linewidth} \vspace{0pt} \centering #3 \end{minipage} #4\efigNoLine}

\newcommand{\sixPlotsNoLine}[8][p]{\bfig[#1] \begin{minipage}[t]{0.48\linewidth} \vspace{0pt} \centering #2 \end{minipage} \hfill \begin{minipage}[t]{0.48\linewidth} \vspace{0pt} \centering #3 \end{minipage}\\ \vspace{20pt}
\begin{minipage}[t]{0.48\linewidth} \vspace{0pt} \centering #4 \end{minipage} \hfill \begin{minipage}[t]{0.48\linewidth} \vspace{0pt} \centering #5 \end{minipage} \\ 
\vspace{20pt}
\begin{minipage}[t]{0.48\linewidth} \vspace{0pt} \centering #6 \end{minipage} \hfill \begin{minipage}[t]{0.48\linewidth} \vspace{0pt} \centering #7 \end{minipage} #8\efigNoLine}

\newcommand{\bq}{\begin{equation}}
\newcommand{\eq}{\end{equation}}

\newcommand{\bqq}{\begin{eqnarray}}
\newcommand{\eqq}{\end{eqnarray}}

\newcommand{\cms}{\sqrt{s}}

\newcommand{\etain}[1]{$|\eta|$~$<$~$#1$}

\newcommand{\degrees}{^{\circ}}
\DeclareMathAlphabet{\mathpzc}{OT1}{pzc}{m}{it} 

\begin{document}%
%
%
\begin{titlepage}
\PHnumber{2011-204}                 
\PHdate{09 Dec 2011}              
%
%
\title{Underlying Event measurements in pp collisions at $\sqrt{s} =$ 0.9 and \unit[7]{TeV} with the ALICE experiment at the LHC}
\ShortTitle{Underlying Event measurements with ALICE}   
%
\Collaboration{ALICE Collaboration%
         \thanks{See Appendix~\ref{app:collab} for the list of collaboration 
                      members}}
\ShortAuthor{ALICE Collaboration}      
\begin{abstract}
\input{abstract.tex}
\end{abstract}
\end{titlepage}
\setcounter{page}{2}
\input{introduction}

\input{detectors}

\input{data_samples}

\input{event_track_sel}

\input{analysis}

\input{corrections}

\input{systematics}

\input{results}

\input{conclusions}

\newpage
\input{plots}

\newenvironment{acknowledgement}{\relax}{\relax}
\begin{acknowledgement}
\section{Acknowledgements}
\input{acknowledgements_Nov2011}    
\end{acknowledgement}

\bibliography{paper}

\newpage
%
%
\appendix
\section{The ALICE Collaboration}
\label{app:collab}
\input{authorlist1}  

\end{document}

%% file: abstract.tex
We present measurements of Underlying Event observables in pp collisions at $\sqrt{s} = 0.9$ and \unit[7]{TeV}.
The analysis is performed as a function of the highest charged-particle transverse momentum $\ptlead$ in the event. 
Different regions are defined with respect to the azimuthal direction of the leading (highest transverse momentum) track: Toward, Transverse and Away. The Toward and Away regions collect the fragmentation products of the hardest  partonic interaction. The Transverse region is expected to be most sensitive to the Underlying Event activity.
The study is performed with charged particles above three different $\pT$ thresholds: 0.15, 0.5 and 1.0 \gmom. 
In the Transverse region we observe an increase in the multiplicity of a factor 2-3 between the  lower and higher collision energies, depending on the track $\pt$ threshold considered.
Data are compared to \textsc{Pythia}~6.4, \textsc{Pythia}~8.1 and \textsc{Phojet}. 
On average, all models considered underestimate the multiplicity and summed $\pT$ in the Transverse region by about 10-30\%.

%% file: introduction.tex
\section{Introduction}
The detailed characterization of hadronic collisions is of great interest for the understanding of the underlying physics. 
The production of particles can be classified according to the energy scale of the process involved. 
At high transverse momentum transfers ($\pT \gtrsim \unit[2]{\gmom}$) perturbative Quantum Chromodynamics (pQCD) is the appropriate 
theoretical framework to describe partonic interactions.
This approach can be used to quantify parton yields and correlations, whereas
the transition from partons to hadrons is a non-perturbative process that has to be treated using phenomenological approaches.
Moreover, the bulk of particles produced in high-energy hadronic collisions originate from low-momentum transfer processes. 
For momenta of the order of the QCD scale, $\mathcal{O}$(100 MeV), a perturbative treatment is no longer feasible.
Furthermore, at the center-of-mass energies of the Large Hadron Collider (LHC), at momentum transfers of a few GeV/$c$, 
the calculated QCD cross-sections for 2-to-2 parton scatterings exceed the total hadronic cross-section \cite{mpi_theory}. 
This result indicates that Multiple Partonic Interactions (MPI) occur in this regime.
The overall event dynamics cannot be derived fully from first principles and must be modeled using phenomenological calculations. 
Measurements at different center-of-mass energies are required to test and constrain these models.

In this paper, we present an analysis of the bulk particle production in pp collisions at the LHC by measuring the so-called Underlying Event (UE) activity \cite{cdf1}.
The UE is defined as the sum of all the processes that build up the final hadronic state in a collision excluding the hardest leading order partonic interaction. 
This includes fragmentation of beam remnants, multiple parton interactions and initial- and final-state radiation (ISR/FSR) associated to each interaction.
Ideally, we would like to study the correlation between the UE and perturbative QCD interactions by isolating the two leading partons with topological cuts and measuring the remaining event activity as a function of the transferred momentum scale ($Q^{2}$).
Experimentally, one can
identify the products of the hard scattering, usually the leading jet, 
and  study the region azimuthally perpendicular to it as a function of the jet energy. 
Results of such an analysis have been published by the CDF~\cite{cdf1, cdf2, cdf3, cdf4} and  STAR~\cite{star} collaborations for pp collisions 
at $\sqrt{s} = 1.8$ and \unit[0.2]{TeV}, respectively.  
Alternatively, the energy scale is given by the leading charged-particle
transverse momentum, circumventing uncertainties related to the jet reconstruction procedure at low $\pT$.
It is clear that this is only an approximation to the original outgoing parton momentum, 
the exact relation depends on the details of the fragmentation mechanism.
The same strategy based on the leading charged particle has recently been applied by the ATLAS~\cite{atlas1} and CMS~\cite{cms1} collaborations.

In the present paper we consider only charged primary particles\footnote{Primary particles are defined as prompt particles produced in the collision and their decay products (strong and electromagnetic decays), 
except products of weak decays of strange particles such as $K^{0}_{S}$ and $\Lambda$.}, due to the limited calorimetric acceptance of the ALICE detector systems in azimuth. 
Distributions are measured for particles in the pseudorapidity range $|\eta| < 0.8$ with  
$\pT > p_{\rm T, min}$, where $p_{\rm T,min} =$ 0.15, 0.5 and \unit[1.0]{\gmom}, and are studied as a function of the leading particle  transverse momentum.

Many Monte Carlo (MC) generators for the simulation of pp collisions are available; see \cite{mc_lhc} for a recent review discussing for example \textsc{Pythia} \cite{pythia6}, \textsc{Phojet} \cite{phojet}, \textsc{Sherpa} \cite{mc_lhc} and \textsc{Herwig} \cite{herwig}. 
These provide different
descriptions of the UE associated with high energy hadron collisions. 
A general strategy is to combine a perturbative QCD treatment of the hard scattering with a phenomenological approach to soft processes. 
This is the case for the two models used in our analysis: \textsc{Pythia} and \textsc{Phojet}. 
In \textsc{Pythia} the simulation starts with a hard LO QCD process of the type 2 $\to$ 2. Multi-jet topologies are generated with the parton shower formalism and hadronization is implemented through the Lund string fragmentation model \cite{lund_string}. 
Each collision is characterized by a different impact parameter $b$. Small $b$ values correspond to a large overlap of the two incoming hadrons  and to an increased probability for MPIs. 
At small $\pT$ values color screening effects need to be taken into account. Therefore a cut-off $p_{\rm T,0}$ is introduced, which damps the QCD cross-section for $\pT \ll p_{\rm T,0}$. This cut-off is one of the main tunable model parameters.\\
In \textsc{Pythia} version 6.4 \cite{pythia6} MPI and ISR have a common transverse momentum evolution scale (called interleaved evolution \cite{interleaved}). Version 8.1 \cite{pythia8} is a natural extension of version 6.4, where the FSR evolution is interleaved with MPI and ISR and parton rescatterings \cite{rescatt} are considered. In addition initial-state partonic fluctuations are introduced, leading to a different amount of color-screening in each event.\\
\textsc{Phojet} is a two-component event generator, where the soft regime is described by the Dual Parton Model (DPM) \cite{dpm} and the high-$\pT$ particle production by perturbative QCD. The transition between the two regimes happens at a $\pT$ cut-off value of 3 \gmom. A high-energy hadronic collision is described by the exchange of effective Pomerons. Multiple-Pomeron exchanges, required by unitarization, naturally introduce MPI in the model.

UE observables allow one to study the interplay of the soft part of the event with particles produced in the hard scattering and are therefore good candidates for Monte Carlo tuning.
A better understanding of the processes contributing to the global event activity will help to improve the predictive power of such models. Further, a good description of the UE is needed to understand backgrounds to other observables, e.g., in the reconstruction of high-$\pt$ jets.

The paper is organized in the following way: the ALICE sub-systems used in the analysis are described in Section~\ref{detectors} 
and the data samples in Section~\ref{data_samples}. 
Section~\ref{event_track_sel} is dedicated to the event and track selection. Section~\ref{analysis} introduces the analysis strategy. 
In Sections~\ref{corrections} and \ref{systematics} we focus on the data correction procedure and systematic uncertainties, respectively. 
Final results are presented in Section~\ref{results} and in Section~\ref{conclusions} we draw  conclusions.

%% file: detectors.tex
\section{ALICE detector}
\label{detectors}

Optimized for the high particle densities encountered in heavy-ion collisions, 
the ALICE detector is also well suited for the study of pp interactions. 
Its high granularity and particle identification capabilities can be exploited for precise measurements 
of global event properties ~\cite{alice1,alice2,alice3,alice4,alice5,alice6,alice7}. 
The central barrel covers the polar angle range $45\degrees-135\degrees$ (\etain{1}) and full azimuth. 
It is contained in the L3 solenoidal magnet which provides a nominal uniform magnetic field of $0.5 \, {\rm T}$. 
In this section we describe only the trigger and tracking detectors used in the analysis, 
while a detailed discussion of all ALICE sub-systems can be found in~\cite{alice_inst}.

The V0A and V0C counters consist of scintillators with a pseudorapidity coverage 
of $-3.7<\eta<-1.7$ and $2.8<\eta<5.1$, respectively. They are used as trigger detectors and to reject beam--gas interactions. 

Tracks are reconstructed combining information from the two main tracking detectors in the ALICE central barrel: the Inner Tracking System (ITS) and 
the Time Projection Chamber (TPC).
The ITS is the innermost detector of the central barrel and consists of six layers of silicon sensors. 
The first two layers, closely surrounding the beam pipe, are equipped with high granularity Silicon Pixel Detectors (SPD). 
They cover the pseudorapidity ranges $|\eta| < 2.0$ and $|\eta| < 1.4$ respectively. 
The position resolution is $12 \, {\rm \mu m}$ in $r \phi$ and about $100 \, {\rm \mu m}$ along the beam direction.
The next two layers are composed of Silicon Drift Detectors (SDD). 
The SDD is an intrinsically 2-dimensional sensor. The position along the beam direction is measured via collection anodes 
and the associated resolution is about $50 \, {\rm \mu m}$. The $r \phi$ coordinate is given by a drift time measurement 
with a spatial resolution of about $60 \, {\rm \mu m}$. Due to drift field non-uniformities, which were not corrected for in the 2010 data, a systematic uncertainty of $300 \, {\rm \mu m}$ is assigned to the SDD points.
Finally, the two outer layers are made of double-sided Silicon micro-Strip Detectors (SSD) with a position
resolution of $20 \, {\rm \mu m}$ in $r \phi$ and about $800 \, {\rm \mu m}$ along the beam direction.
The material budget of all six layers including support and services amounts to 7.7\% of a radiation length.

The main tracking device of ALICE is the Time Projection Chamber that covers the pseudorapidity range of about $|\eta| < 0.9$ for tracks traversing the maximum radius.
In order to avoid border effects, the fiducial region has been restricted in this analysis to $|\eta| <  0.8$. 
The position resolution along the $r \phi$ coordinate varies from $1100 \, {\rm \mu m}$ 
at the inner radius to $800\, {\rm \mu m}$ at the outer. 
The resolution along the beam axis ranges from $1250\, {\rm \mu m}$ to $1100\, {\rm \mu m}$. 

For the evaluation of the detector performance we use events generated with the \textsc{Pythia} 6.4 \cite{pythia6} Monte Carlo with tune Perugia-0 \cite{perugia0} passed through a full detector simulation based on \textsc{Geant3} \cite{geant3}. The same reconstruction algorithms are used for simulated and real data.

%% file: data_samples.tex
\section{Data samples}
\label{data_samples}
The analysis uses two data sets which were taken at the center-of-mass energies of $\sqrt{s} =$ 0.9 and \unit[7]{TeV}.
In May 2010, ALICE recorded about 6 million good quality minimum-bias events at $\sqrt{s} =$ \unit[0.9]{TeV}. 
The luminosity was of the order of 10$^{26}$~cm$^{-2}$~s$^{-1}$ and, thus, the probability 
for pile-up events in the same bunch crossing was negligible. 
The $\sqrt{s} =$ \unit[7]{TeV} \, sample of about 25 million events was collected in April 2010 with a luminosity of $10^{27}$~cm$^{-2}$~s$^{-1}$. 
In this case the mean number of interactions per bunch crossing $\mu$ ranges from 0.005 to 0.04. 
A set of high pile-up probability runs ($\mu=0.2-2$) was analysed in order to study our pile-up rejection procedure and determine its related uncertainty. 
Those runs are excluded from the analysis.

Corrected data are compared to three Monte Carlo models: \textsc{Pythia} 6.4 (tune Perugia-0), \textsc{Pythia} 8.1  (tune 1 \cite{pythia8}) and \textsc{Phojet} 1.12.

%% file: event_track_sel.tex

	\btabonecol[t]
	\begin{tabular}{|c|c|c|}
	\hline
	\multicolumn{3}{|c|}{\textbf{Collision energy: 0.9 TeV}}\\
    \hline
	& \textbf{Events} &  \textbf{\% of all } \\
	\hline
	Offline trigger & 5,515,184 & 100.0 \\
	\hline
	Reconstructed vertex & 4,482,976 & 81.3 \\
	\hline
	Leading track $\pT>0.15$ \gmom & 4,043,580 & 73.3 \\
	\hline	
	Leading track $\pT>0.5$ \gmom & 3,013,612 & 54.6 \\
	\hline	
	Leading track $\pT>1.0$ \gmom  & 1,281,269 & 23.2\\
	\hline	
	\hline
	\multicolumn{3}{|c|}{\textbf{Collision energy: 7 TeV}}\\
	\hline
	& \textbf{Events} &  \textbf{\% of all } \\
	\hline
	Offline trigger & 25,137,512 & 100.0 \\
	\hline
	Reconstructed vertex & 22,698,200 &  90.3\\
	\hline
	Leading track $\pT>0.15$ \gmom & 21,002,568 & 83.6 \\
	\hline
		Leading track $\pT>0.5$ \gmom & 17,159,249 & 68.3 \\
	\hline		
	    Leading track $\pT>1.0$ \gmom & 9,873,085 & 39.3 \\
	\hline		
	\end{tabular}	
	
	\caption{\textit{Events remaining after each event selection step.}}
	\label{events}
	\etabonecol

\section{Event and track selection}
\label{event_track_sel}

\subsection{Trigger and offline event selection}
\label{event_sel}
Events are recorded if either of the three triggering systems, V0A, V0C or SPD, has a signal. The arrival time of particles in the V0A and V0C are used to reject beam--gas interactions that occur outside the nominal interaction region.
A more detailed description of the online trigger can be found in~\cite{alice3}. An additional offline selection is made following the same criteria but considering reconstructed information instead of online trigger signals.

For each event a reconstructed vertex is required. The vertex reconstruction procedure is based on tracks as well as signals in the SPD. Only vertices within 
$\pm 10$ cm of the nominal interaction point along the beam axis are considered.
Moreover, we require at least one track with $\pT >$ $ p_{\rm T,min} =$ 0.15, 0.5 or \unit[1.0]{\gmom} \ in the acceptance $| \eta|<0.8$.

A pile-up rejection procedure is applied to the set of data taken at $\sqrt{s} =$ \unit[7]{TeV}: events with more than one distinct reconstructed primary vertex are rejected. 
This cut has a negligible effect on simulated events without pile-up: only 0.06\% of the events are removed. 
We have compared a selection of high pile-up probability runs (see Section~\ref{data_samples}) with a sample of low pile-up probability runs. The UE distributions differ by 20-25\% between the two samples. After the above mentioned rejection procedure, the difference is reduced to less than 2\%. Therefore, in the runs considered in the analysis, the effect of pile-up is negligible.

No explicit rejection of cosmic-ray events is applied since cosmic particles are efficiently suppressed by our track selection 
cuts~\cite{alice6}. 
This is further confirmed by the absence of a sharp enhanced correlation at $\Delta \phi = \pi$ from the leading track which would be caused by almost straight high-$\pt$ tracks crossing the detector.

Table~\ref{events} summarizes the percentage of events remaining after each event selection step.
We do not explicitly select non-diffractive events, although the above mentioned event selection significantly reduces the amount of diffraction in the sample. 
Simulated events show that the event selections reduce the fraction of diffractive events from 18-33\% to 11-16\% (\textsc{Pythia} 6.4 and \textsc{Phojet} at 0.9 and \unit[7]{TeV}). We do not correct for this contribution.


	\btabonecol[t!]
	\begin{tabular}{|c|c|}
	\hline
	\textbf{Selection criteria} &  \textbf{Value} \\
	\hline
	Detectors required & ITS,TPC\\
	\hline
	Minimum number of TPC clusters & 70\\
	\hline
	Maximum $\chi^{2}$ per TPC cluster & 4\\
	\hline
	Minimum number of ITS clusters & 3\\
	\hline
	Minimum number of SPD or $1^{st}$ layer SDD clusters & 1\\
	\hline
	Maximum $DCA_{Z}$ & 2~cm\\
	\hline
	Maximum $DCA_{XY}(\pT)$ & 7$\sigma$\\
	\hline
	\end{tabular}	
	
	\caption{\textit{Track selection criteria.}}
	\label{track_cuts}
	\etabonecol
 
\subsection{Track cuts}
The track cuts are optimized to minimize the contamination from secondary tracks. 
For this purpose a track must have at least 3 ITS clusters, one of which has to be in the first 3 layers.
Moreover, we require at least 70 (out of a maximum of 159) clusters in the TPC drift volume. 
The quality of the track fitting measured in terms of the $\chi^{2}$ per space point is required to be lower than 4 (each space point having 2 degrees of freedom). 
We require the distance of closest approach of the track to the primary vertex along the beam axis (DCA$_{Z}$) to be smaller than 2~cm. 
In the transverse direction we apply a $\pT$ dependent DCA$_{XY}$ cut, corresponding to 7~standard deviations of its inclusive probability distribution.
These cuts are summarized in Table~\ref{track_cuts}.

%% file: analysis.tex
\section{Analysis strategy}
\label{analysis}
The Underlying Event activity is characterized by the following observables \cite{cdf1}:
\begin{itemize}
        \item average charged particle density vs. leading track transverse momentum $p_{\rm T,LT}$:
        \begin{equation}  
        \label{eq_numbdens}      
        \frac{1}{\Delta\eta \cdot \Delta\Phi} \frac{1}{N_{\rm ev}(p_{\rm T,LT})}N_{\rm ch}(p_{\rm T,LT})
        \end{equation}
        \item average summed $\pT$ density vs. leading track $p_{\rm T,LT}$:
		\begin{equation} 
		\label{eq_sumpt}       
        \frac{1}{\Delta\eta \cdot \Delta\Phi} \frac{1}{N_{\rm ev}(p_{\rm T,LT})}\sum \pT( p_{\rm T,LT})
        \end{equation}
        \item $\Delta \phi$-correlation between tracks and the leading track:
        \begin{equation}
        \label{eq_deltaphi}  
        \frac{1}{\Delta\eta}\frac{1}{N_{\rm ev}(p_{\rm T,LT})}\frac{{\rm d} N_{\rm ch}}{{\rm d}\Delta\phi }
         \end{equation} 
         (in bins of leading track $p_{\rm T,LT}$).
    \end{itemize} 
$N_{\rm ev}$ is the total number of events selected and $N_{\rm ev}(p_{\rm T,LT})$ is the number of events 
in a given leading-track transverse-momentum bin. 
The first two variables are evaluated in three distinct regions. 
These regions, illustrated in Fig.~\ref{regions}, are defined with respect to the leading track azimuthal angle:
    \begin{itemize}
        \item Toward: $|\Delta \phi| < 1/3$ $\pi$
        \item Transverse: $1/3$ $\pi< |\Delta \phi| < 2/3$ $\pi$
        \item Away: $|\Delta \phi| > 2/3$ $\pi$
    \end{itemize}
where $\Delta \phi = \phi_{LT} - \phi$ is defined in $\pm \pi$.
In Eq. (\ref{eq_numbdens})-(\ref{eq_deltaphi}) the normalization factor $\Delta\Phi$ is equal to $2/3 \pi$, which is the size of each region.
$\Delta \eta =$~1.6 
corresponds to the acceptance in pseudorapidity. 
The leading track is not included in the final distributions.

    \bfig[t!]
    	
       \includegraphics[width=10cm]{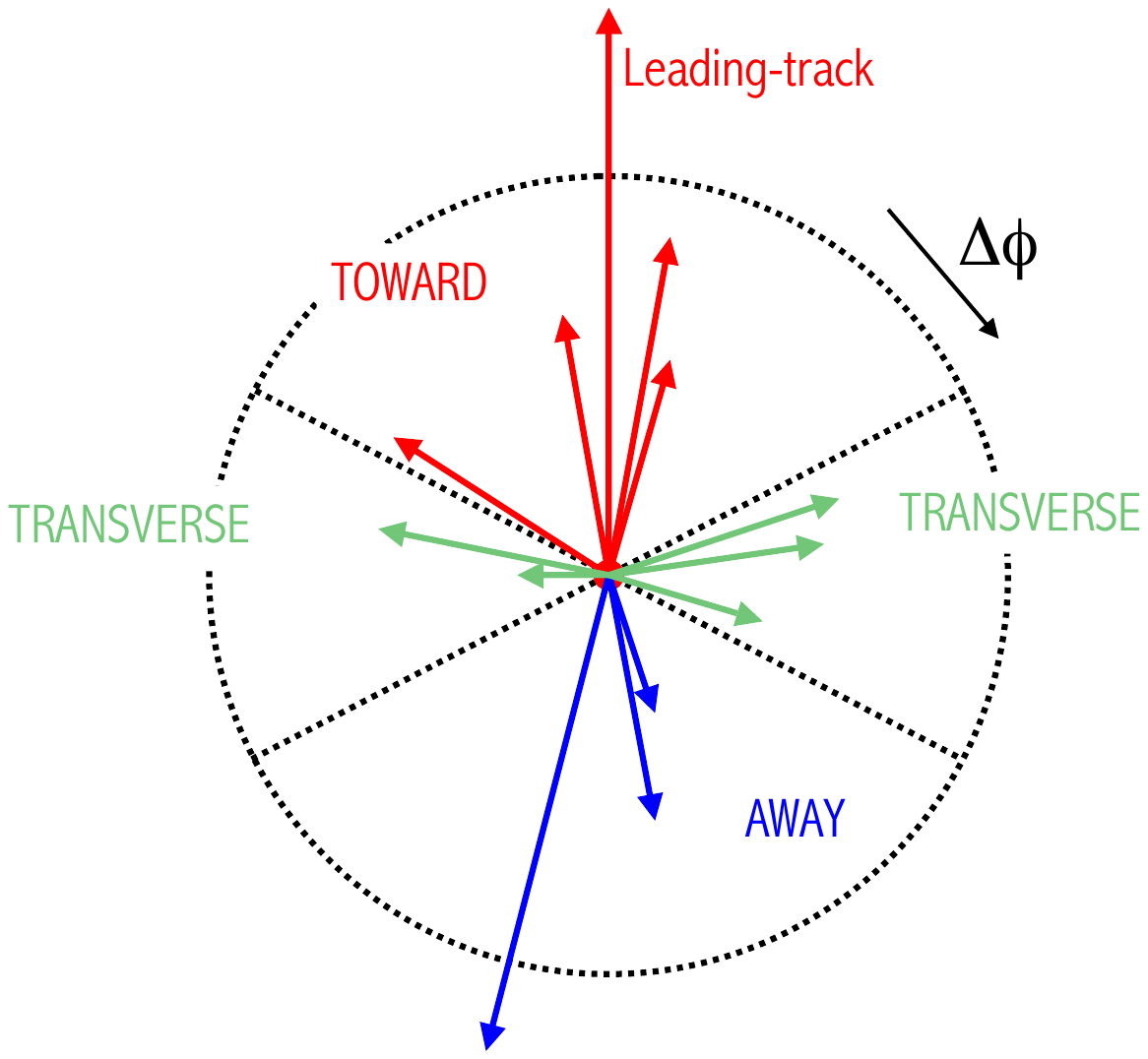}
            
       \caption{\textit{Definition of the regions Toward, Transverse and Away w.r.t. leading track direction.}}
       \label{regions}
    \efigNoLine

%% file: corrections.tex

\section{Corrections}
\label{corrections}
We correct for the following detector effects: vertex reconstruction efficiency, tracking efficiency, 
contamination from secondary particles and leading-track misidentification bias. 
The various corrections are explained in more detail in the following subsections.
We do not correct for the trigger efficiency since its value is basically 100\% for events which have at least one particle with $\pT>$ \unit[0.15]{\gmom} in the range $|\eta|<0.8$.
In Table ~\ref{table_corrections} we summarize the maximum effect of each correction on  the measured
final observables at the two collision energies for $p_{\rm T,min} = \unit[0.5]{\gmom}$.

\paragraph{Vertex reconstruction}
The correction for finite vertex reconstruction efficiency is performed as a function of the measured multiplicity. 
Its value is smaller than 0.7\% and 0.3\% at $\sqrt{s}=$ 0.9 and $\sqrt{s}=\unit[7]{TeV}$, respectively.
\paragraph{Tracking efficiency}
The tracking efficiency depends on the track level observables $\eta$ and $\pT$. The projections of the tracking efficiency on the $\pT$ and $\eta$~axes are shown in Fig.~\ref{efficiency}.
In the pseudorapidity projection we observe a dip of about 1\% at  $\eta=0$ due to the central TPC cathode. The slight asymmetry between positive and negative $\eta$  is due to a different number of active SPD and SDD modules in the two halves of the detector. The number of active modules also differs between the data-taking periods at the two collision energies.
Moreover, the efficiency decreases by 5\% in the range \unit[1-3]{\gmom}. 
This is due to the fact that above about \unit[1]{\gmom} tracks are almost straight and can be contained completely in the 
dead areas between TPC sectors. Therefore, at high $\pT$ the efficiency is dominated by geometry 
and has a constant value of about 80\% at both collision energies. 
To avoid statistical fluctuations, the estimated efficiency is fitted with a constant for $\pT > \unit[5]{\gmom}$ (not shown in the figure).
\paragraph{Contamination from secondaries}
\label{cont_corr}
We correct for secondary tracks that pass the track selection cuts. 
Secondary tracks are predominantly produced by weak decays of strange particles (e.g. $K^{0}_{S}$ and $\Lambda$), 
photon conversions or hadronic interactions in the detector material, and decays of charged pions. 
The relevant track level observables for the contamination correction are transverse momentum and pseudorapidity. 
The correction is determined from detector simulations and is found to be 15-20\% for tracks with  
$\pT < $ \unit[0.5]{\gmom} and saturates at about 2\% for higher transverse momenta (see Fig.~\ref{contamination}). 
\btabonecol[t]
	\begin{tabular}{|c|c|c|}
	\hline
	\textbf{Correction} & \textbf{$\sqrt{s}=0.9$ TeV} & \textbf{$\sqrt{s}= 7$ TeV}\\
	\hline
	Leading track misidentification & $< 5$\% & $< 8$\% \\
	\hline
	Contamination & $< 3$\% & $< 3$\% \\
	\hline
	Efficiency & $< 19$\% & $< 19$\% \\
	\hline
	Vertex reconstruction & $< 0.7$\% & $< 0.3$\% \\
	\hline
	\end{tabular}	
	
	\caption{\textit{Maximum effect of corrections on final observables for $p_{\rm T,min} = \unit[0.5]{\gmom}$.}}
	\label{table_corrections}
	\etabonecol

We multiply the contamination estimate by a data-driven coefficient to take into account 
the  low strangeness yield in the Monte Carlo compared to data \cite{alice7}. The coefficient is derived from a fit of the discrepancy between data and Monte Carlo strangeness yields in the tails of the DCA$_{XY}$ distribution which are predominantly populated by secondaries.
The factor has a maximum value of 1.07 for tracks with  $\pT < $ \unit[0.5]{\gmom} 
and is equal to 1 for $\pT > $ \unit[1.5]{\gmom}. 
This factor is included in the Contamination entry in Table ~\ref{table_corrections}.
\paragraph{Leading-track misidentification}
Experimentally, the real leading track can escape detection because of tracking inefficiency and the detector's finite acceptance. 
In these cases another track (i.e. the sub-leading or sub-sub-leading etc.) will be selected as the leading one, thus
biasing the analysis in two possible ways.
Firstly, the sub-leading track will have a different transverse momentum than the leading one. 
We refer to this as leading-track $\pT$ bin migration. 
It has been verified with Monte Carlo that this effect is negligible due to the weak dependence of the final distributions
on $p_{\rm T,LT}$.
Secondly, the reconstructed leading track might have a significantly different orientation with respect to 
the real one, resulting in a rotation of the overall event topology. 
The largest bias occurs when the misidentified leading track falls in the Transverse region defined by the real leading track.

We correct for leading-track misidentification with a data-driven procedure.
Starting from the measured distributions, for each event the track loss due to
inefficiency is applied a second time to the data 
(having been applied the first time naturally by the detector) by rejecting tracks randomly. 
If the leading track is considered reconstructed it is used as before to define the  different regions. 
Otherwise the sub-leading track is used. 
Since the tracking inefficiency is quite small (about 20\%) applying it on the reconstructed data a second time 
does not alter the results significantly. 
To verify this statement we compared our results with a two step procedure. 
In this case the inefficiency is applied two times on measured data, half of its value at a time. 
The correction factor obtained in this way is compatible with the one step procedure.
Furthermore, the data-driven procedure has been tested on simulated data where the true leading particle
is known. 
We observed a discrepancy between the two methods, especially at low leading-track $\pT$ values, which is taken into account in 
the systematic error. The maximum leading-track misidentification correction is 8\% on the final distributions.
\twoPlotsOneCol[t]{\includegraphics[width=8cm]{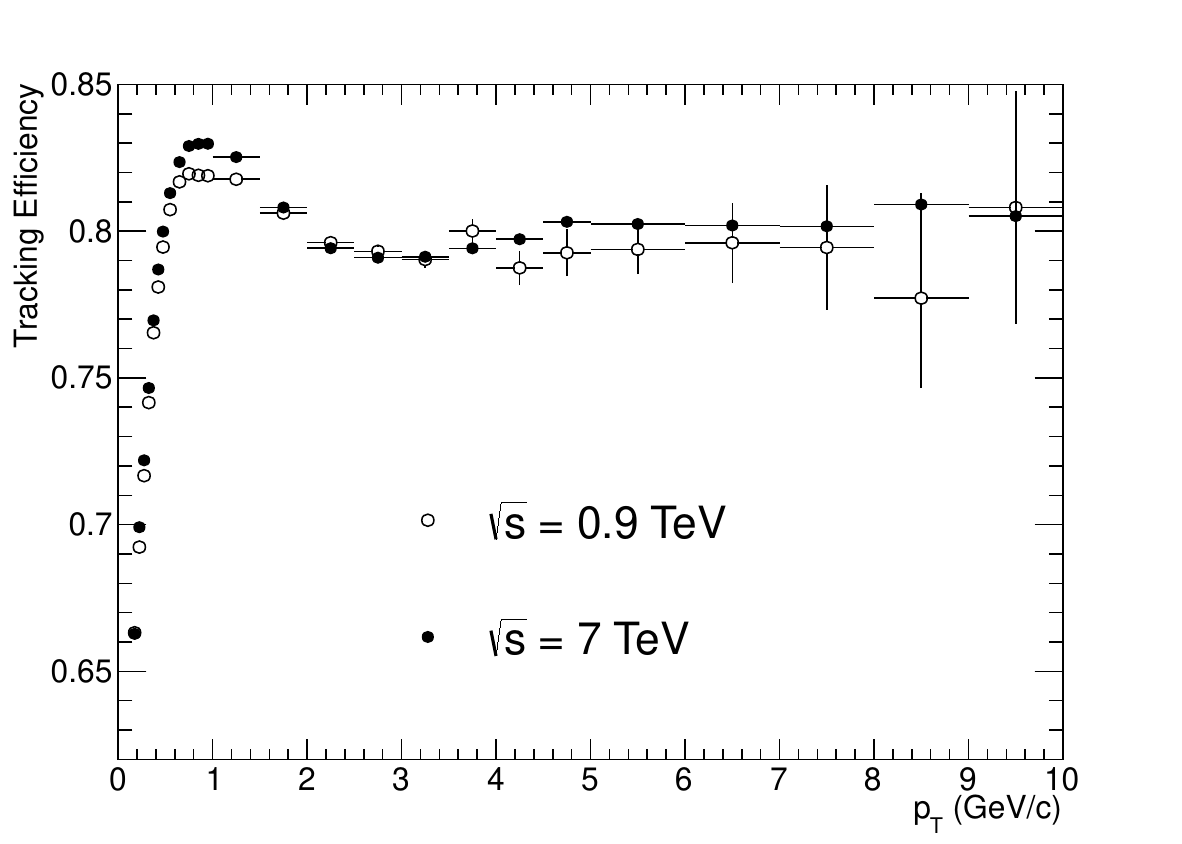}} {\includegraphics[width=8cm]{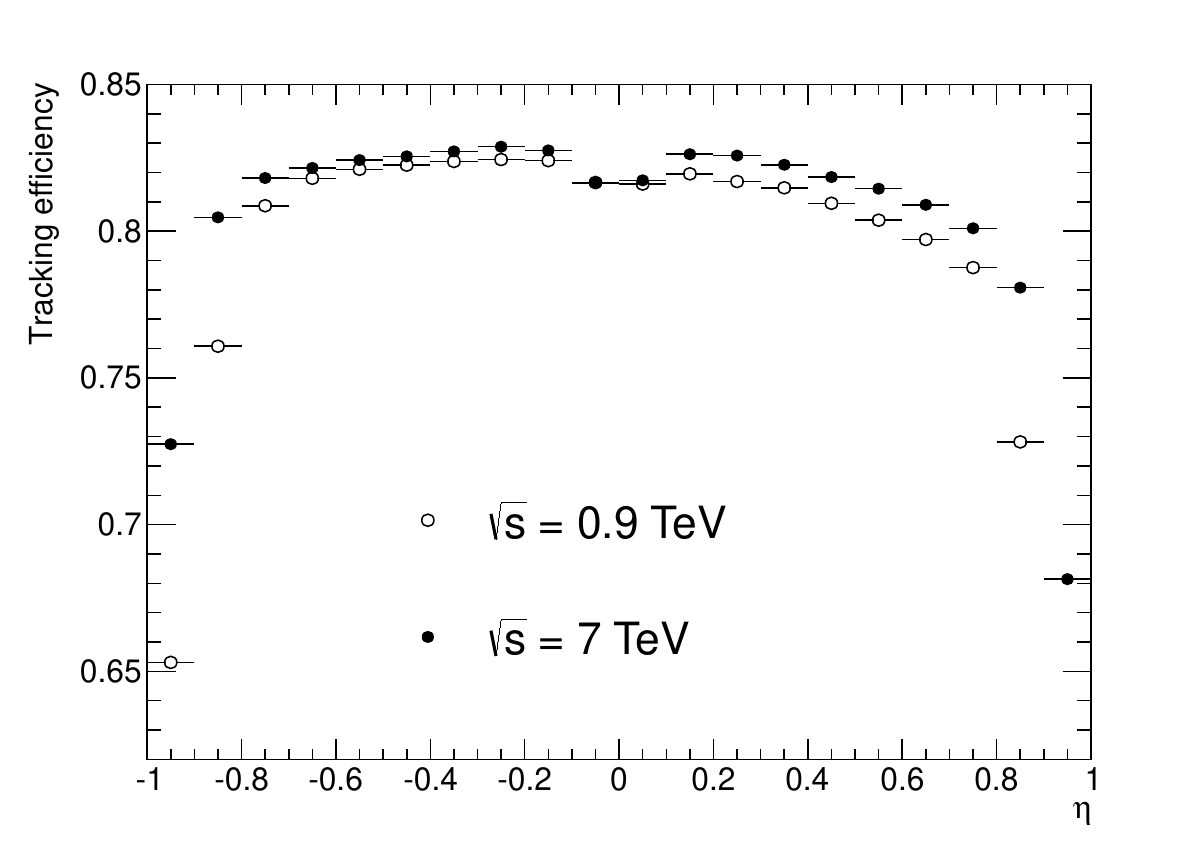}}
{\caption{\textit{Tracking efficiency vs. track $\pT$ (left, $|\eta|<0.8$) and $\eta$ (right, $\pT > \unit[0.5]{\gmom}$) from a \textsc{Pythia}~6.4 and \textsc{Geant3} simulation.}} \label{efficiency} }
\paragraph{Two-track effects}
By comparing simulated events corrected for single-particle efficiencies with the input Monte Carlo, we observe a 0.5\% discrepancy around $\Delta \phi=0$. This effect is called non-closure in Monte Carlo (it  will be discussed further in Section \ref{systematics}) and in this case is related to small two-track resolution effects. Data are corrected for this discrepancy.

%% file: systematics.tex
\section{Systematic uncertainties}
\label{systematics}

In Tables~\ref{syst_table_const}, \ref{syst_table_900} and \ref{syst_table_7000} we summarize the systematic 
uncertainties evaluated in the analysis for the three  track thresholds: 
$\pT > 0.15$, 0.5 and \unit[1.0]{\gmom}. 
Each uncertainty is explained in more detail in the following subsections.
Uncertainties which are constant as a function of
 leading-track $\pT$ are listed in Table~\ref{syst_table_const}.
Leading-track $\pT$ dependent uncertainties are summarized in Tables~\ref{syst_table_900} and \ref{syst_table_7000} for $\sqrt{s}=\unit[0.9]{TeV}$ and $\unit[7]{TeV}$, respectively. Positive and negative uncertainties are propagated separately, resulting in asymmetric final uncertainties.
\paragraph{Particle composition}
The tracking efficiency and contamination corrections depend slightly on the particle species 
mainly due to their decay length and absorption in the material. 
To assess the effect of an incorrect description of the particle abundances in the Monte Carlo, we varied the relative yields 
of pions, protons, kaons, and other particles by 30\% relative to the default Monte Carlo predictions.
The maximum variation of the final values is 0.9\% and 
represents the systematic uncertainty related to the particle composition (see Table~\ref{syst_table_const}).

Moreover, we have compared our assessment of the underestimation of strangeness  yields with a direct measurement from the ALICE collaboration \cite{alice7}. Based on the discrepancy between the two estimates, we assign a systematic uncertainty of 0-2.3\% depending on the $\pT$ threshold and collision energy, see Tables~\ref{syst_table_900} and~\ref{syst_table_7000}.
\twoPlotsOneCol [t]{\includegraphics[width=8cm]{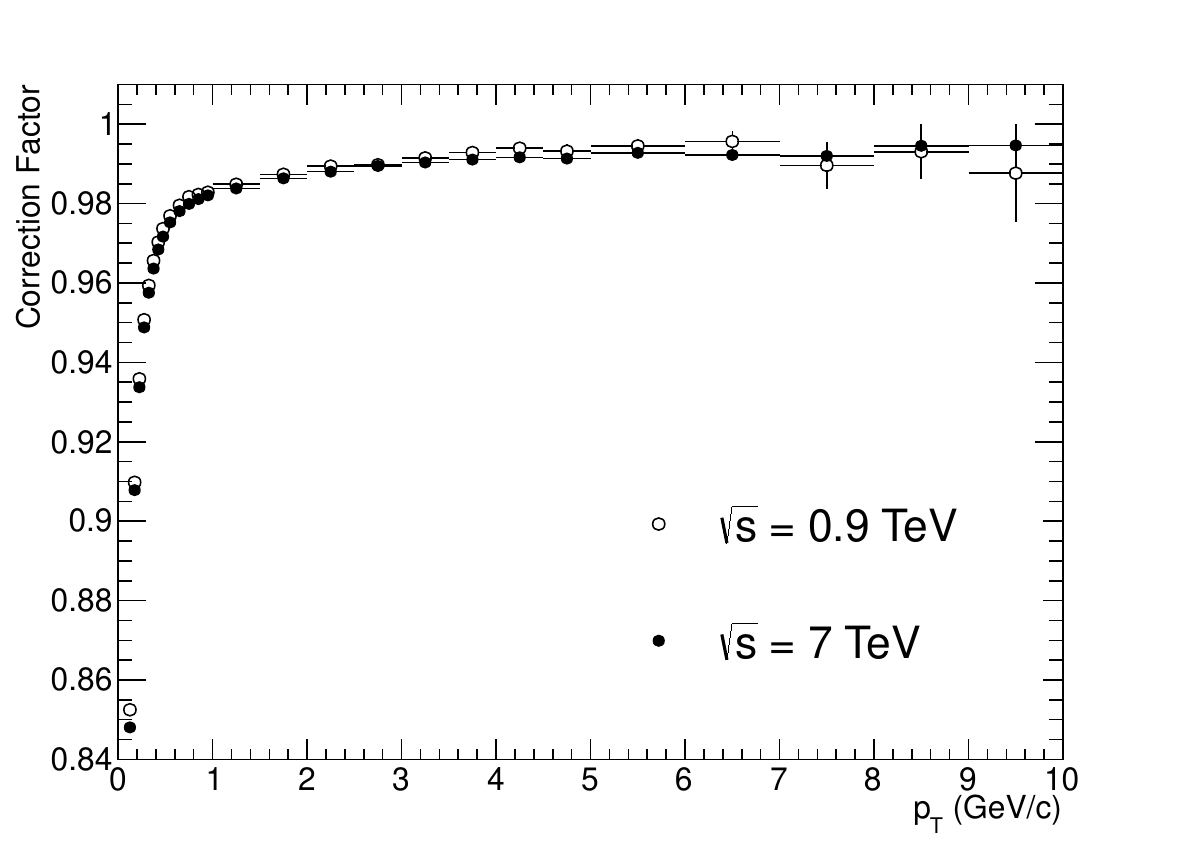}} {\includegraphics[width=8cm]{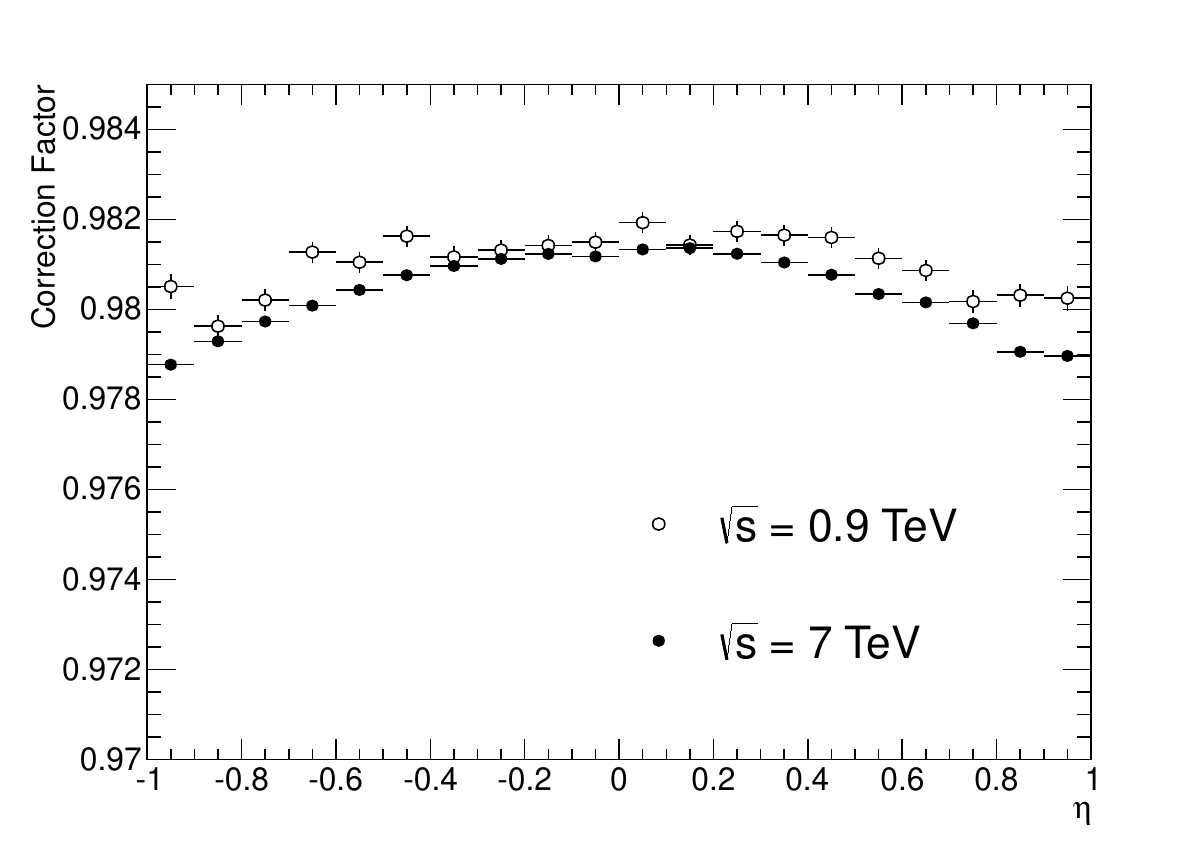}}
{\caption{\textit{Contamination correction: correction factor vs. track $\pT$ (left, $|\eta|<0.8$) and $\eta$ (right, $\pT > \unit[0.5]{\gmom}$) from a  \textsc{Pythia}~6.4 and \textsc{Geant3} simulation.}} \label{contamination} }	
	
\paragraph{ITS and TPC efficiency}
The tracking efficiency 
depends on the level of precision of the description of the ITS and TPC detectors in the simulation and the modeling of their response. 
After detector alignment with survey methods, cosmic-ray events and pp collision events \cite{alignment}, the uncertainty on the  efficiency due to the ITS description is estimated to be below 2\% and  affects only tracks with $\pT <$ \unit[0.3]{\gmom}. 
The uncertainty due to the TPC reaches 4.5\% at very low $\pT$ and is smaller than 1.2\% for tracks with $\pT >$ \unit[0.5]{\gmom}. 
The resulting maximum uncertainty on the final distributions is below 1.9\%.
Moreover, an uncertainty of 1\% is included to account for uncertainties in the MC description of the matching between TPC and ITS tracks (see Table~\ref{syst_table_const}).
	\begin{table*}[t!]
	\begin{center}
	\footnotesize
	\begin{tabular}{|c||c|c|c|}
	\hline
	 &\multicolumn{3}{|c|}{\textbf{ $\sqrt{s}=0.9$ TeV}}\\
	\hline
	 &$\pT > 0.15 \,\gmom$ & $\pT > 0.5 \, \gmom$ & $\pT > 1.0 \, \gmom$  \\
	\hline
	\textbf{Particle composition} & $\pm$ 0.9\% & $\pm$ 0.7\%  & $\pm$ 0.4\%   	\\
	\hline
	\textbf{ITS efficiency}   & $\pm$ 0.6\%  & -- & --     \\
	\hline
	\textbf{TPC efficiency}  & $\pm$ 1.9\%  & $\pm$ 0.8\%  & $\pm$ 0.4\%        \\
	\hline
	\textbf{Track cuts}     & $^{+\ 3.0\%}_{-\ 1.1\%}$ & $^{+\ 2.0\%}_{-\ 1.1\%}$ & $^{+\ 0.9\%}_{-\ 1.5\%}$       \\
	\hline
	\textbf{ITS/TPC matching}  & $\pm$ 1.0\%  & $\pm$ 1.0\%  &  $\pm$ 1.0\%   \\
	\hline
    \textbf{MC dependence} & $+$ 1.1\% ,  $+$ 1.1\% ,  $+$ 1.6\% & $+$ 0.9\% & $+$ 0.9\% ,  $+$ 0.9\% ,  $+$ 1.3\%        \\
	\hline	
	    \textbf{Material budget}   & $\pm$ 0.6\% &  $\pm$ 0.2\% & $\pm$ 0.2\%        \\
	\hline	
	\hline
	 &\multicolumn{3}{|c|}{\textbf{ $\sqrt{s}=7$ TeV}}\\
	\hline
	 &$\pT > 0.15 \, \gmom$ &$\pT > 0.5 \, \gmom$ & $\pT > 1.0 \, \gmom$  \\
	\hline
	\textbf{Particle composition} & $\pm$ 0.9\% & $\pm$ 0.7\%  & $\pm$ 0.5\%   	\\
	\hline
	\textbf{ITS efficiency}   & $\pm$ 0.5\%  & -- & --     \\
	\hline
	\textbf{TPC efficiency}  & $\pm$ 1.8\%  & $\pm$ 0.8\%  & $\pm$ 0.5\%        \\
	\hline
	\textbf{Track cuts}      & $^{+\ 2.1\%}_{-\ 2.3\%}$ & $^{+\ 1.6\%}_{-\ 3.2\%}$ & $^{+\ 2.5\%}_{-\ 3.5\%}$       \\
	\hline
	\textbf{ITS/TPC matching}  & $\pm$ 1.0\%  & $\pm$ 1.0\%  &  $\pm$ 1.0\%   \\
	\hline
    \textbf{MC dependence}  & $+$ 0.8\% ,  $+$ 0.8\% ,  $+$ 1.2\% & $+$ 0.8\% & $+$ 1.0\%    \\
	\hline	
		    \textbf{Material budget}  & $\pm$ 0.6\% &  $\pm$ 0.2\% & $\pm$ 0.2\%        \\
	\hline	
	\end{tabular}	
	
	\caption{\textit{Constant systematic uncertainties at both collision energies. When more than one number is quoted, separated by a comma, the first value refers to the number density distribution, the second to the summed $\pT$ and the third to the azimuthal correlation. Some of the uncertainties are quoted asymmetrically.}}
	\label{syst_table_const}
	\end{center}
	\end{table*}

\paragraph{Track cuts}
By applying the efficiency and contamination corrections we correct for those particles which are lost due to detector effects 
and for secondary tracks which have not been removed by the selection cuts.
These corrections rely on detector simulations and therefore, one needs to estimate the systematic uncertainty 
introduced in the correction procedure by one particular choice of track cuts. 
To do so, we repeat the analysis with different values of the track cuts, both for simulated and real data. 
The variation of the final distributions with different track cuts is a measure of the systematic uncertainty.
The overall effect, considering all final distributions, is smaller than 3.5\% at both collision energies (see Table~\ref{syst_table_const}).
\paragraph{Misidentification bias}
The uncertainty on the leading-track misidentification correction is estimated  from the discrepancy between the data-driven correction 
used in the analysis and that based on simulations. 
The effect influences only the first two leading-track $\pT$~bins at both collision energies. The maximum uncertainty ($\sim 18\%$) affects the first leading-track $\pT$~bin for the track $\pT$ cut-off of \unit[0.15]{GeV/$c$}. In all other bins this uncertainty is of the order of few percent.
As summarized in Tables~\ref{syst_table_900} and~\ref{syst_table_7000}, the uncertainty has slightly different values 
for the various UE distributions. 
\paragraph{Vertex-reconstruction efficiency}
The analysis accepts reconstructed vertices with at least one contributing track. 
We repeat the analysis requiring at least two contributing tracks. 
The systematic uncertainty related to the vertex reconstruction efficiency is given by the maximum variation 
in the final distributions between the cases of one and two contributing tracks. Its value is 2.4\% for $p_{\rm T, min}=$ \unit[0.15]{GeV/$c$} and below 1\% for the other cut-off values  (see Tables~\ref{syst_table_900} and~\ref{syst_table_7000}).
The effect is only visible in the first leading-track $\pT$~bin.
\paragraph{Non-closure in Monte Carlo}
By correcting a Monte Carlo prediction after full detector simulation with corrections extracted from the same generator, 
we expect to obtain the input Monte Carlo prediction within the statistical uncertainty. 
This consideration holds true only if each correction is evaluated  with respect to all the variables 
to which the given correction is sensitive. 
Any statistically significant difference between input and corrected distributions is referred to as 
\textit{non-closure in Monte Carlo}. 

The overall non-closure effect is sizable ($\sim 17\%$) in the first leading-track $\pT$ bin and is 0.6-5.3\% in all other bins at both collision energies.
\paragraph{Monte-Carlo dependence}
The difference in final distributions when correcting the data with \textsc{Pythia}~6.4 or \textsc{Phojet} generators is of the order of 1\% and equally affects all the leading-track $\pT$  bins.

\paragraph{Material budget}
The material budget has been measured by reconstructing photon conversions which allows a precise $\gamma$-ray tomography of the ALICE detector. For the detector regions important for this analysis the remaining uncertainty on the extracted material budget is less than 7\%.
Varying the material density in the detector simulation, the effect on the observables presented is determined to be 0.2-0.6\% depending on the $\pT$ threshold considered.
 

	\begin{table}[t]
	\begin{center}
    \footnotesize
	\begin{tabular}{|c|c||c|c|c|}
	\hline
	& &\multicolumn{3}{|c|}{\textbf{ $\sqrt{s}=0.9$ TeV}}\\
	\hline
	& &\multicolumn{3}{|c|}{\textbf{Number density}}\\
    \hline	
    & $p_{\rm T,LT}$& $\pT > 0.15 \, \gmom$ & $\pT > 0.5 \, \gmom$ & $\pT > 1.0 \, \gmom$ \\    
    \hline
    \textbf{Lead. track misid. }& $1^{st}$ bin & $+$ (17.8,  16.3, 16.3)\%   &  $+$ (4.6, 3.5, 3.5)\%  & $+$  (4.2, 2.9, 1.7)\% \\
    \hline
                       & $2^{nd}$ bin & $+$ 2.9\%  & $+$ 1.3\% & --\\
    \hline
            \textbf{MC non closure}& $1^{st}$ bin & $-$ 17.2\%  &  $-$ 3.6\%  & $-$ 1.2\%\\
        \hline
        				& $2^{nd}$ bin & $-$ 3.2\%  &  $-$ 0.8\%  & $-$ 1.2\%\\
        \hline
                       & others & $-$ 0.6\%  & $-$ 0.8\% & $-$ 1.2\%\\
    \hline
    \textbf{Strangeness }& $1^{st}$ bin & $\pm$ 1.9\%  &  $\pm$ 0.2\%  & --\\
    \hline
                       & others & $\pm$ 1.0\%   &  $\pm$ 0.2\%  & --\\
    \hline 
    \textbf{Vertex reco.}& $1^{st}$ bin & $-$ 2.4\%  &  $-$ 0.7\%  &  $-$ 0.5\% \\
    \hline 
	& &\multicolumn{3}{|c|}{\textbf{Summed $\pT$}}\\
    \hline	
    &$p_{\rm T,LT}$ & $\pT > 0.15 \, \gmom$ & $\pT > 0.5 \, \gmom$ 
 & $\pT > 1.0\, \gmom$ \\
    \hline   
    \textbf{Lead. track misid. }& $1^{st}$ bin & $+$ (20.0, 18.1, 18.1)\% &  $+$ (5.3, 4.1, 4.1)\%  & $+$ (4.8, 3.4, 3.4)\% \\
    \hline
                       & $2^{nd}$ bin & $+$ 3.7\%  & $+$ 1.6\% & --\\
    \hline 
        \textbf{MC non closure }& $1^{st}$ bin & $-$ 17.0\%  &  $-$ 2.8\%  &  $-$ 1.1\%\\
        \hline
        				& $2^{nd}$ bin & $-$ 3.0\%  &  $-$ 1.0\%  &  $-$ 1.1\%\\
        \hline
                       & others & $-$ 0.7\%  & $-$ 1.0\% & $-$ 1.1\%\\
    \hline
    \textbf{Strangeness }& $1^{st}$ bin & $\pm$ 1.9\%  &  $\pm$ 0.2\%  & --\\
    \hline
                       & others & $\pm$ 1.0\%   &  $\pm$ 0.2\%  & --\\
    \hline    
    \textbf{Vertex reco.}& $1^{st}$ bin & $-$ 2.4\%  &  $-$ 0.7\%  &  $-$ 0.5\% \\
    \hline         
	& &\multicolumn{3}{|c|}{\textbf{Azimuthal correlation}}\\
    \hline	
    &$p_{\rm T,LT}$ & $\pT > 0.15 \, \gmom$   & $\pT > 0.5 \, \gmom$  & $\pT > 1.0 \, \gmom$ \\    
    \hline	
        \textbf{Lead. track misid. }& $1^{st}$ bin & $+$ 12.0\%  &  $+$ 3.9\%  & $+$ 2.5\%\\
    \hline
                       & $2^{nd}$ bin & $+$ 2.6\%  & $+$ 1.1\% & --\\
    \hline 
            \textbf{MC non closure  }& $1^{st}$ bin & $-$ 17.1\%  &  $-$ 3.3\%  &  $-$ 1.6\%\\
                    \hline
                       & $2^{nd}$ bin & $-$ 3.5\%  & $-$ 3.0\% & $-$ 1.6\%\\
        \hline
                       & others & $-$ 2.4\%  & $-$ 3.0\% & $-$ 1.6\%\\
    \hline
    \textbf{Strangeness }& $1^{st}$ bin & $\pm$ 1.9\%  &  $\pm$ 0.2\%  & --\\
    \hline
                       & others & $\pm$ 1.0\%   &  $\pm$ 0.2\%  & --\\
    \hline 
    \textbf{Vertex reco.}& $1^{st}$ bin & $-$ 2.4\%  &  $-$ 0.4\%  & -- \\
    \hline
                       & others &  $-$ 0.5\%   &  $-$ 0.4\% & --\\
    \hline                
	\end{tabular}	
	
	\caption{\textit{Systematic uncertainties vs. leading track $\pT$ at $\sqrt{s}= \unit[0.9]{TeV}$. When more than one number is quoted, separated by a comma, the first value refers to the Toward, the second to the Transverse and the third to the Away region. The second column denotes the leading track $\pT$ bin for which the uncertainty applies. The numbering starts for each case from the first bin above the track $\pT$ threshold.}}
	\label{syst_table_900}
	\end{center}
	\end{table}

	\begin{table}[t]
	\begin{center}
	\footnotesize
	\begin{tabular}{|c|c||c|c|c|}
	\hline
	& &\multicolumn{3}{|c|}{\textbf{ $\sqrt{s}=7$ TeV}}\\
	\hline
	& &\multicolumn{3}{|c|}{\textbf{Number density}}\\
    \hline	
    &$p_{\rm T,LT}$ &$\pT > 0.15 \, \gmom$ & $\pT > 0.5 \, \gmom$
 & $\pT > 1.0 \, \gmom$ \\    
    \hline
    \textbf{Lead. track misid. }& $1^{st}$ bin & $+$ (17.9, 16.3, 16.3)\%  &  $+$ (4.0, 3.2, 3.2)\% & $+$ (2.5, 1.2, 1.2)\% \\
    \hline
                       & $2^{nd}$ bin & $+$ 2.7\%  &   -- &   $+$ 0.7\%  \\
    \hline
                \textbf{MC non closure }& $1^{st}$ bin & $-$ 16.8\%  &  $-$ 2.6\%  & $-$ 1.9\%\\
        \hline
                		& $2^{nd}$ bin & $-$ 2.9\%  &  $-$ 1.4\%  & $-$ 1.9\%\\
        \hline        
                       & others & $-$ 0.6\%  & $-$ 1.0\% & $-$ 1.9\%\\
    \hline
    \textbf{Strangeness }& $1^{st}$ bin & $\pm$ 1.8\%  &  $\pm$ 2.3\%  & --\\
    \hline
                       & others & $\pm$ 1.0\%   &  $\pm$ 2.3\%  & --\\
    \hline 
    \textbf{Vertex reco.}& $1^{st}$ bin & $-$ 2.4\%  &  $-$ 0.7\%  &  $-$ 0.5\% \\
    \hline 
	& &\multicolumn{3}{|c|}{\textbf{Summed $\pT$}}\\
    \hline	
    &$p_{\rm T,LT}$ & $\pT > 0.15 \, \gmom$   & $\pT > 0.5 \, \gmom$ 
 & $\pT > 1.0 \, \gmom$ \\
    \hline   
    \textbf{Lead. track misid. }& $1^{st}$ bin & $+$ (20.0, 17.9, 17.9)\%  &  $+$ (4.9, 3.8,  3.8)\%    & $+$ (3.4, 1.9, 1.9)\% \\
    \hline
                       & $2^{nd}$ bin & $+$ 3.4\%  & $+$ 0.8\%    &  $+$ 1.1\%  \\
    \hline 
                    \textbf{MC non closure }& $1^{st}$ bin & $-$ 16.7\%  &  $-$ 2.7\%  & $-$ 1.5\%\\
        \hline
                       & $2^{nd}$ bin & $-$ 2.6\%  &  $-$ 1.2\%  & $-$ 1.5\%\\
        \hline
                       & others & $-$ 0.8\%  & $-$ 1.0\% & $-$ 1.5\%\\
    \hline
    \textbf{Strangeness }& $1^{st}$ bin & $\pm$ 1.8\%  &  $\pm$ 2.3\%  & --\\
    \hline
                       & others & $\pm$ 1.0\%   &  $\pm$ 2.3\%  & --\\
    \hline    
    \textbf{Vertex reco.}& $1^{st}$ bin & $-$ 2.4\%  &  $-$ 0.7\%  &  $-$ 0.5\% \\
    \hline         
	& &\multicolumn{3}{|c|}{\textbf{Azimuthal correlation}}\\
    \hline	
    & $p_{\rm T,LT}$& $\pT > 0.15 \, \gmom$  & $\pT > 0.5 \, \gmom$ 
 & $\pT > 1.0 \, \gmom$ \\    
    \hline	
        \textbf{Lead. track misid. }& $1^{st}$ bin & $+$ 16.8\%  &  $+$ 3.4\%  & $+$ 0.9\%\\
    \hline
                       & $2^{nd}$ bin & $+$ 2.5\%  & --  &-- \\
    \hline 
               \textbf{MC non closure  }& $1^{st}$ bin & $-$ 25.3\%  &  $-$ 4.3\%  &  $-$ 1.2\%\\
                    \hline
                       & $2^{nd}$ bin & $-$ 5.3\%  & $-$ 2.1\% & $-$ 1.2\%\\
        \hline
                       & others & $-$ 2.1\%  & $-$ 2.1\% & $-$ 1.2\%\\
    \hline
    \textbf{Strangeness }& $1^{st}$ bin & $\pm$ 1.8\%  &  $\pm$ 2.3\%  & --\\
    \hline
                       & others & $\pm$ 1.0\%   &  $\pm$ 2.3\%  & --\\
    \hline 
    \textbf{Vertex reco.}& $1^{st}$ bin & $-$ 2.4\%  &  $-$ 0.4\%  & -- \\
    \hline
                       & others &  $-$ 0.5\%   &  $-$ 0.4\% & --\\
    \hline                
	\end{tabular}	
	
	\caption{\textit{Systematic uncertainties vs. leading track $\pT$ at $\sqrt{s}= \unit[7]{TeV}$. When more than one number is quoted, separated by a comma, the first value refers to the Toward, the second to the Transverse and the third to the Away region. The second column denotes the leading track $\pT$ bin for which the uncertainty applies. The numbering starts for each case from the first bin above the track $\pT$ threshold.}}
	\label{syst_table_7000}
	\end{center}
	\end{table}


%% file: results.tex
\section{Results}
\label{results}
In this section we present and discuss the corrected results for the three UE distributions in all regions 
at the two collision energies.
The upper part of each plot shows the relevant measured distribution (black points) compared to a set of Monte Carlo 
predictions (coloured curves). 
Shaded bands represent the systematic uncertainty only. 
Error bars along the $x$ axis indicate the bin width.
The lower part shows the ratio between Monte Carlo and data. 
In this case the shaded band is the sum in quadrature of statistical and systematic uncertainties.

The overall agreement of data and simulations is of the order of 10-30\%
and we were not able to identify a preferred model that can reproduce all measured observables. 
In general, all three generators underestimate the event activity in the Transverse region.
Nevertheless, an agreement of the order of 20\% has to be considered a success, considering the complexity of the system under study. Even though an exhaustive comparison of data with the latest models available is beyond the scope of this paper, in the next sections we will indicate some general trends observed in the comparison with the chosen models. 

In the following discussion we define the leading track $\pT$ range from 4 to \unit[10]{\gmom} \  at $\sqrt{s}=\unit[0.9]{TeV}$  and from 10 to \unit[25]{\gmom}   at $\sqrt{s}=\unit[7]{TeV}$ as the \textit{plateau}.
	\btabonecol[t]
	\begin{tabular}{|c|c|c|c|c|}
   	\hline
   	&\multicolumn{4}{|c|}{\textbf{$\sqrt{s}=0.9$ TeV}} \\
   	\hline
   	& \multicolumn{2}{|c|}{\textbf{Number density }}& \multicolumn{2}{|c|}{\textbf{Summed $\pT$}} \\  	
   	\hline
   	& \textbf{Slope $(\gmom)^{-1}$} &\textbf{Mean}  & \textbf{Slope } &\textbf{Mean $(\gmom)$}  \\  
   	   	\hline
   	$\pT  >$ 0.15 GeV/c &  0.00 $\pm$ 0.02 &   1.00 $\pm$ 0.04  & 0.00	$\pm$ 0.01 & 0.62	$\pm$ 0.02 \\
   	\hline
   	$\pT >$ 0.5 GeV/c &   0.00 $\pm$ 	0.01  & 0.45 $\pm$	 0.02 &  0.01 $\pm$ 	0.01 & 0.45 $\pm$	0.02 \\
   	\hline
   	$\pT >$ 1.0 GeV/c & 0.003 $\pm$ 0.003 & 0.16	$\pm$ 0.01 & 0.006	$\pm$ 0.005 & 0.24	$\pm$ 0.01 \\
   	\hline
   	   	\hline
   	&\multicolumn{4}{|c|}{\textbf{$\sqrt{s}=7$ TeV}} \\
   	\hline
   	& \multicolumn{2}{|c|}{\textbf{Number density }}& \multicolumn{2}{|c|}{\textbf{Summed $\pT$}} \\  	
   	\hline
   	& \textbf{Slope $(\gmom)^{-1}$} &\textbf{Mean} &   \textbf{Slope } &\textbf{Mean $(\gmom)$} \\  
   	   	\hline
   	$\pT >$ 0.15 GeV/c & 0.00	 $\pm$ 0.01 & 1.82 $\pm$ 	0.06 &  0.01	$\pm$  0.01 & 1.43 $\pm$ 	0.05 \\
   	\hline
   	$\pT >$ 0.5 GeV/c &  0.005	$\pm$ 0.007 & 0.95 $\pm$ 	0.03 &  0.01 $\pm$ 	0.01 & 1.15	$\pm$  0.04 \\
   	\hline
   	$\pT >$ 1.0 GeV/c &  0.001 $\pm$	0.003 & 0.41 $\pm$	0.01 & 0.008 $\pm$	0.006 & 0.76 $\pm$	0.03\\
   	\hline   
   	   	\hline
   	&\multicolumn{4}{|c|}{\textbf{$\sqrt{s}=1.8$ TeV (CDF)}} \\
   	\hline
   	&\multicolumn{4}{|c|}{ \textbf{Number density (at leading charged jet $\pT=\unit[20]{\gmom}$)}}\\  	
   	\hline
   	$\pT >$ 0.5 GeV/c &\multicolumn{4}{|c|}{ 0.60 }\\
   	\hline   		
    \end{tabular} 

	\caption{\textit{Saturation values in the Transverse region for the two collision energies. The result from CDF is also given, for details see text.}}
    \label{saturation}
	\etabonecol

\subsection{Number density}
In Fig.~\ref{numbdens_1_away}-\ref{numbdens_3_away} we show the multiplicity density as a function of leading track $\pT$ 
in the three regions: Toward, Transverse and Away. 
Toward and Away regions are expected to collect the fragmentation products of the two back-to-back outgoing partons 
from the elementary hard scattering. 
We observe that the multiplicity density in these regions increases monotonically with the $p_{\rm T,LT}$ scale.
In the Transverse region, after  a monotonic increase at low leading track $\pT$, 
the distribution tends to flatten out. 
The same behaviour is observed at both collision energies and all values of $p_{\rm T,min}$.

The rise with $p_{\rm T,LT}$ has been interpreted as evidence for an impact parameter dependence in the hadronic collision \cite{mpi_impact_param}.
More central collisions have an increased probability for MPI, leading to a larger transverse multiplicity. 
Nevertheless, we must be aware of a trivial effect also contributing to the low $p_{\rm T,LT}$ region. For instance
for any probability distribution, the maximum value 
per randomized sample averaged over many samples
rises steadily with the sample size $M$.
In our case, the conditional probability density $\mathcal{P}(p_{\rm T,LT}|M)$ shifts towards larger $p_{\rm T,LT}$ with increasing $M$.
Using Bayes' theorem one expects the conditional probability density
$\mathcal{P} (M|p_{\rm T,LT})$ to shift towards larger $M$ with rising $p_{\rm T,LT}$:
\begin{equation}
\mathcal{P} (M|p_{\rm T,LT}) \sim \mathcal{P} (p_{\rm T,LT} | M) \mathcal{P}(M).
\end{equation}

The saturation of the distribution at higher values of  $p_{\rm T,LT}$ indicates the onset
of the event-by-event partitioning into azimuthal regions containing the particles from the hard scattering
and the UE region. The bulk particle production becomes independent of the hard scale.

The plateau range is fitted with a line.  The fit slopes, consistent with zero, and mean values for the three  $\pT$ thresholds are reported in Table~\ref{saturation}. In the fit, potential correlations of the systematic uncertainties in different $\pT$ bins are neglected.

ATLAS has published a UE measurement where the hard scale is given by the leading track $\pT$, with a $\pT$ threshold for particles of \unit[0.5]{\gmom}  and an acceptance of $|\eta| < 2.5$ \cite{atlas1}. Given the different acceptance with respect to our measurement, the results in the Toward and Away regions are not comparable. On the other hand the mean values of the Transverse plateaus from the two measurements are in good agreement, indicating an independence of the UE activity on the pseudorapidity range.
The CDF collaboration measured the UE as a function of charged particle jet $\pT$ at a collision energy of \unit[1.8]{TeV}\cite{cdf1}. The particle $\pT$ threshold is \unit[0.5]{\gmom}  and the acceptance $|\eta| < 1$. In the Transverse region CDF measures 3.8 charged particles per unit pseudorapidity above the $\pT$ threshold at leading-jet $\pT=\unit[20]{\gmom} $.
This number needs to be divided by $2\pi$ in order to be compared with  the average number of particles in the plateau from Table~\ref{saturation} at the same threshold value.  The scaled CDF result is 0.60, also shown in Table~\ref{saturation} for comparison. As expected it falls between our two measurements at $\sqrt{s}=\unit[0.9]{\tev}$ and $\sqrt{s}=\unit[7]{\tev}$. The values do not scale linearly with the collision energy, in particular the increase is higher from 0.9 to \unit[1.8]{\tev} than from 1.8 to \unit[7]{\tev}. Interpolating between our measurements assuming a logarithmic dependence on $\sqrt{s}$ results in 0.62 charged particles per unit area at \unit[1.8]{\tev}, consistent with the CDF result.

For illustration, Figure~\ref{mb_vs_ue} presents the number density in the plateau of the Transverse region for $\pT  >\unit[0.5]{\gmom}$ (our measurement as well as the value measured by CDF at \unit[1.8]{TeV}) compared with 
\linebreak 
$dN_{\rm ch}/d\eta|_{\eta=0}$ of charged particles with $\pT >$~\unit[0.5]{\gmom} in minimum-bias events \cite{atlasdndeta} (scaled by $1/2\pi$).\footnote{These data are for events that have at least one charged particle in $|\eta| < 2.5$.}
The UE activity in the plateau region is more than a factor 2 larger than the $dN_{\rm ch}/d\eta$.
Both can be fitted with a logarithmic dependence on $s$ ($a+b\ln{s}$). 
The relative increase from 0.9 to \unit[7]{TeV} for the UE is larger than that for the $dN_{\rm ch}/d\eta$: about 110\% compared to about 80\%, respectively.

In Fig.~\ref{enscaling} (left) we show the ratio between the number density distribution at $\sqrt{s}=\unit[7]{\tev}$  and $\sqrt{s}=\unit[0.9]{\tev}$. Most of the systematic uncertainties are expected to be correlated between the two energies, therefore we consider only statistical uncertainties. The ratio saturates for leading track $\pT > \unit[4]{\gmom}$. The results of a constant fit in the range $4 < \pTlead < \unit[10]{\gmom}$ are reported in Table \ref{enscaling_tab}. The measured scaling factor for a $\pT$ threshold of \unit[0.5]{\gmom}  is in agreement with the observations of ATLAS \cite{atlas1, atlas2} and CMS \cite{cms2}.

For the track threshold $\pT>\unit[0.15]{\gmom}$  all models underestimate the charged multiplicity in the Transverse and Away regions. In particular at $\sqrt{s}=\unit[7]{\tev}$ PHOJET predictions largely underestimate the measurement in the Transverse region (up to $\sim 50\%$), the discrepancy being more pronounced with increasing $\pT$ cut-off value. \textsc{Pythia} 8 correctly describes the Toward region at both collision energies and \textsc{Phojet} only at  $\sqrt{s}=\unit[0.9]{\tev}$. For  track $\pT>$ \unit[1]{\gmom},  \textsc{Pythia} 8 systematically overestimates the event activity in the jet fragmentation regions (Toward and Away).
	\btabonecol[t]
	\begin{tabular}{|c|c|c|}
	\hline
	&\textbf{Number density}&\textbf{Summed $\pT $ }\\
    \hline
	$\pT  >\unit[0.15]{\gmom}$  & 1.76	$\pm$ 0.02 & 2.00	$\pm$ 0.03\\
		\hline
	$\pT  >\unit[0.5]{\gmom}$  &	1.97	$\pm$ 0.03 & 2.16 $\pm$	0.03\\
			\hline
	$\pT  >\unit[1.0]{\gmom}$ &2.32	$\pm$ 0.04 & 2.48	 $\pm$ 0.05\\
			\hline
	\end{tabular}	
	\caption{\textit{Constant fit in $4 < \ptlead < \unit[10]{GeV/\textit{c}}$ to the ratio between $\sqrt{s}=\unit[0.9]{TeV}$ and $\sqrt{s}=\unit[7]{TeV}$ for number density (left) and summed $\pT$ (right) distributions in the Transverse region. The shown uncertainties are based on statistical and systematic uncertainties summed in quadrature.}}
	\label{enscaling_tab}
	\etabonecol
	
\subsection{Summed $\pT$}
In Fig.~\ref{sumpt_1_away}-\ref{sumpt_3_away} we show the summed $\pT$  density as a function of leading track $\pT$ 
in the three topological regions. 
The shape of the distributions follows a trend similar to that discussed above for the  number density. 

The general trend of \textsc{Pythia} 8 is to overestimate the fragmentation in the Toward region at all  $\pT$ cut-off values. Also in this case at $\sqrt{s}=\unit[7]{\tev}$ PHOJET largely underestimates the measurement in the Transverse region (up to $\sim 50\%$), especially at higher values of $\pT$ cut-off. Other systematic trends are not very pronounced.

In Table~\ref{saturation} we report the mean value of a linear fit in the plateau range. Our results agree with the ATLAS measurement in the Transverse plateau.

In Fig.~\ref{enscaling} (right) we show the ratio between the distribution at $\sqrt{s}=\unit[7]{\tev}$  and $\sqrt{s}=\unit[0.9]{\tev}$, considering as before only statistical errors. The results of a constant fit in the range $4 < \pTlead < \unit[10]{\gmom}$ are reported in Table~\ref{enscaling_tab}. 
Also in this case the scaling factor is in agreement with ATLAS and CMS results. 

The summed~$\pT$ density in the Transverse region can be interpreted as a measurement of the UE activity 
in a given leading track $\pT$ bin. 
Therefore, its value in the plateau can be used, for example, to correct jet spectra.

\subsection{Azimuthal correlation}
In Fig.~\ref{azimuth_1}-\ref{azimuth_11} 
azimuthal correlations between tracks and the leading track are shown in different ranges of leading track $\pT$. 
The range $1/3 \pi < |\Delta \phi| < 2/3 \pi$ corresponds to the Transverse region. 
The regions $-1/3 \pi < \Delta \phi < 1/3 \pi $ (Toward) and $2/3 \pi < |\Delta \phi| < \pi $ (Away) 
collect the fragmentation products of the leading and sub-leading jets.
In general, all Monte Carlo simulations considered fail to reproduce the shape of the measured distributions. \textsc{Pythia} 8 provides the best prediction for the Transverse activity in all 
leading track $\pT$ ranges considered. 
Unfortunately the same model significantly overestimates the jet fragmentation regions.

%% file: conclusions.tex
\section{Conclusions}
\label{conclusions}
We have characterized the Underlying Event in pp collisions at $\sqrt{s}=$ 0.9  and \unit[7]{\tev}    by measuring the number density, the summed $\pt$ distribution and the azimuthal correlation of charged particles with respect to the leading particle.
The analysis is based on about  $6 \cdot 10^{6}$ minimum bias events at $\sqrt{s}=$ \unit[0.9]{\tev}   and $25 \cdot 10^{6}$ events at $\sqrt{s}=$ \unit[7]{\tev}  collected during the data taking periods from April to July 2010.
Measured data have been corrected for detector related effects; in particular we applied a data-driven correction to account for the misidentification of the leading track. The fully corrected final distributions are compared with \textsc{Pythia} 6.4, \textsc{Pythia} 8 and \textsc{Phojet}, showing that pre-LHC tunes have difficulties describing the data. These results are an important ingredient in the required retuning of those generators.

Among the presented distributions, the Transverse region is particularly sensitive to the Underlying Event. We find that the ratio between the distributions at $\sqrt{s}=$ 0.9 and \unit[7]{\tev} in this region saturates at a value of about 2 for track  $\pT> $ \unit[0.5]{\gmom}.
The summed $\pT$ distribution rises slightly faster as a function of $\sqrt{s}$ than the number density distribution, indicating that the available energy tends to increase the particle's transverse momentum in addition to the multiplicity. This is in qualitative agreement with an increased relative contribution of hard processes to the Underlying Event with increasing $\cms$. Moreover, the average number of particles at large $p_{\rm T,LT}$ in the Transverse region seems to scale logarithmically with the collision energy. In general our results are in good qualitative and quantitative agreement with measurements from other LHC experiments (ATLAS and CMS) and show similar trends to that of the Tevatron (CDF). 

Our results show that the activity in the Transverse region increases logarithmically and faster than $dN_{\rm ch}/d\eta$ in minimum-bias events. Models aiming to correctly reproduce these minimum-bias and underlying event distributions need a precise description of the interplay of the hard process, the associated initial and final-state radiation and multiple parton interactions.

%% file: plots.tex
\section*{Number Density - track $\pT > 0.15 \, \gmom$}
\label{density_015}

\sixPlotsNoLine[h!]{\includegraphics[width=7.6cm]{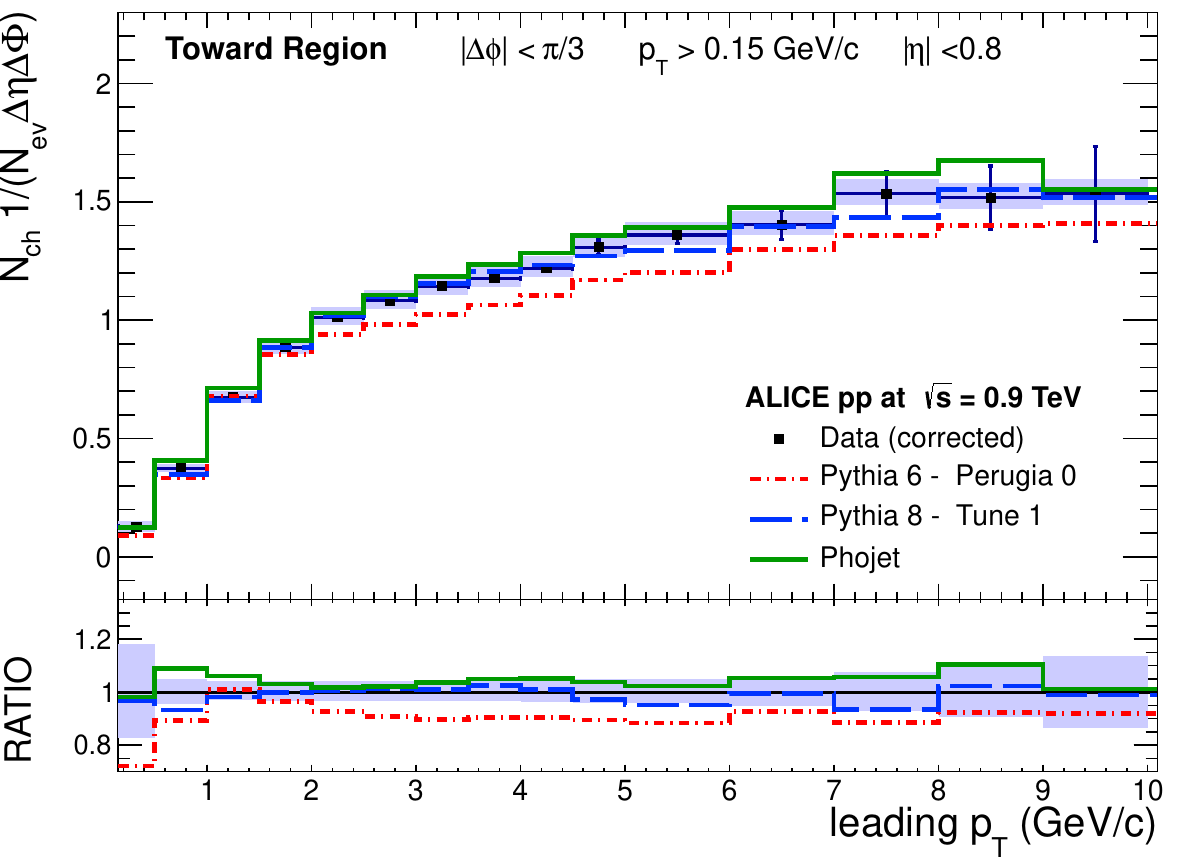}} 
{\includegraphics[width=7.6cm]
{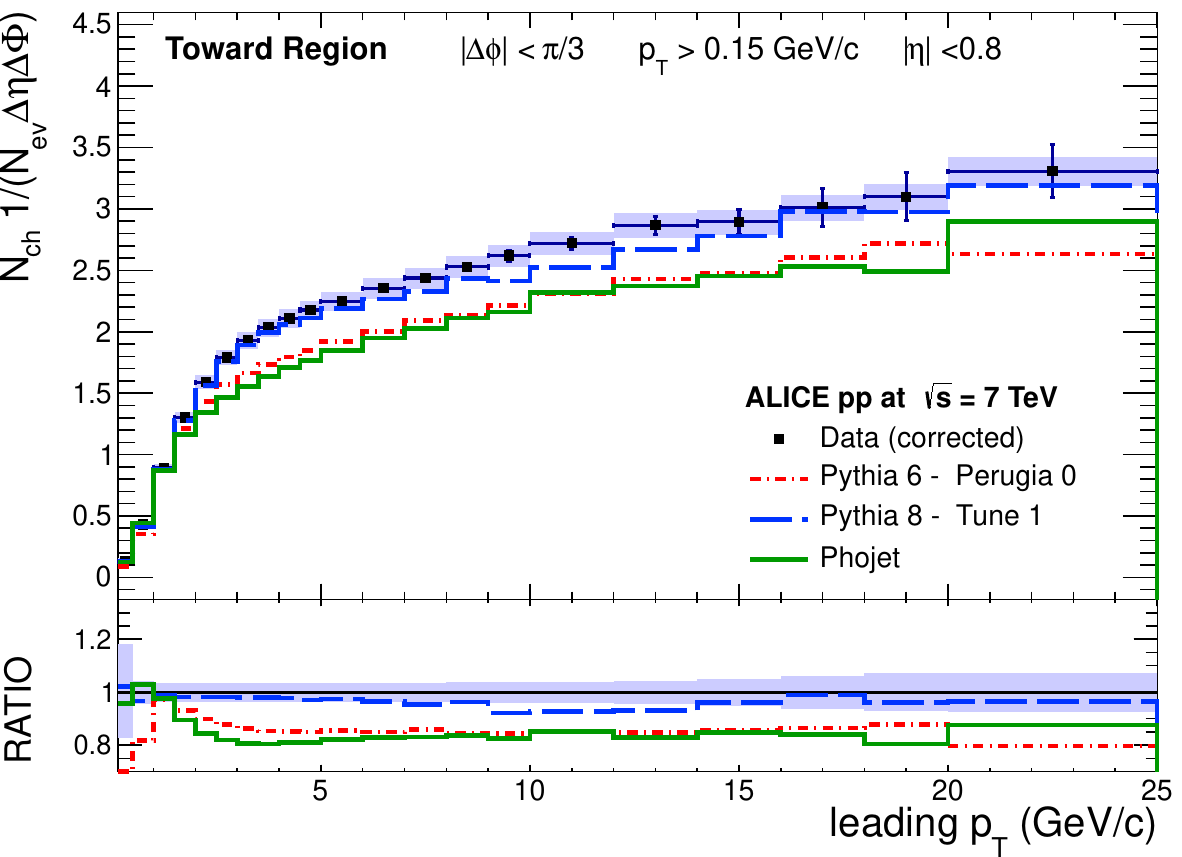}}
{\includegraphics[width=7.6cm]{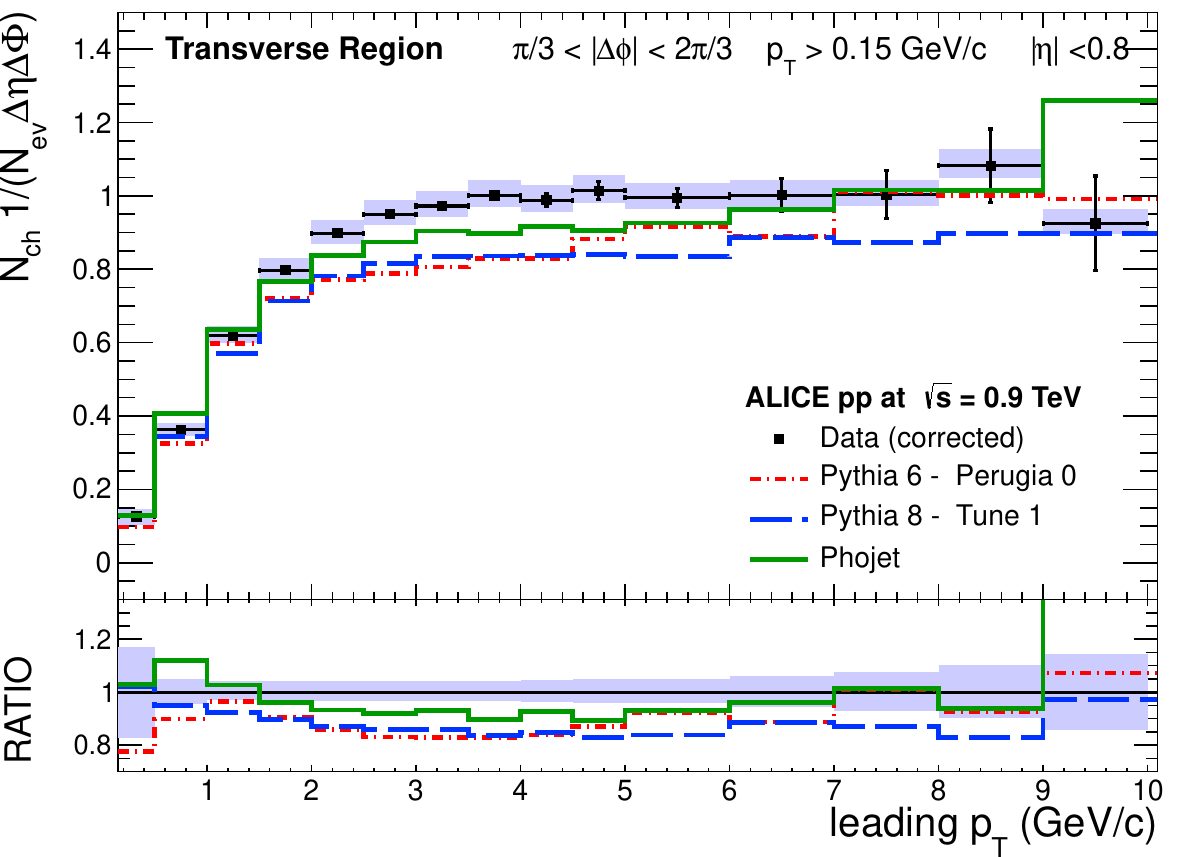}} 
{\includegraphics[width=7.6cm]
{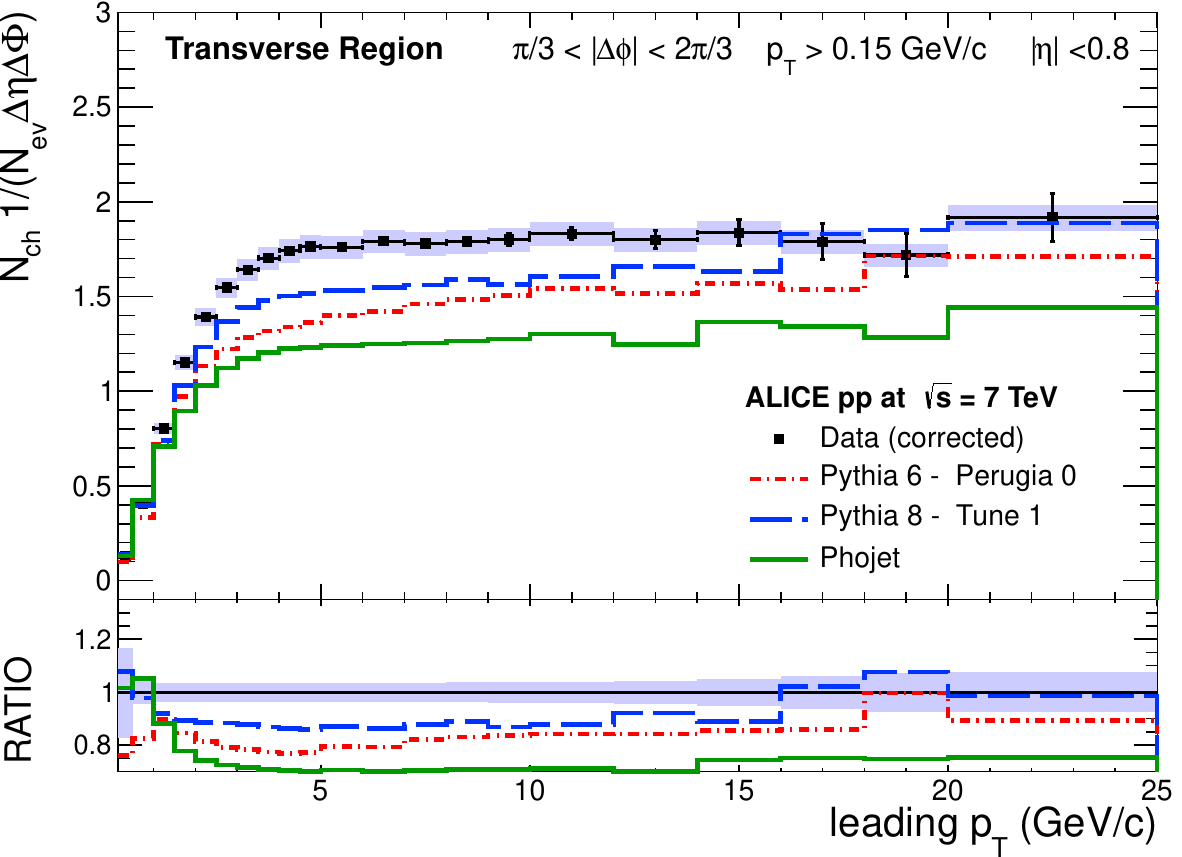}}
{\includegraphics[width=7.6cm]{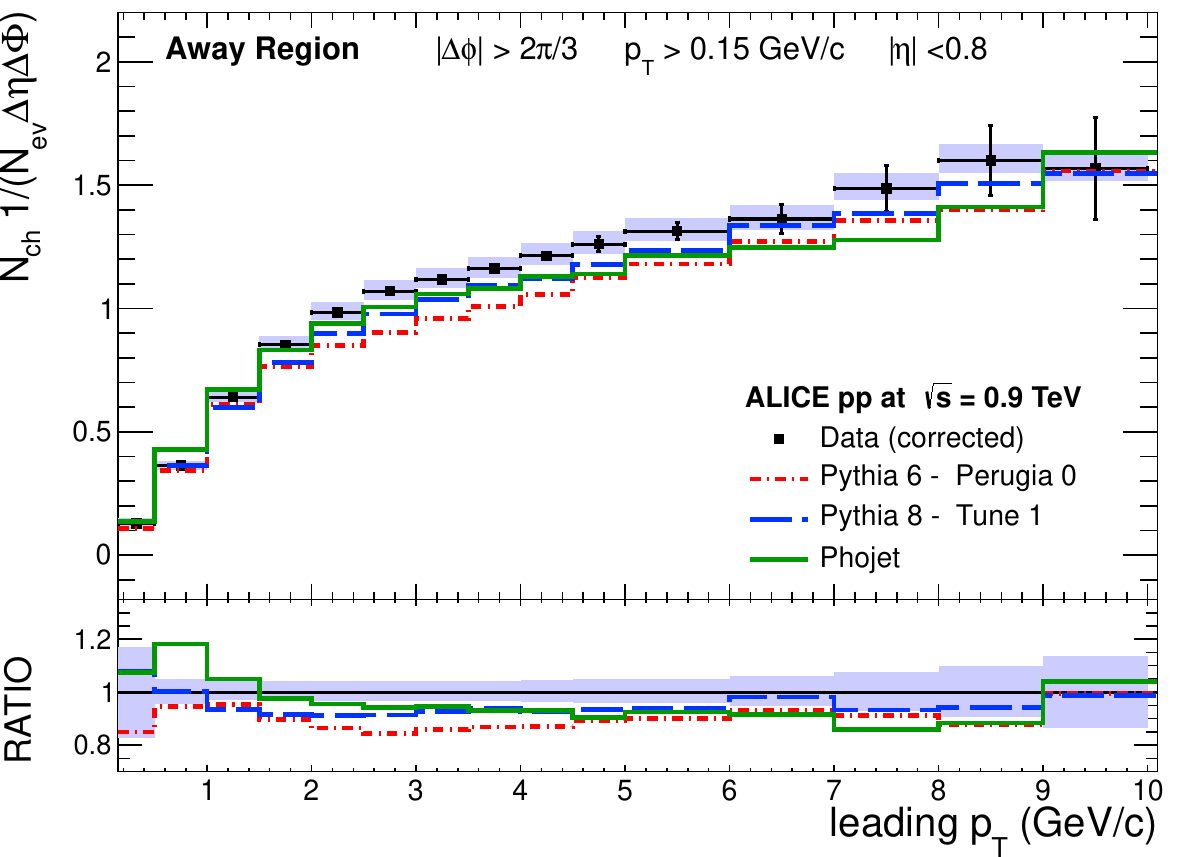}} 
{\includegraphics[width=7.6cm]
{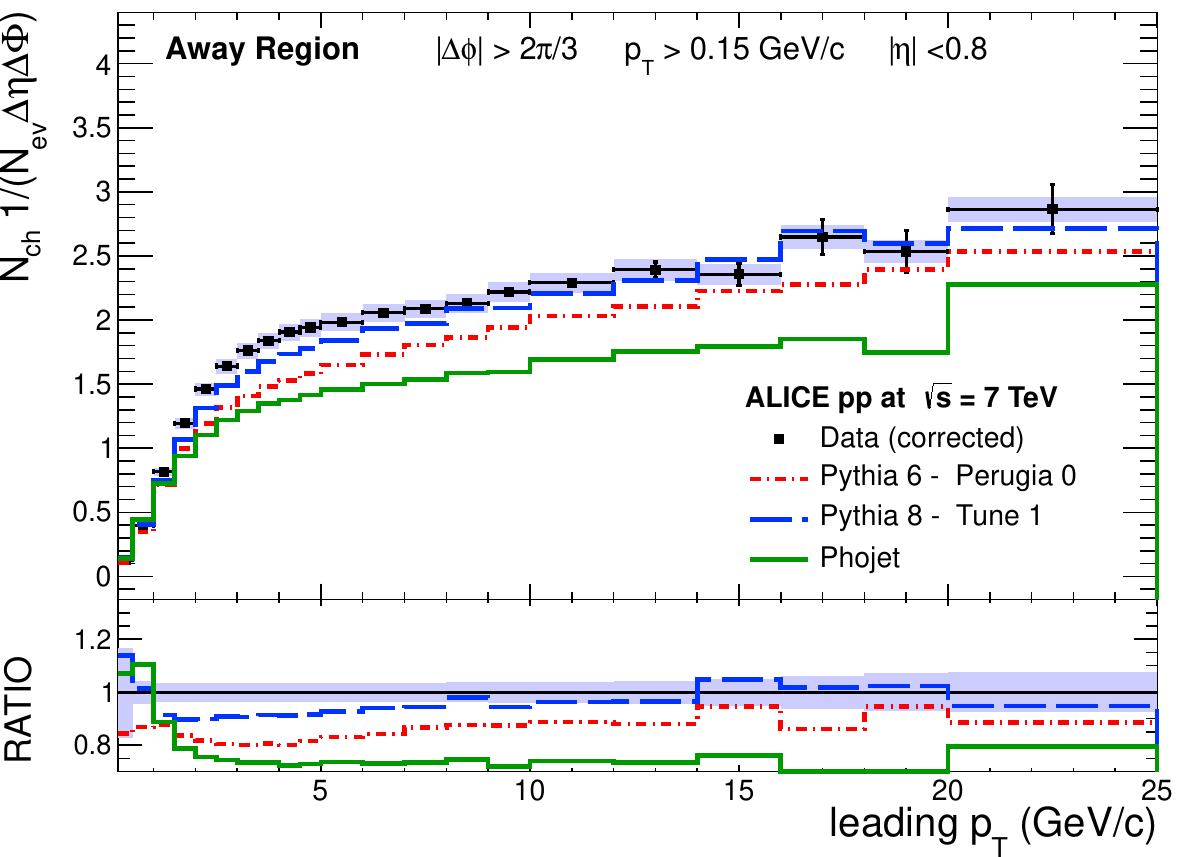}}
{\caption{\textit{Number density in Toward (top), Transverse (middle) and Away (bottom) regions at $\sqrt{s}=0.9$ TeV (left) and $\sqrt{s}=7$ TeV (right). Right and left vertical scales differ by a factor 2. Shaded area in upper plots: systematic uncertainties. Shaded areas in bottom plots: sum in quadrature of statistical and systematic uncertainties. Horizontal error bars: bin width.}}
\label{numbdens_1_away}}

\clearpage
\section*{Number Density - track $\pT > 0.5 \, \gmom $}
\label{density_05}

\sixPlotsNoLine[h!]{\includegraphics[width=7.6cm]{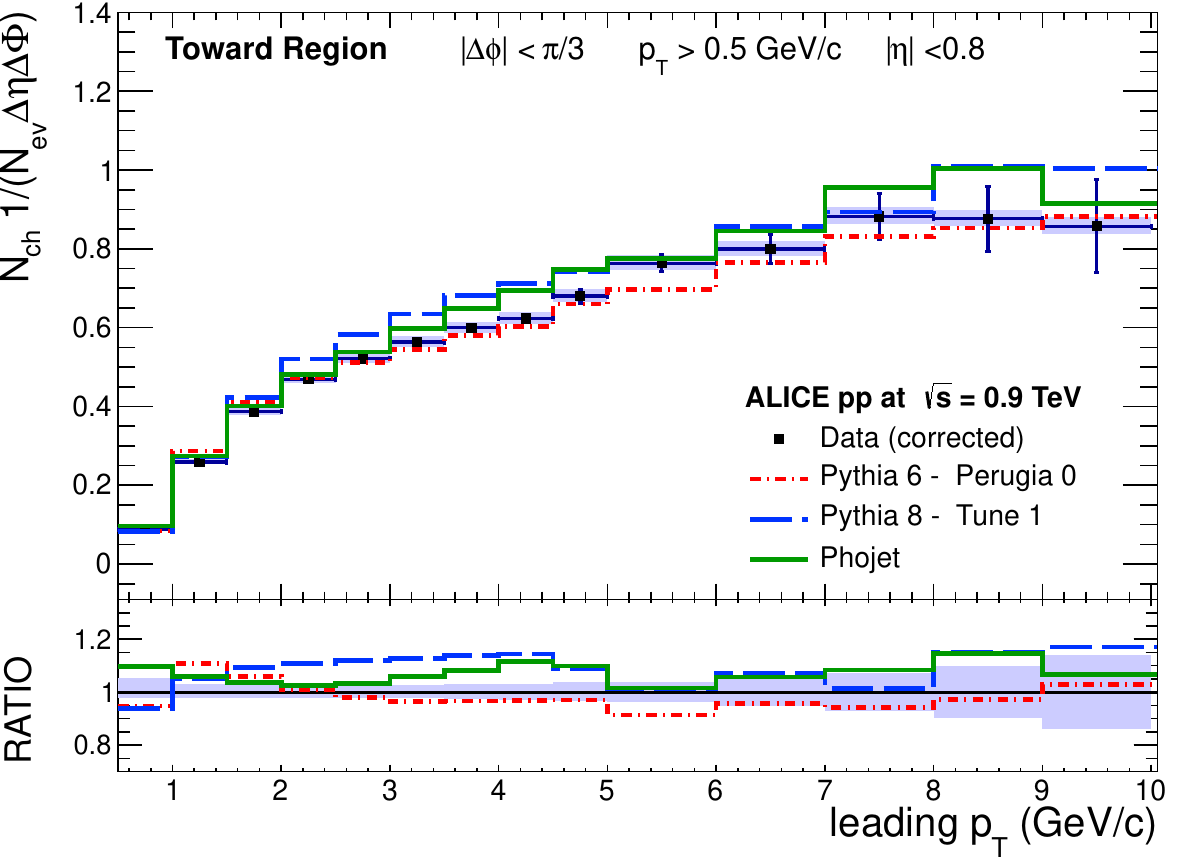}} 
{\includegraphics[width=7.6cm]
{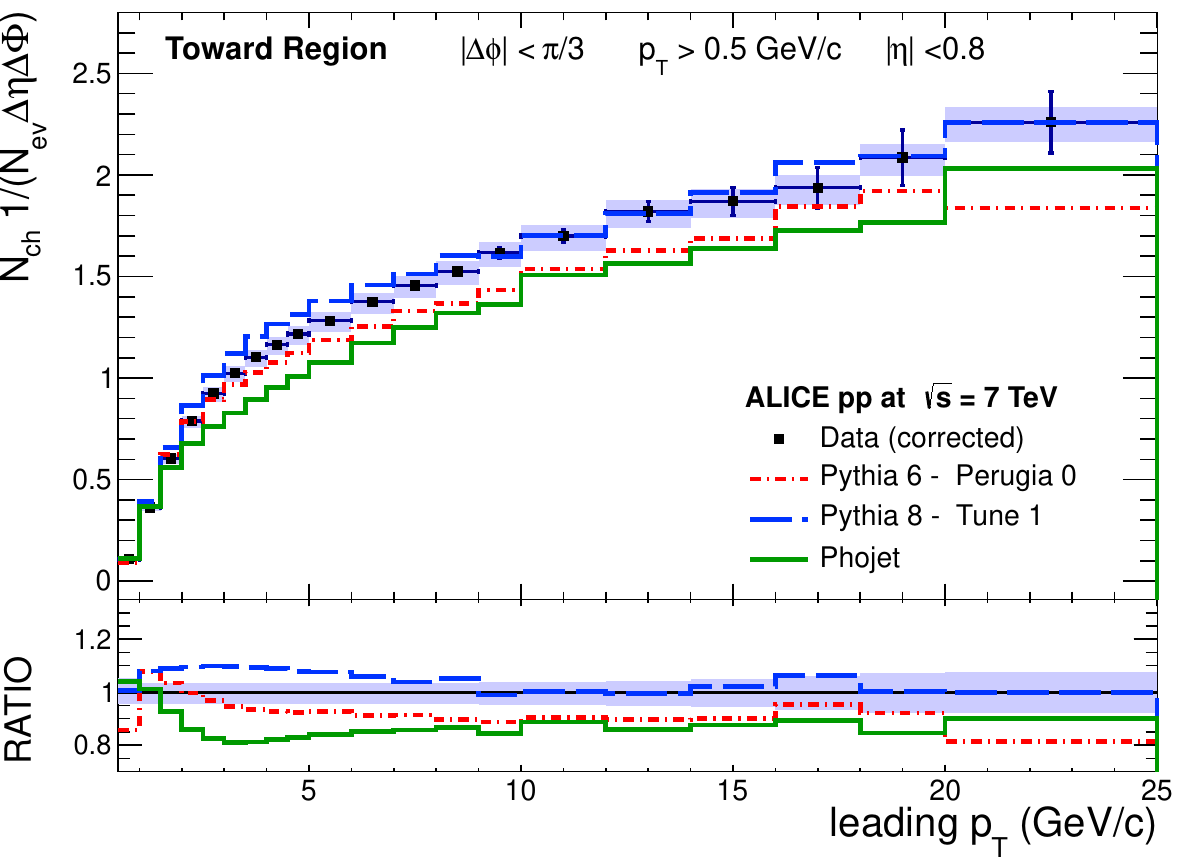}}
{\includegraphics[width=7.6cm]{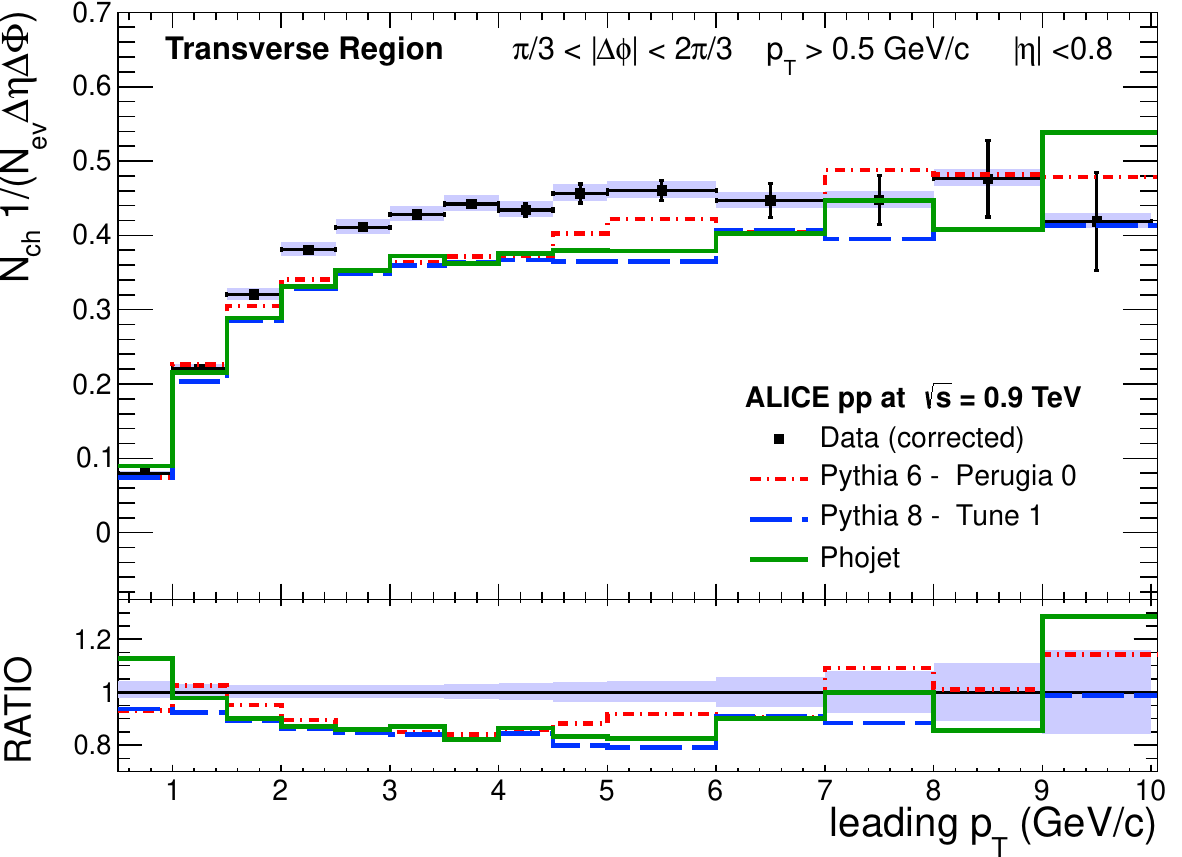}} 
{\includegraphics[width=7.6cm]
{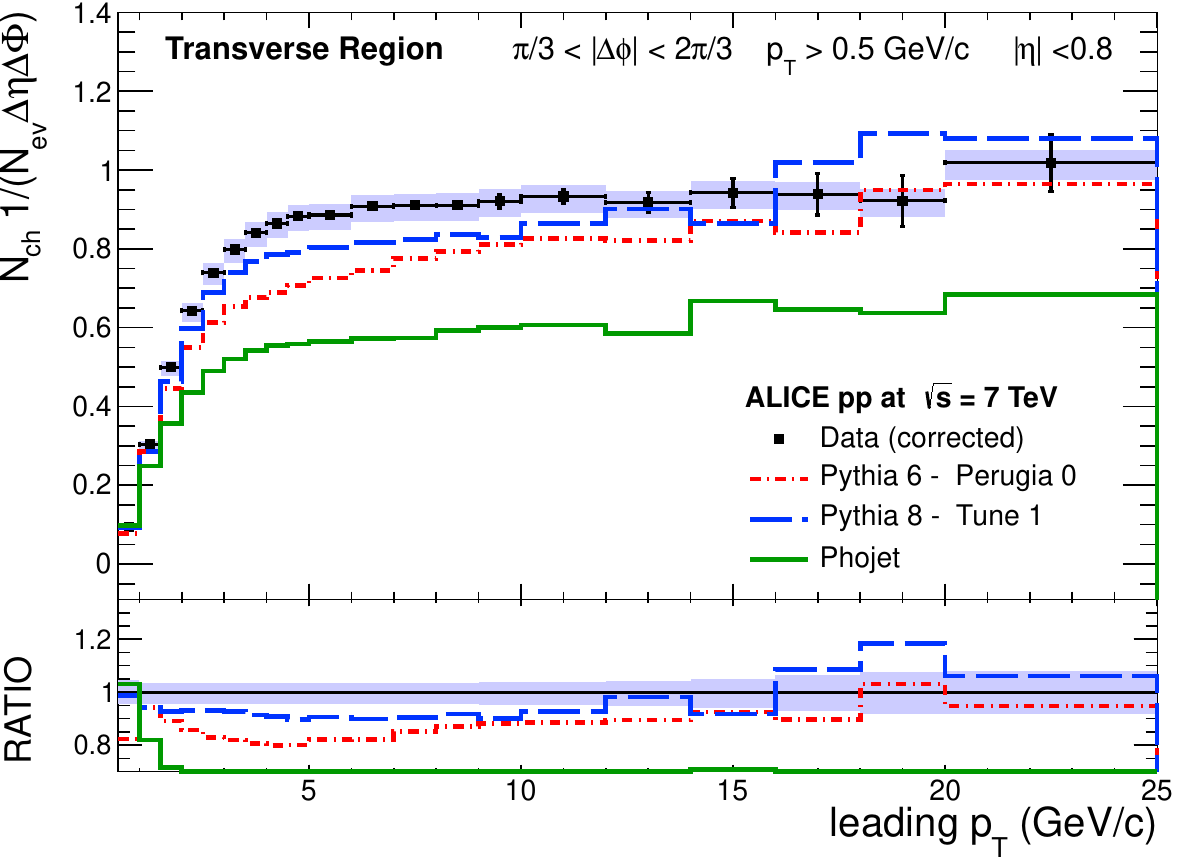}}
{\includegraphics[width=7.6cm]{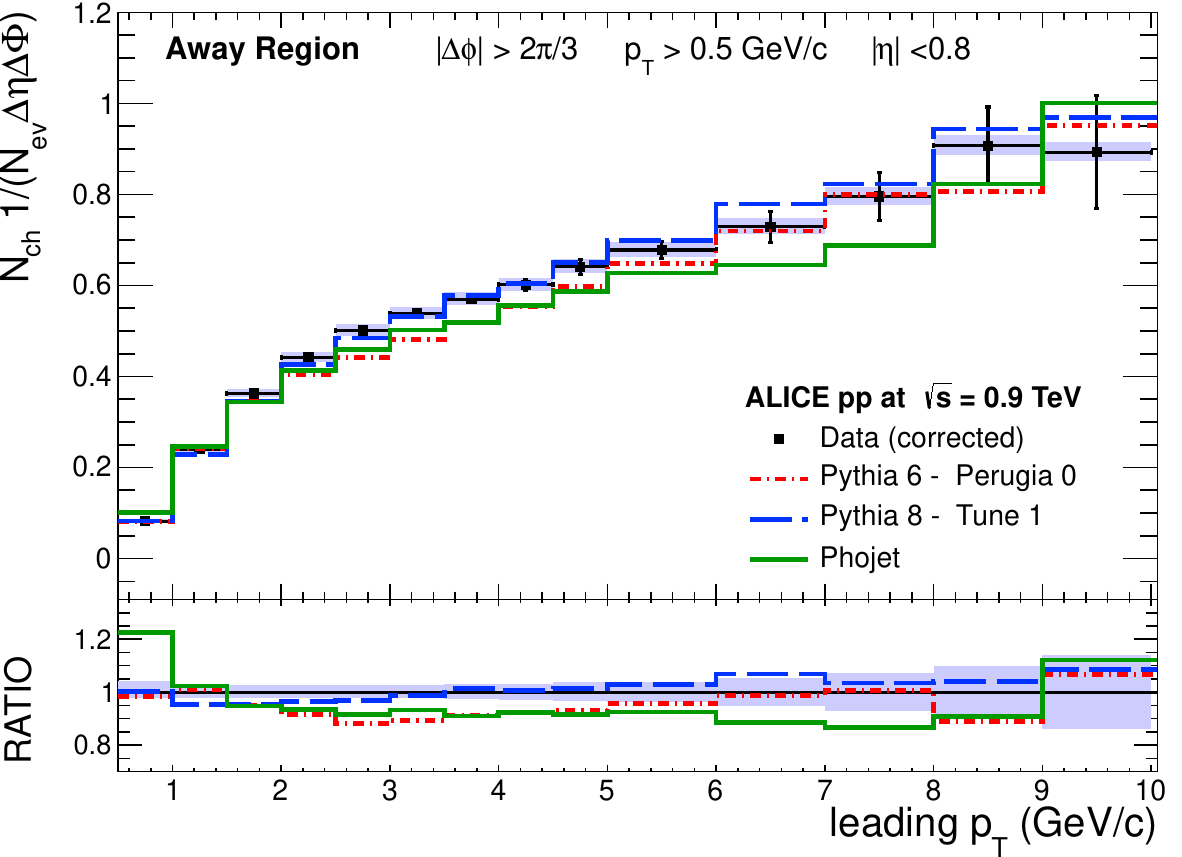}} 
{\includegraphics[width=7.6cm]
{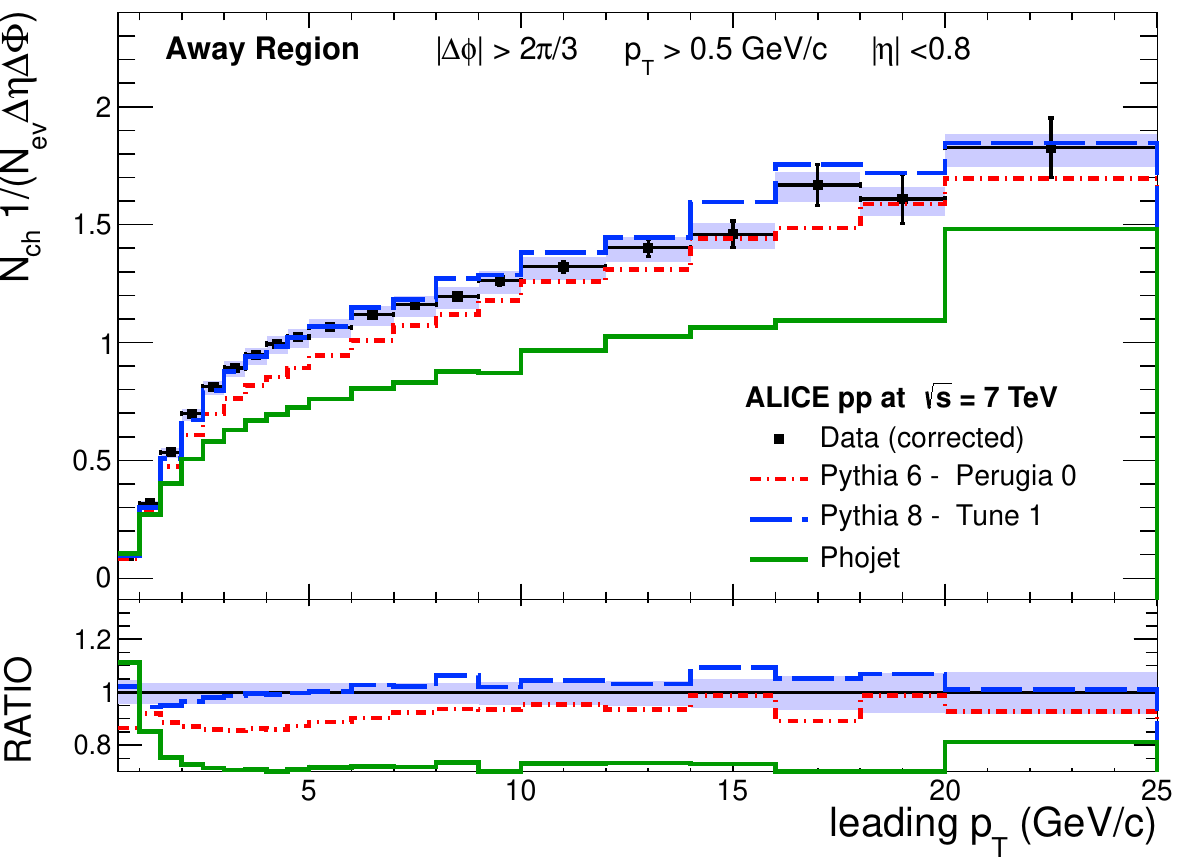}}
{\caption{\textit{Number density in Toward (top), Transverse (middle) and Away (bottom) regions at $\sqrt{s}=0.9$ TeV (left) and $\sqrt{s}=7$ TeV (right).  Right and left vertical scales differ by a factor 2. Shaded area in upper plots: systematic uncertainties. Shaded areas in bottom plots: sum in quadrature of statistical and systematic uncertainties. Horizontal error bars: bin width.}}
\label{numbdens_2_away}}

\clearpage
\section*{Number Density - track $\pT >  1.0 \, \gmom $}
\label{density_1}

\sixPlotsNoLine[h!]{\includegraphics[width=7.6cm]{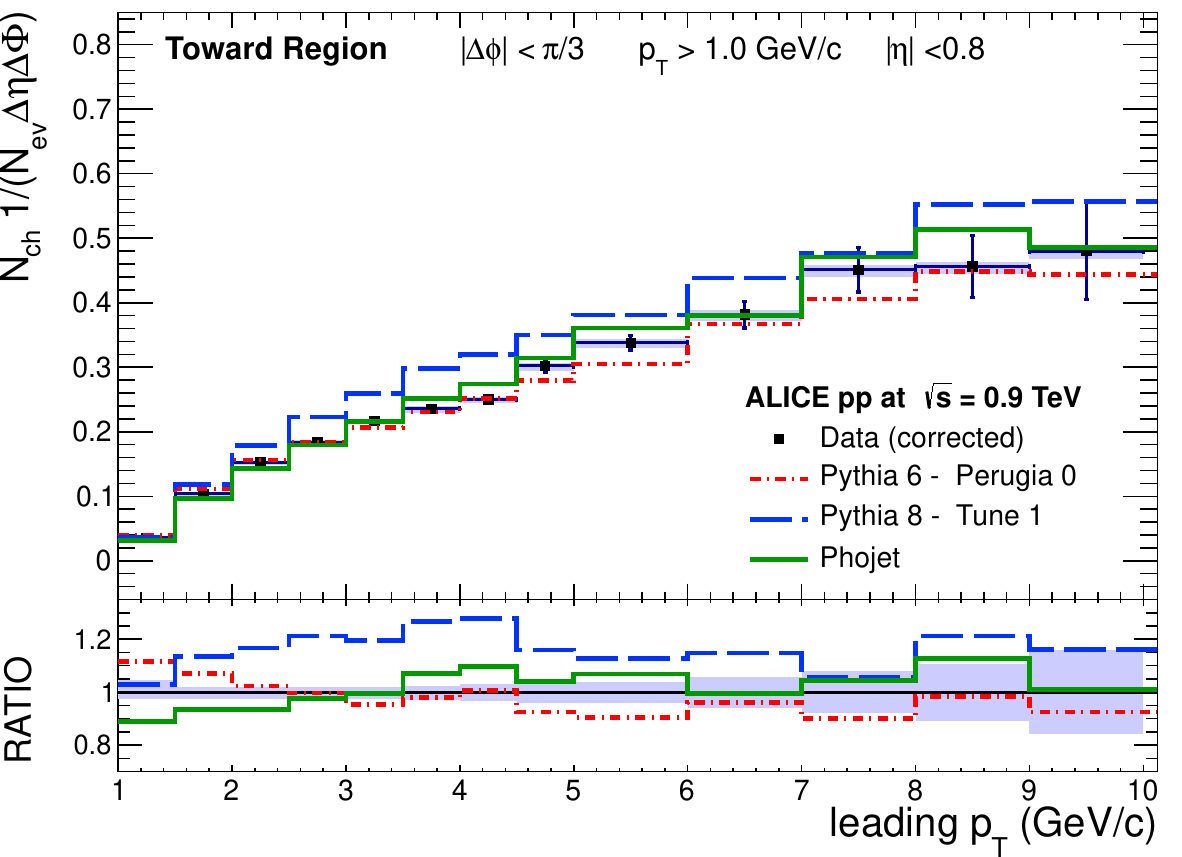}} 
{\includegraphics[width=7.6cm]
{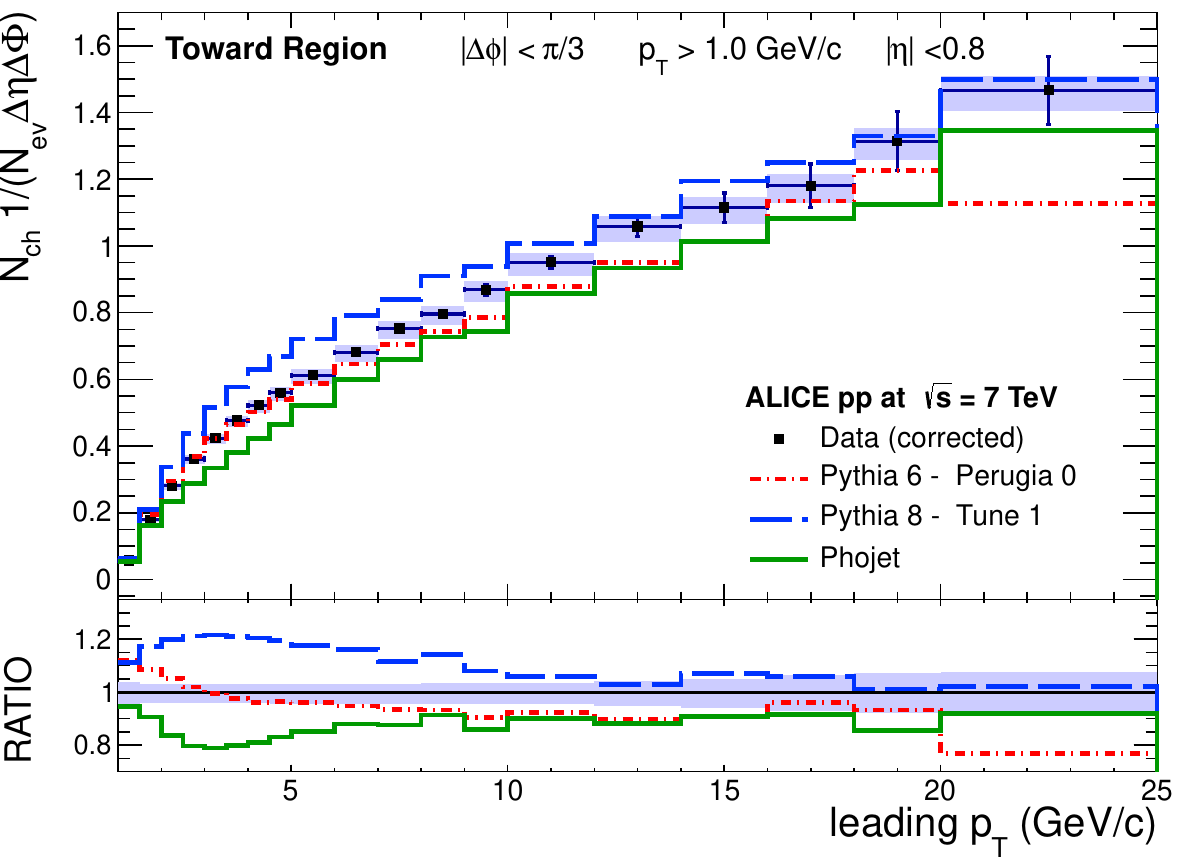}}
{\includegraphics[width=7.6cm]{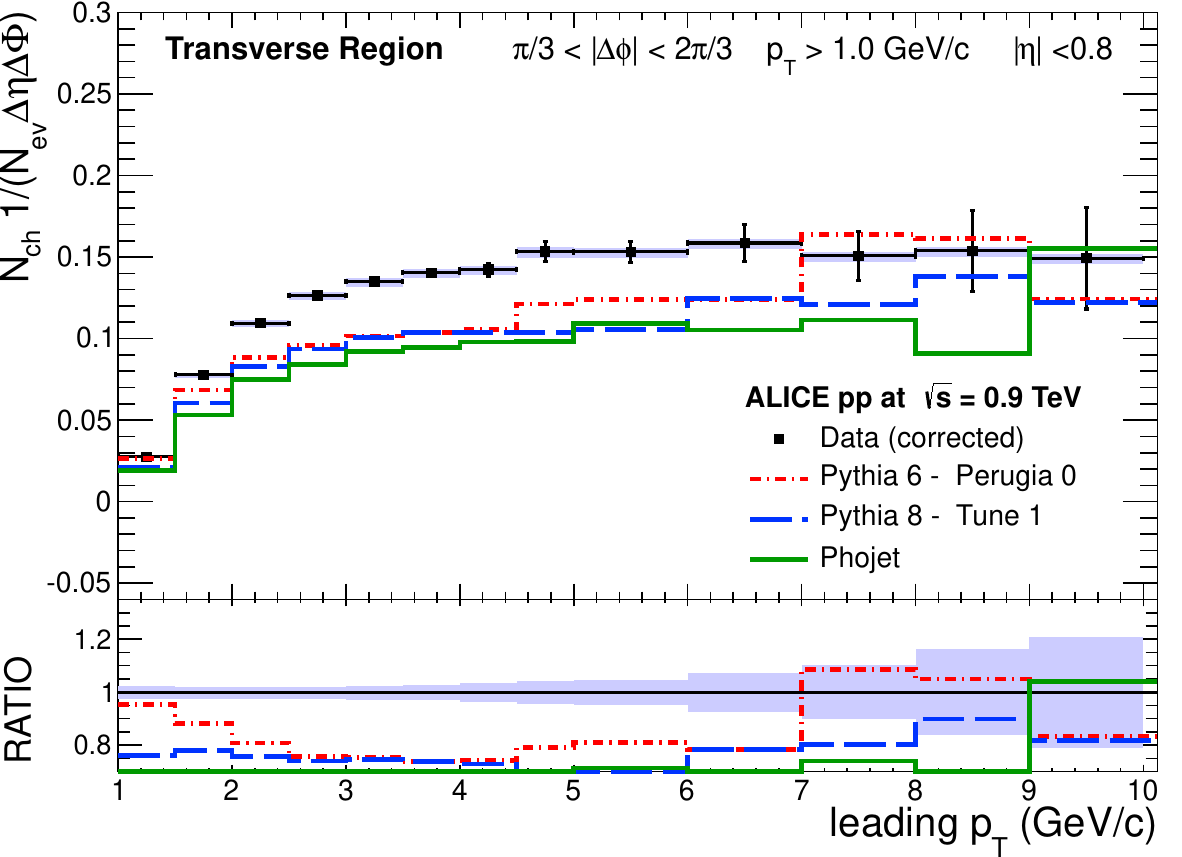}} 
{\includegraphics[width=7.6cm]
{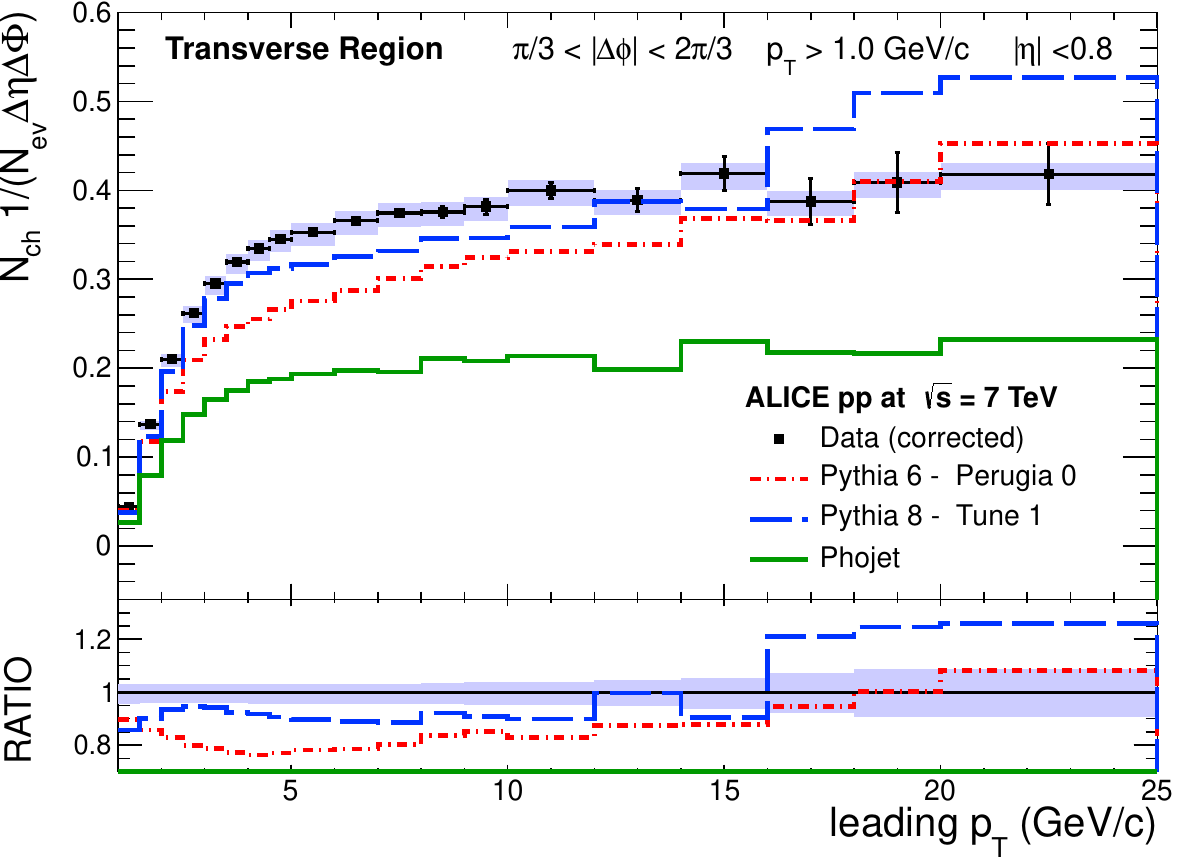}}
{\includegraphics[width=7.6cm]{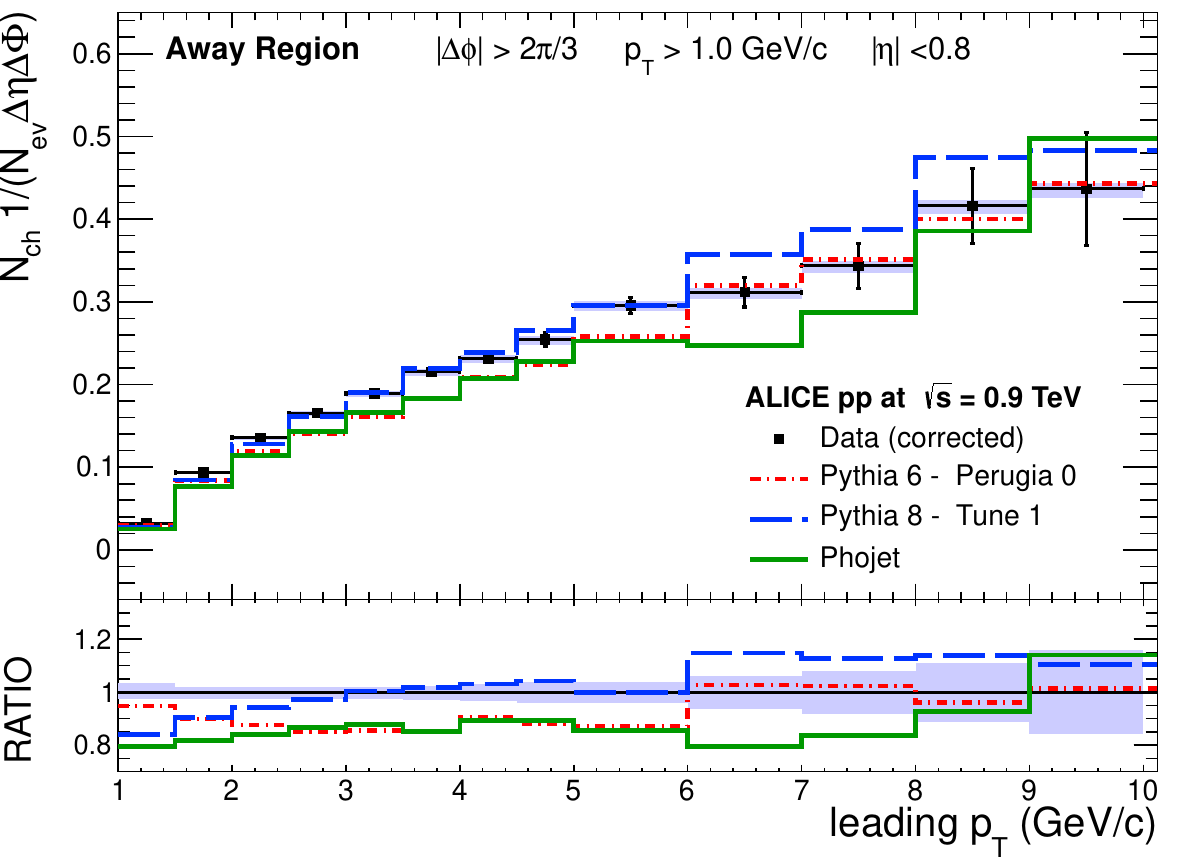}} 
{\includegraphics[width=7.6cm]
{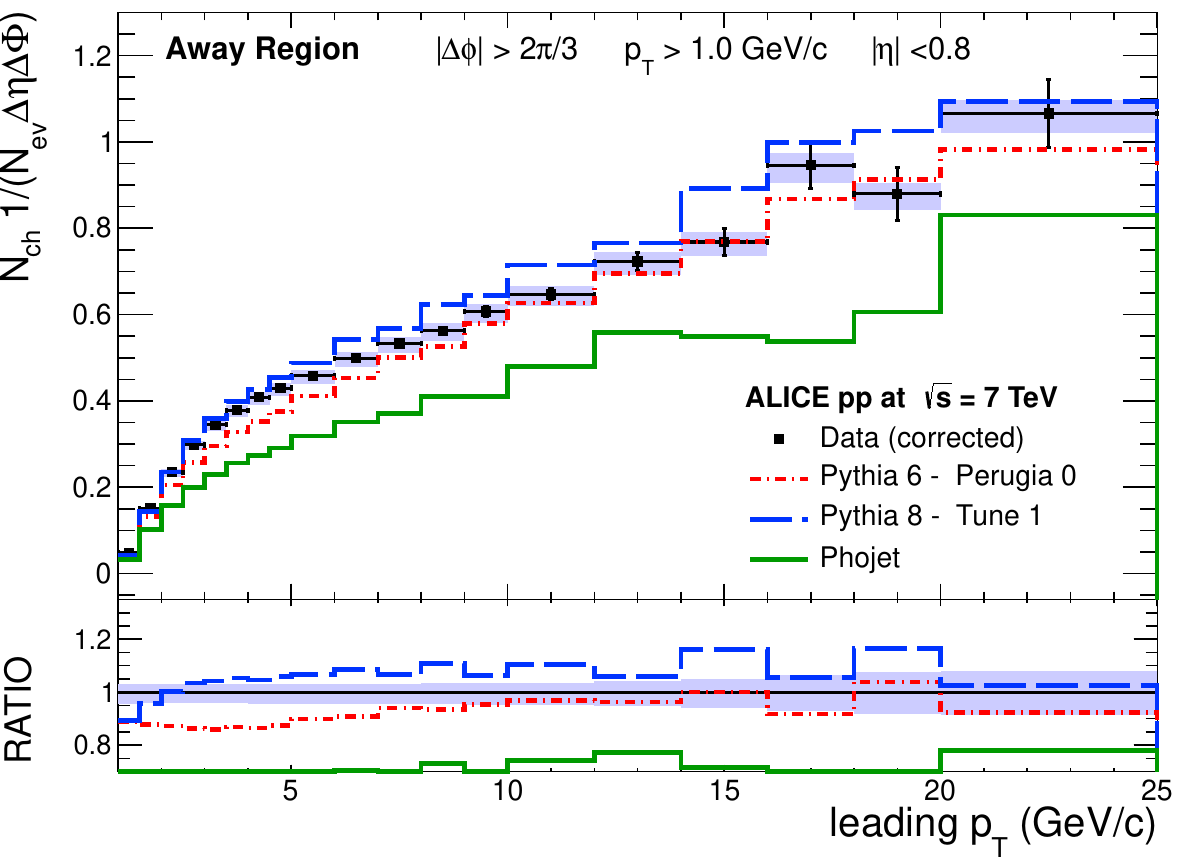}}
{\caption{\textit{Number density in Toward (top), Transverse (middle) and Away (bottom) regions at $\sqrt{s}=0.9$ TeV (left) and $\sqrt{s}=7$ TeV (right). Right and left vertical scales differ by a factor 2. Shaded area in upper plots: systematic uncertainties. Shaded areas in bottom plots: sum in quadrature of statistical and systematic uncertainties. Horizontal error bars: bin width.}}
\label{numbdens_3_away}}

\clearpage

\bfig[t]
  \includegraphics[width=12cm]{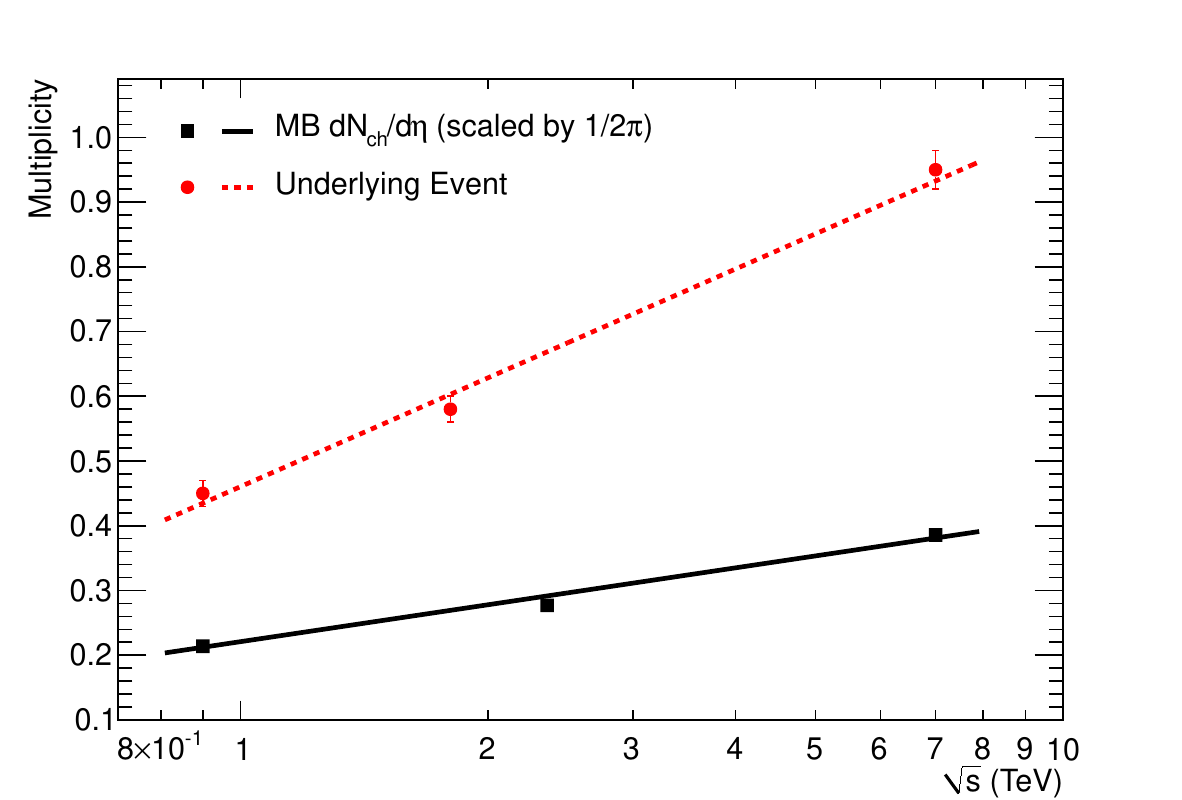}
  \caption{\textit{Comparison of number density in the plateau of the Transverse region (see Table~\ref{enscaling_tab}) and $dN_{\rm ch}/d\eta$ in minimum-bias events (scaled by $1/2\pi$) \cite{atlasdndeta}. Both are for charged particles with $\pT  >\unit[0.5]{\gmom}$. For this plot, statistical and systematic uncertainties have been summed in quadrature. The lines show fits with the functional form $a+b\ln{s}$.} \label{mb_vs_ue}}
\efigNoLine

\twoPlotsNoLine[b!]{\includegraphics[width=8cm]{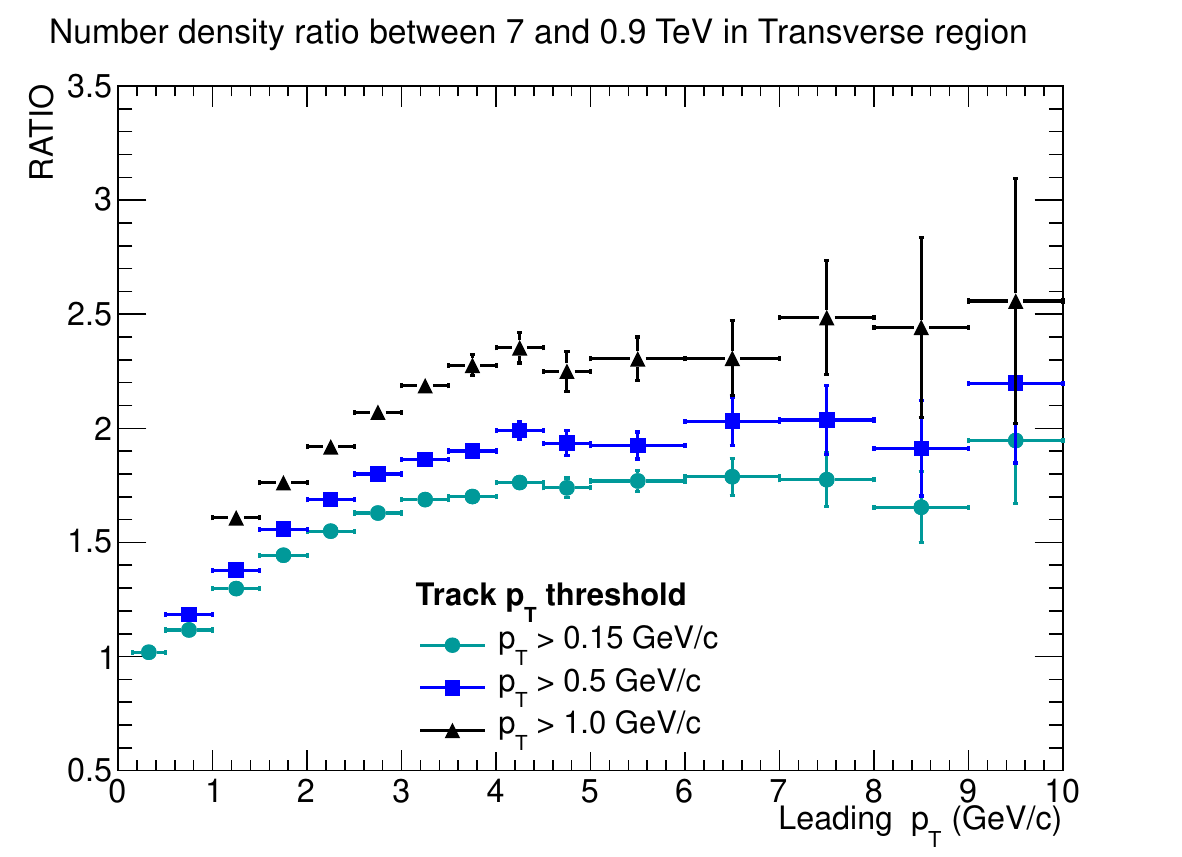}} {\includegraphics[width=8cm]{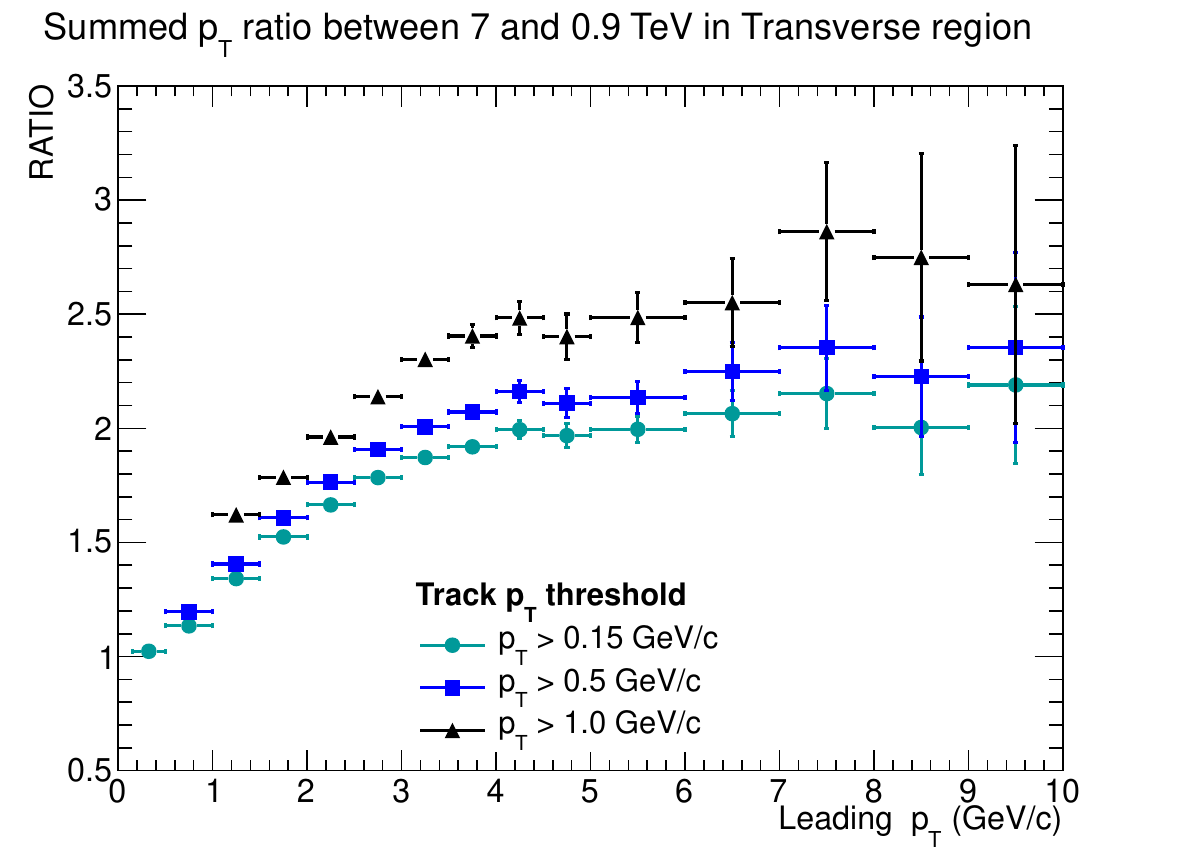}}
{\caption{\textit{Ratio between $\sqrt{s}=\unit[0.9]{TeV}$ and $\sqrt{s}=\unit[7]{TeV}$ for number density (left) and summed $\pT$ (right) distributions in the Transverse region. Statistical uncertainties only.}} \label{enscaling} }

\clearpage
\section*{Summed $\pT$ - track $\pT > 0.15 \, \gmom$}
\label{sumpt_015}

\sixPlotsNoLine[h!]{\includegraphics[width=7.6cm]{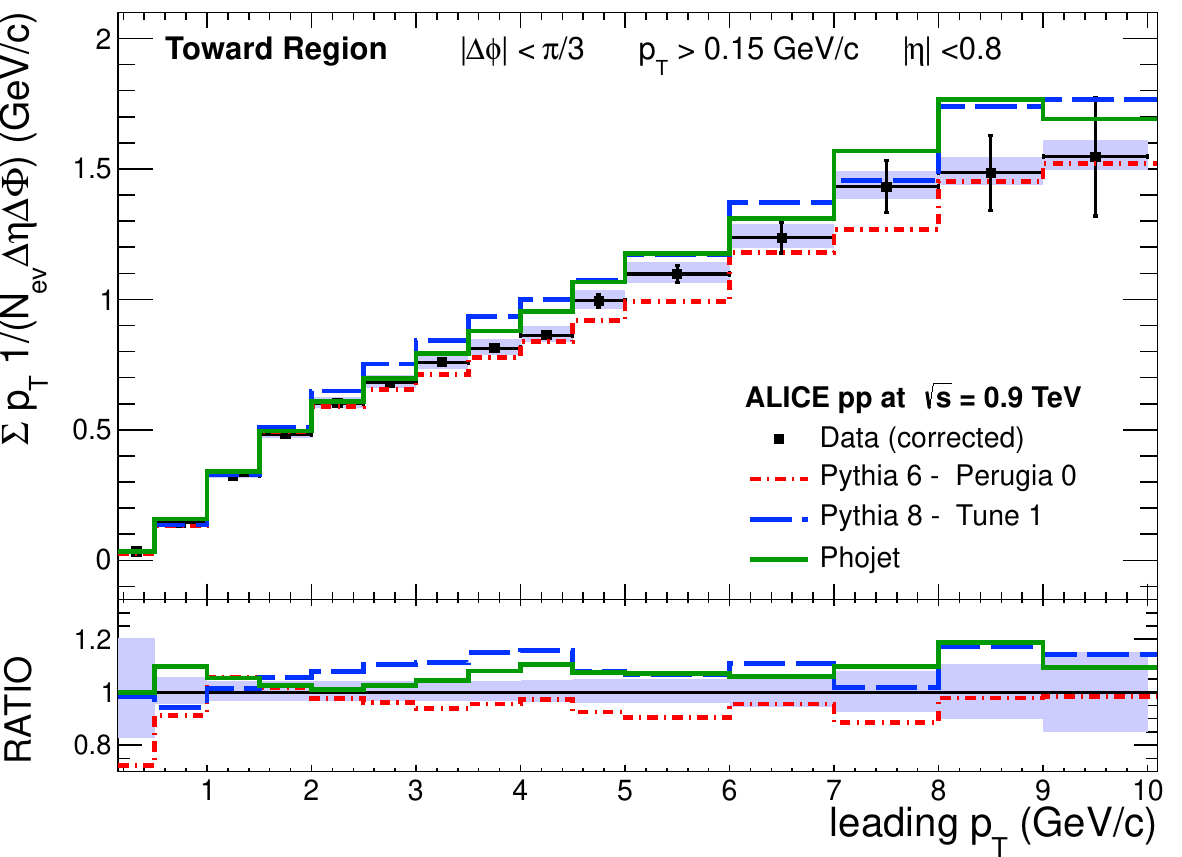}} 
{\includegraphics[width=7.6cm]
{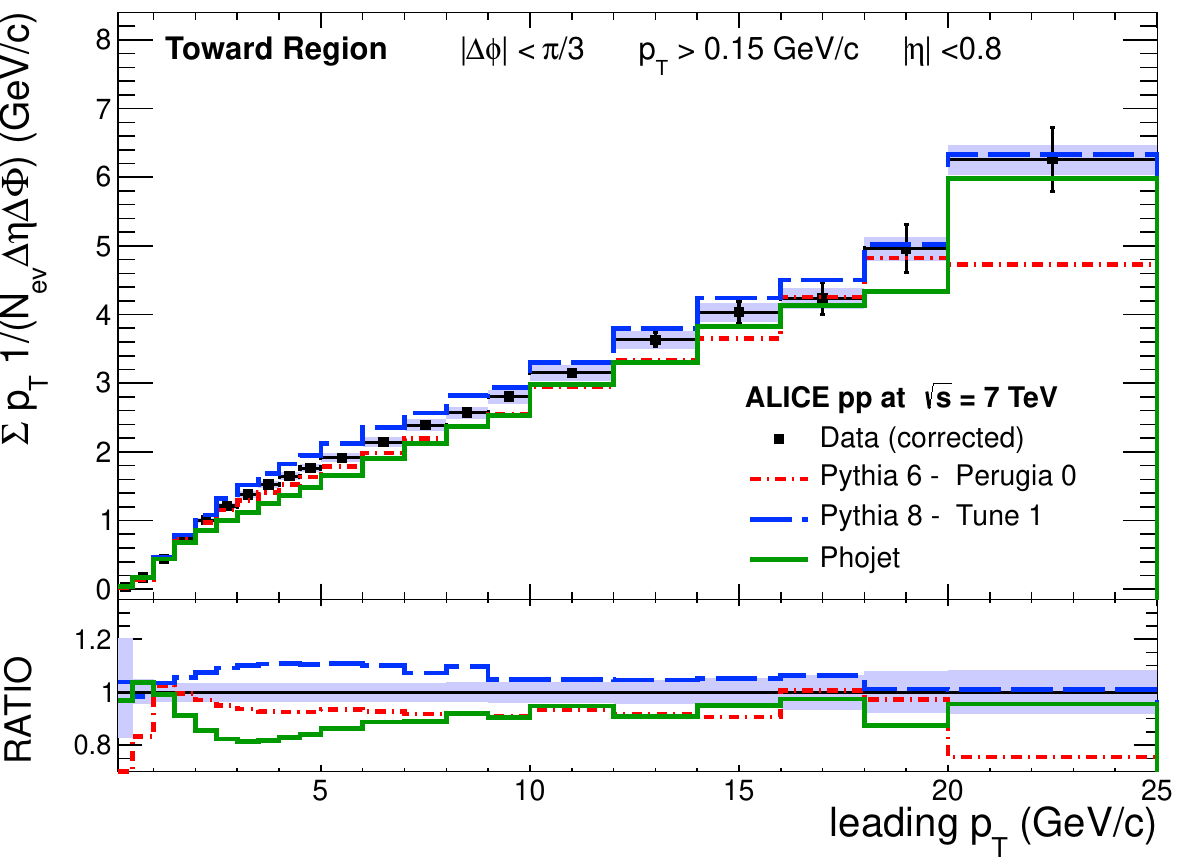}}
{\includegraphics[width=7.6cm]{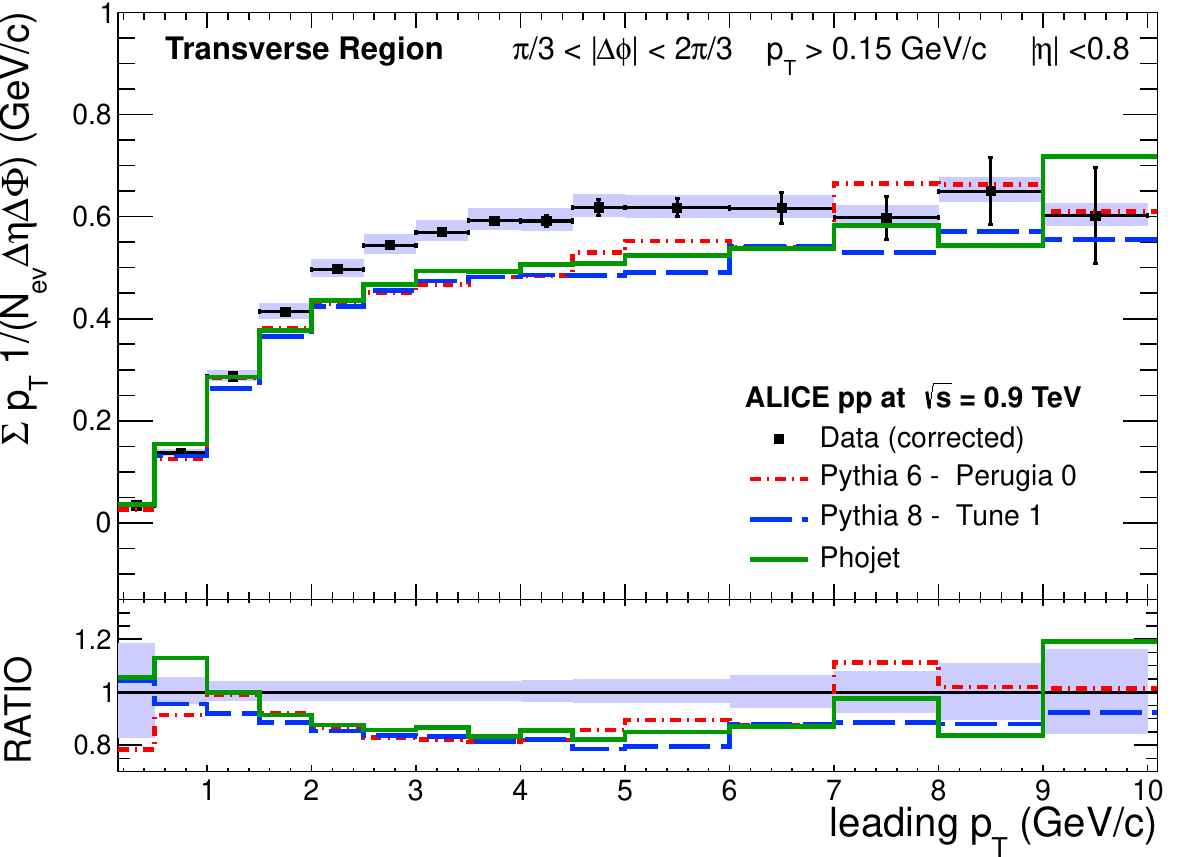}} 
{\includegraphics[width=7.6cm]
{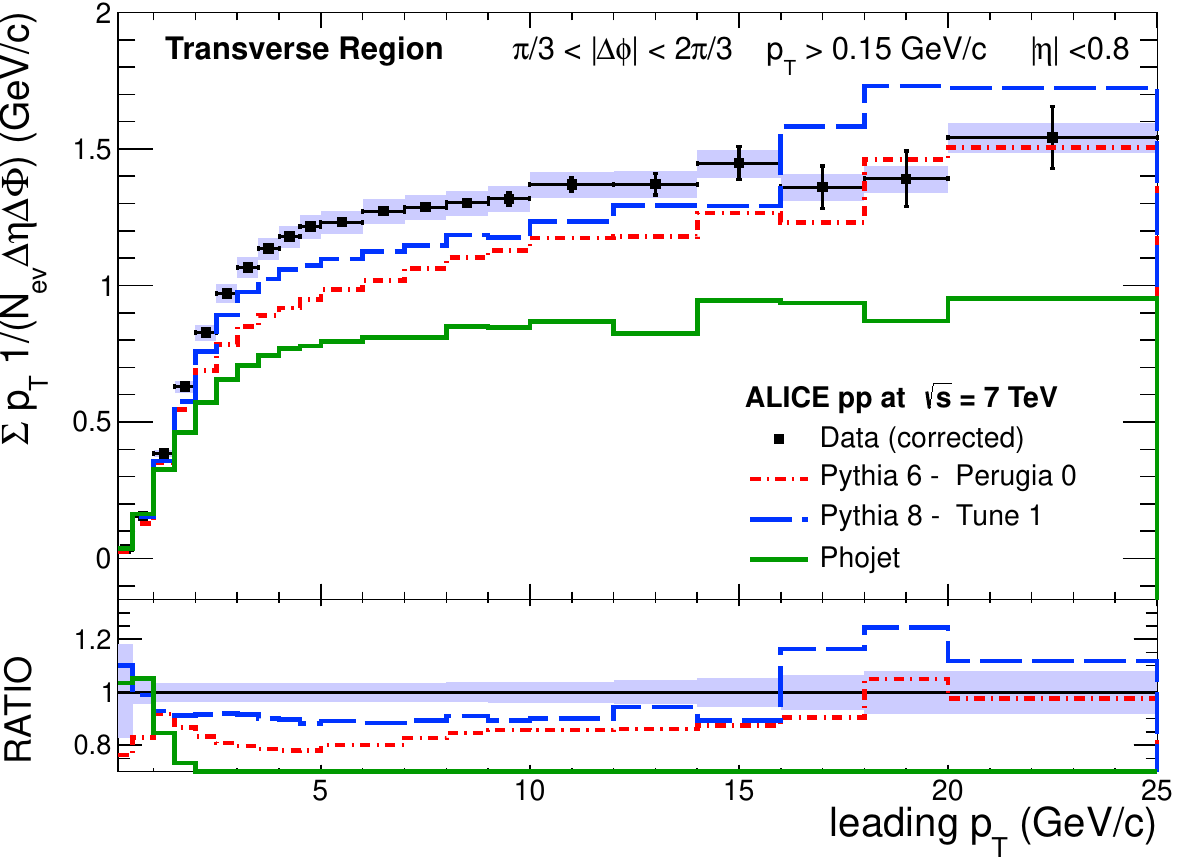}}
{\includegraphics[width=7.6cm]{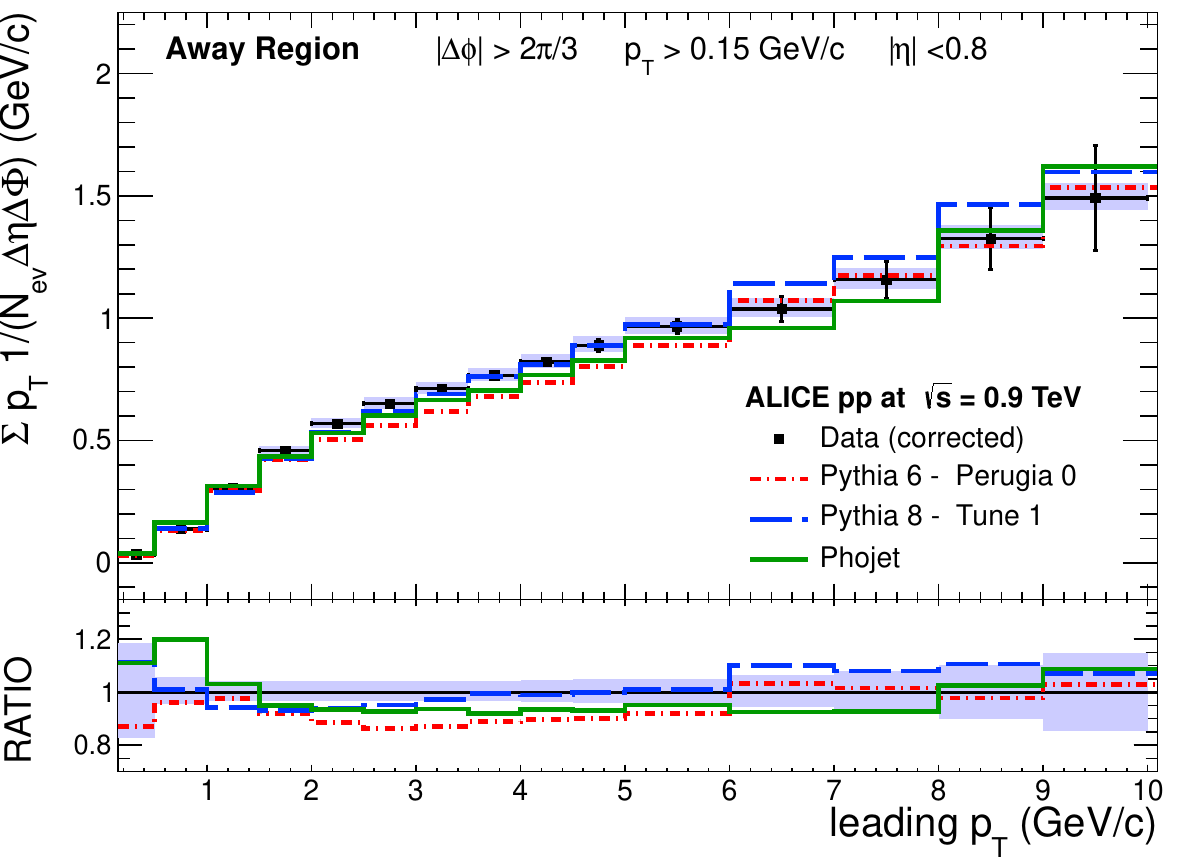}} 
{\includegraphics[width=7.6cm]
{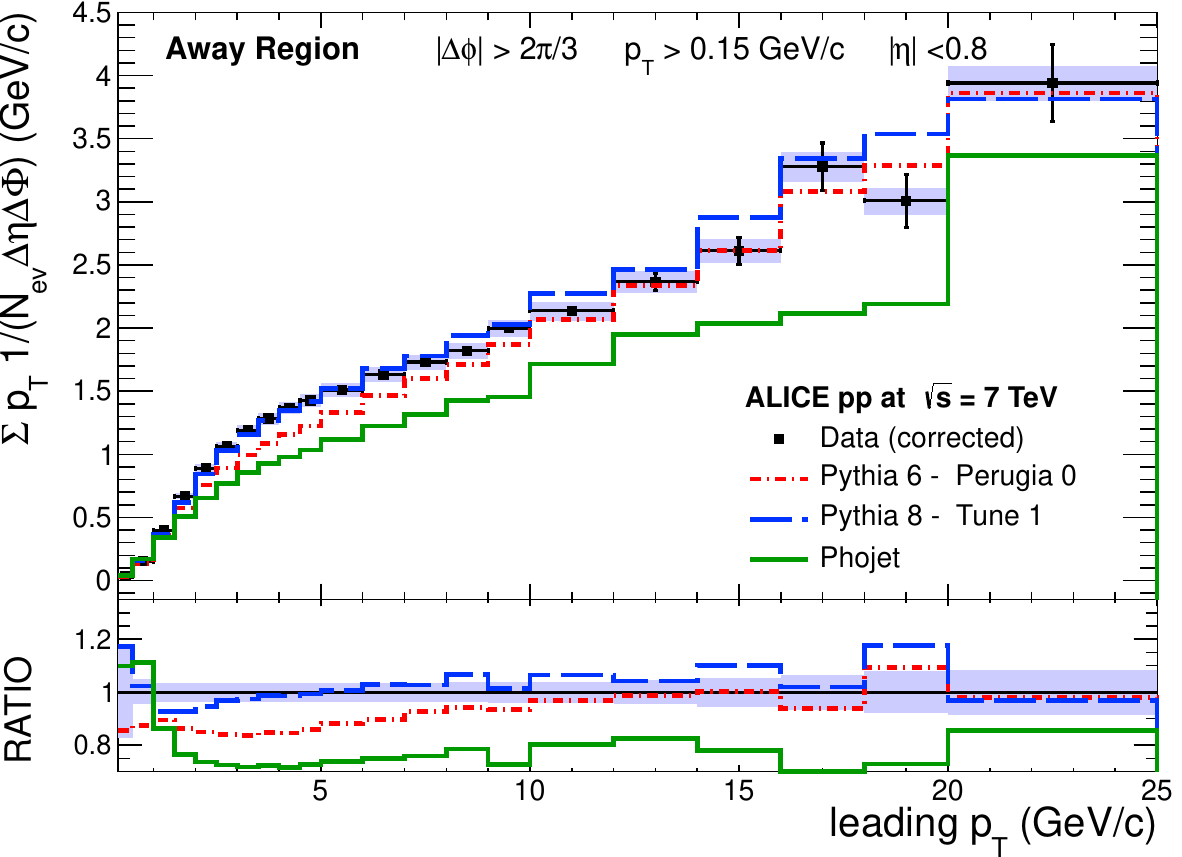}}
{\caption{\textit{Summed  $\pT$ in Toward (top), Transverse (middle) and Away (bottom) regions at $\sqrt{s}=0.9$ TeV (left) and $\sqrt{s}=7$ TeV (right). Right and left vertical scales differ by a factor 4 (2) in the top (middle and bottom) panel. Shaded area in upper plots: systematic uncertainties. Shaded areas in bottom plots: sum in quadrature of statistical and systematic uncertainties. Horizontal error bars: bin width.}}
\label{sumpt_1_away}}

\clearpage
\section*{Summed $\pT$ - track $\pT > 0.5 \, \gmom $}
\label{sumpt_05}

\sixPlotsNoLine[h!]{\includegraphics[width=7.6cm]{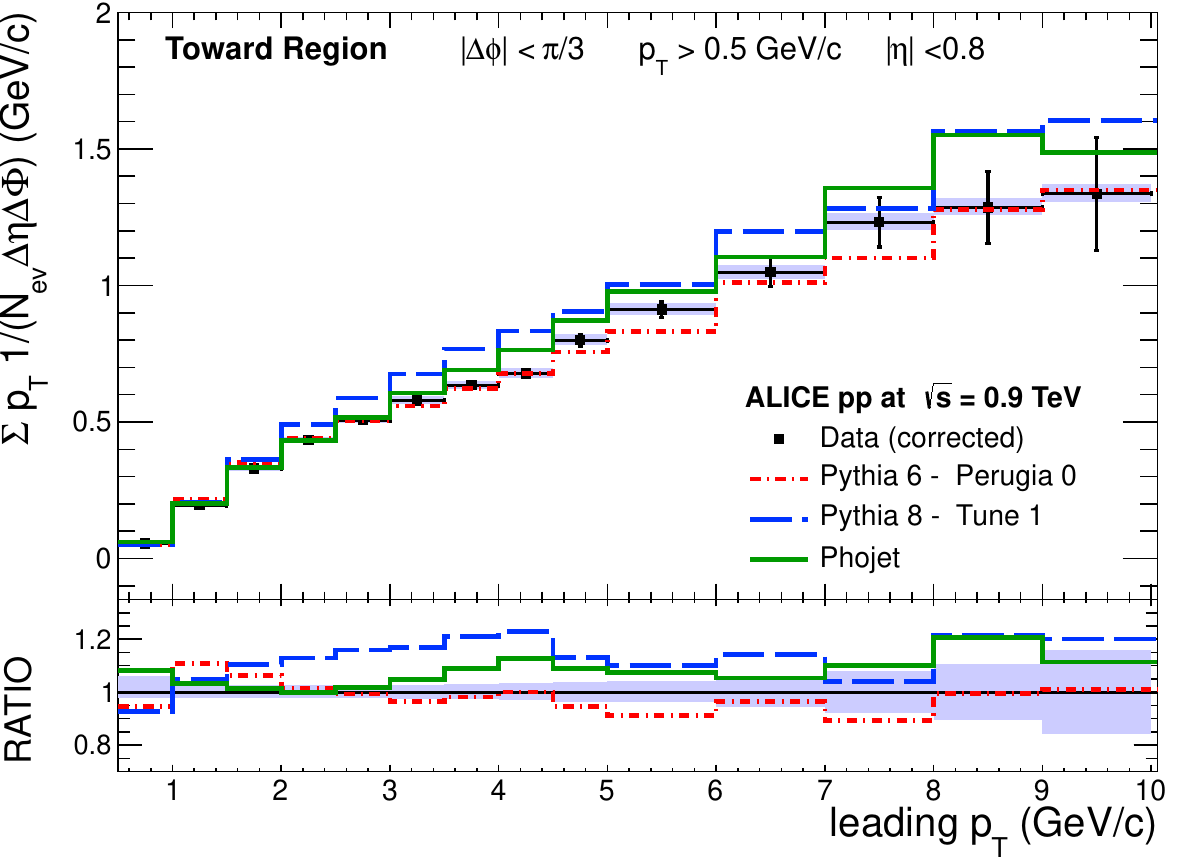}} 
{\includegraphics[width=7.6cm]
{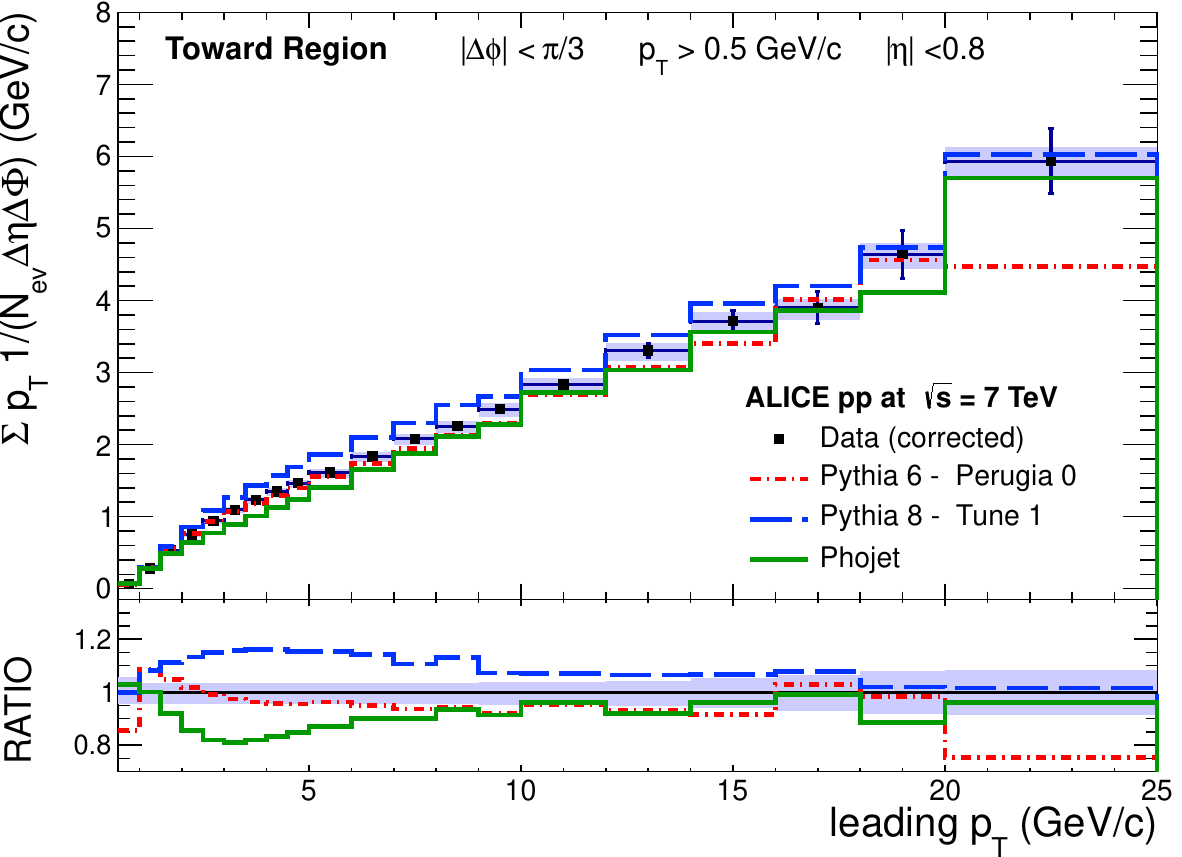}}
{\includegraphics[width=7.6cm]{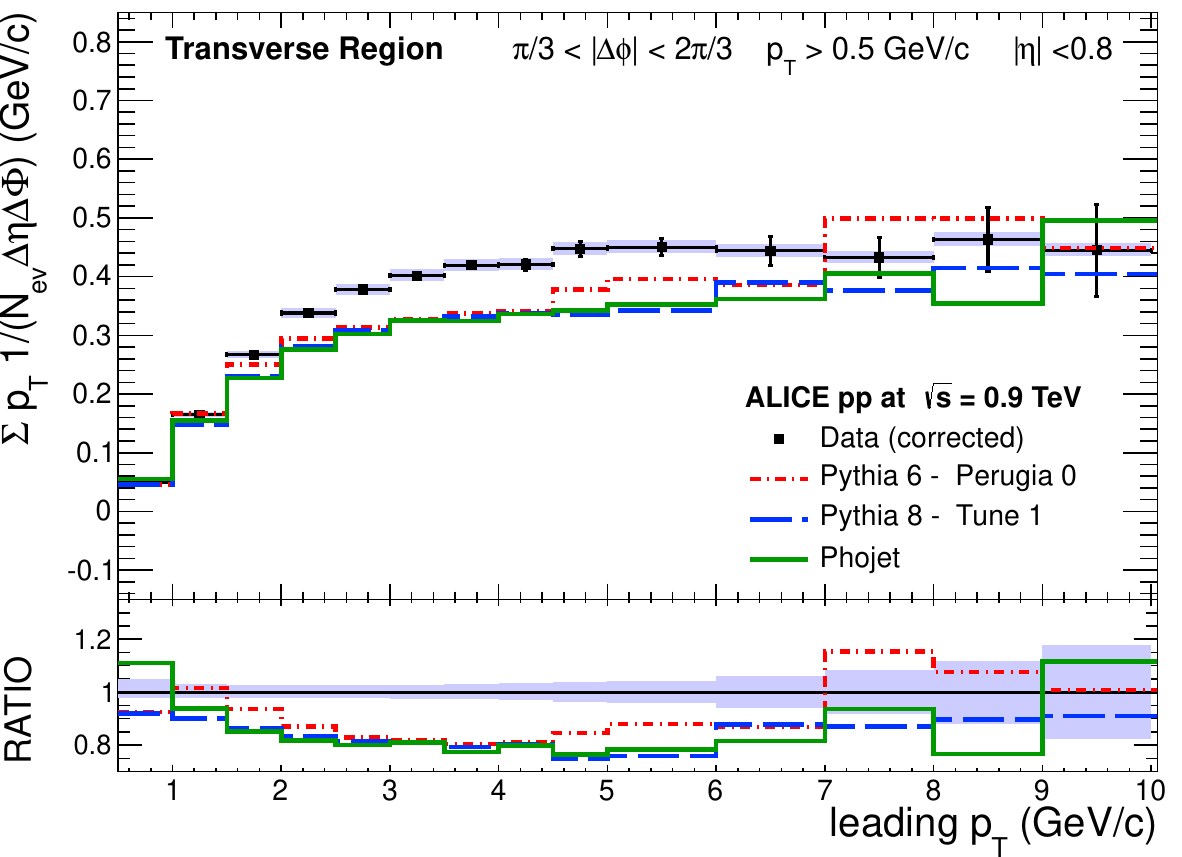}} 
{\includegraphics[width=7.6cm]
{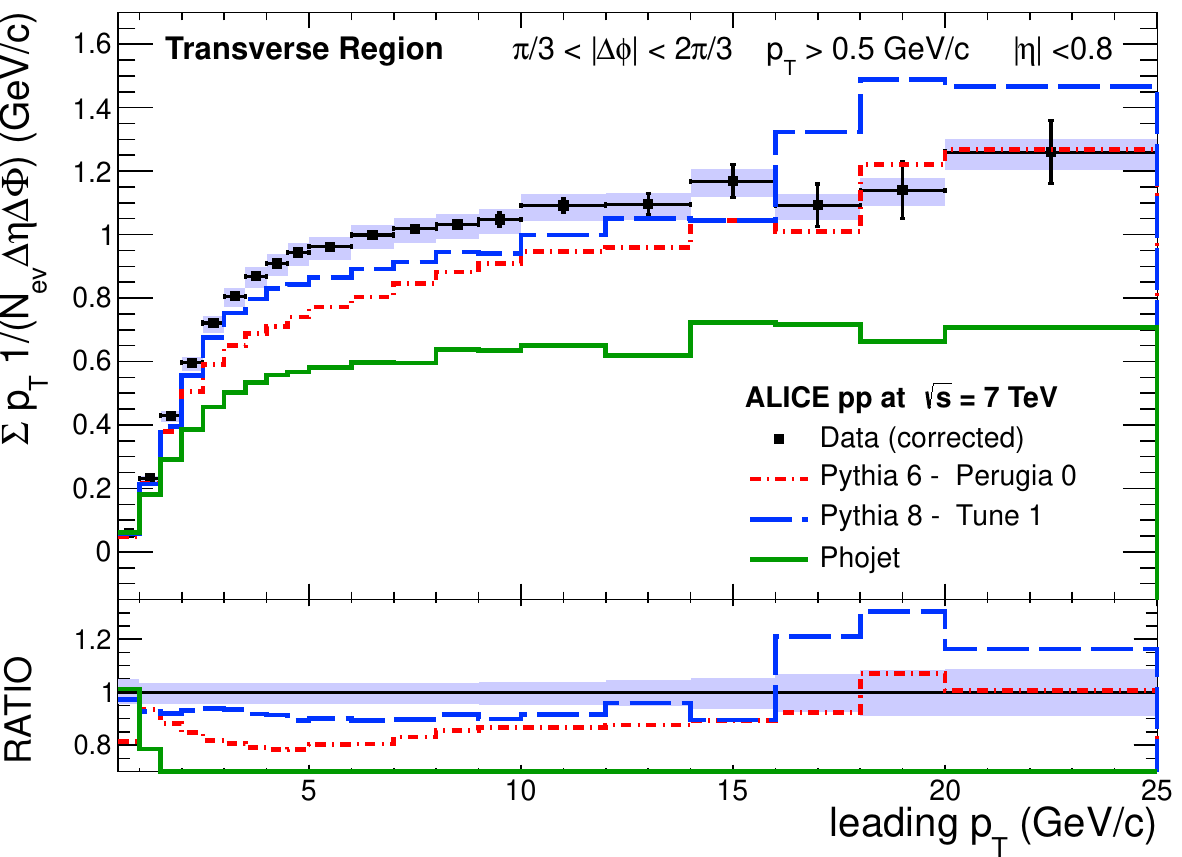}}
{\includegraphics[width=7.6cm]{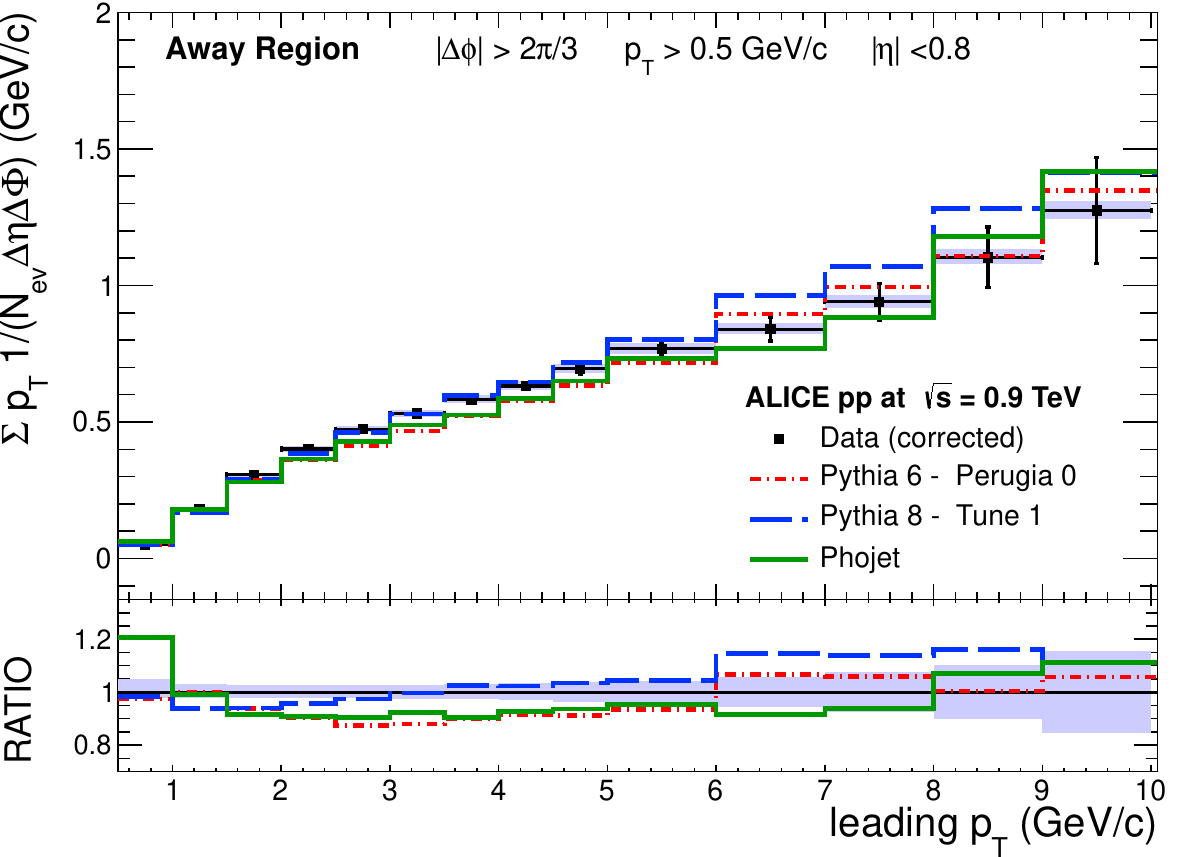}} 
{\includegraphics[width=7.6cm]
{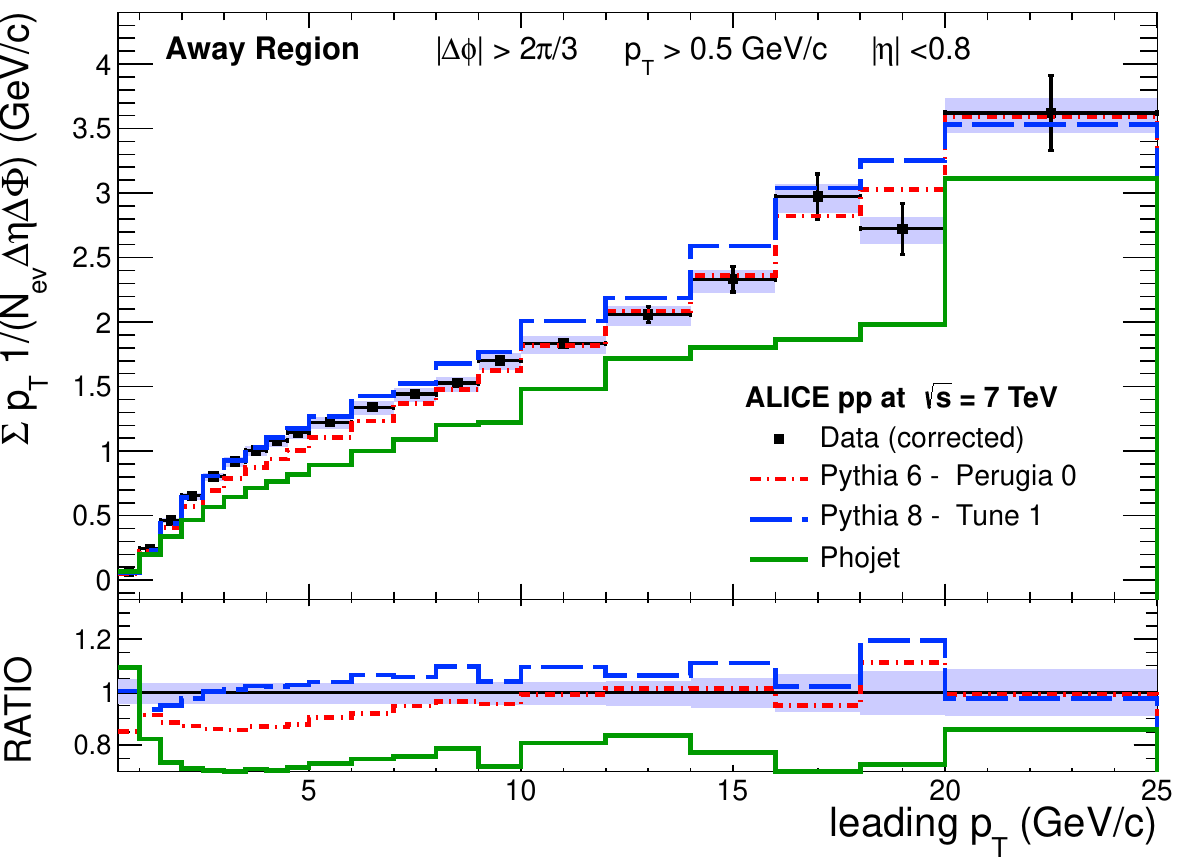}}
{\caption{\textit{Summed  $\pT$ in Toward (top), Transverse (middle) and Away (bottom) regions at $\sqrt{s}=0.9$ TeV (left) and $\sqrt{s}=7$ TeV (right). Right and left vertical scales differ by a factor 4 (2) in the top (middle and bottom) panel. Shaded area in upper plots: systematic uncertainties. Shaded areas in bottom plots: sum in quadrature of statistical and systematic uncertainties. Horizontal error bars: bin width.}}
\label{sumpt_2_away}}

\clearpage
\section*{Summed $\pT$ - track $\pT >  1.0 \, \gmom$}
\label{sumpt_10}

\sixPlotsNoLine[h!]{\includegraphics[width=7.6cm]{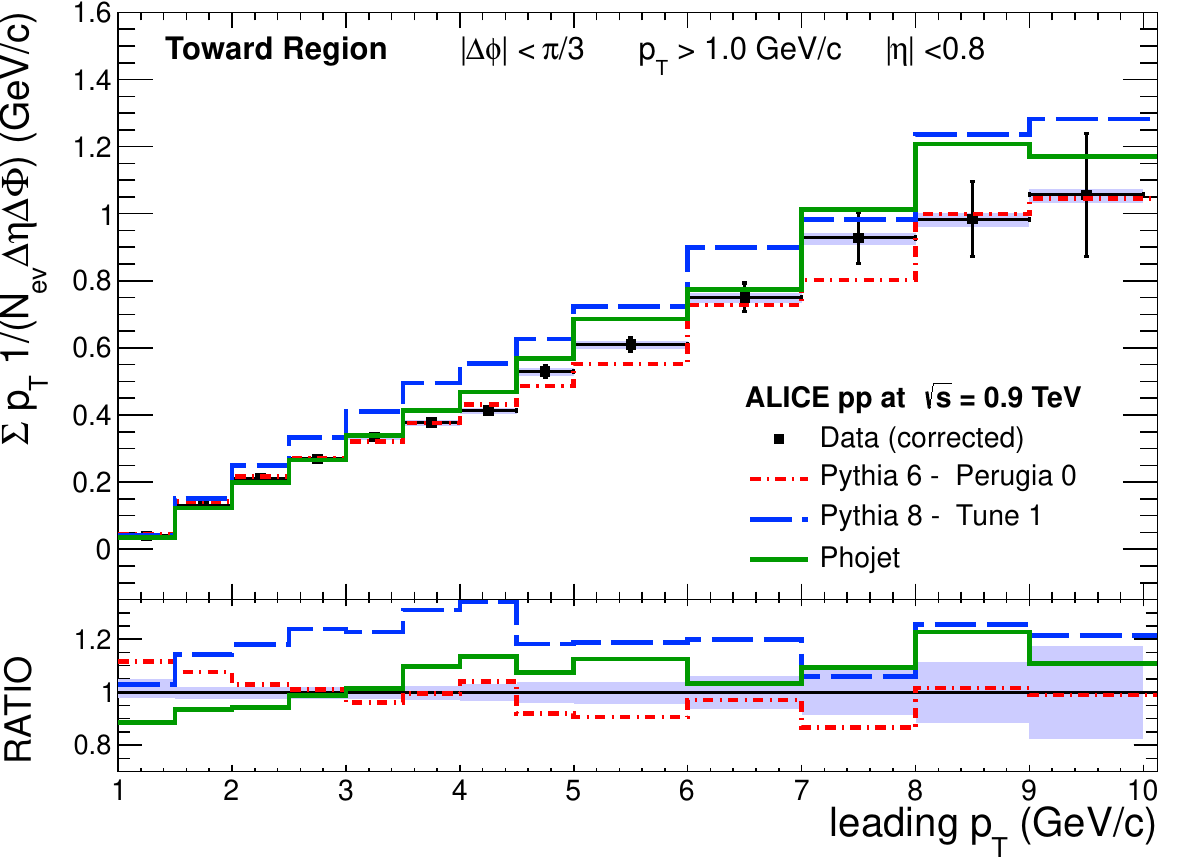}} 
{\includegraphics[width=7.6cm]
{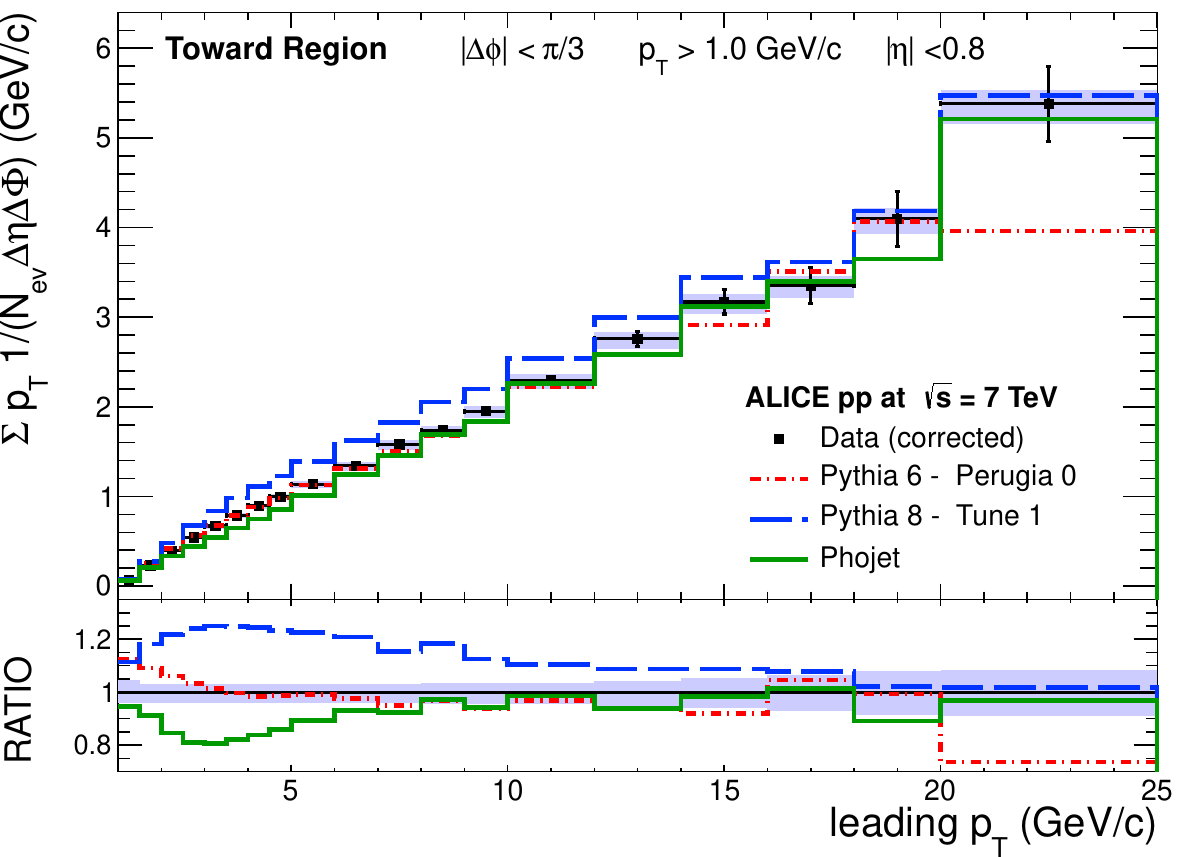}}
{\includegraphics[width=7.6cm]{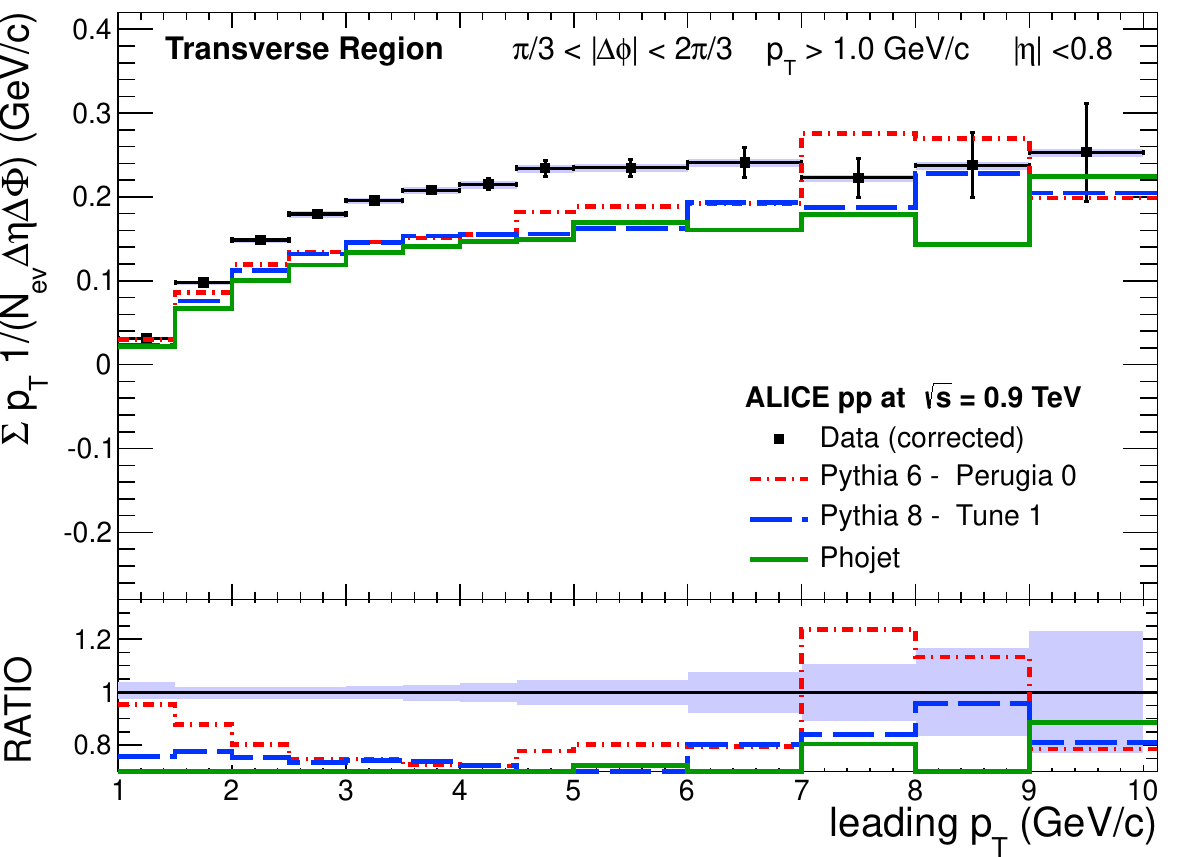}} 
{\includegraphics[width=7.6cm]
{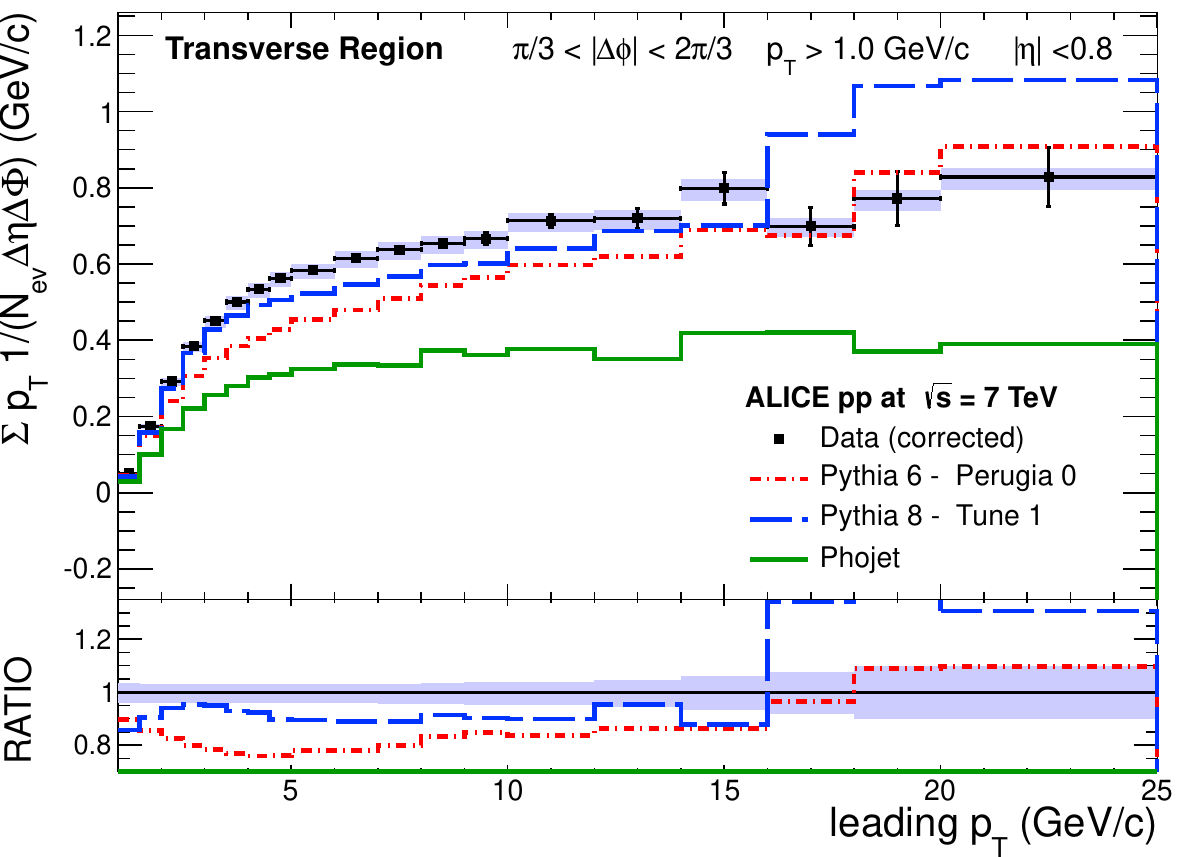}}
{\includegraphics[width=7.6cm]{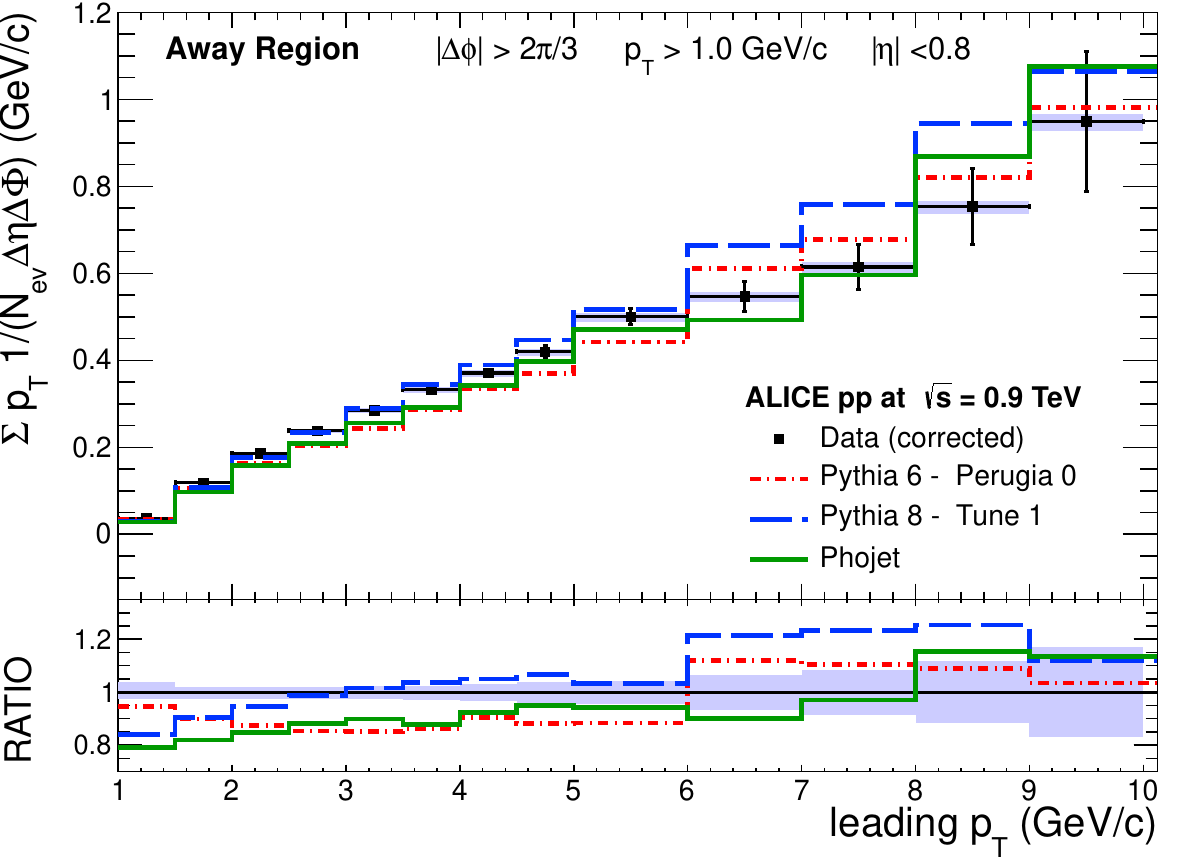}} 
{\includegraphics[width=7.6cm]
{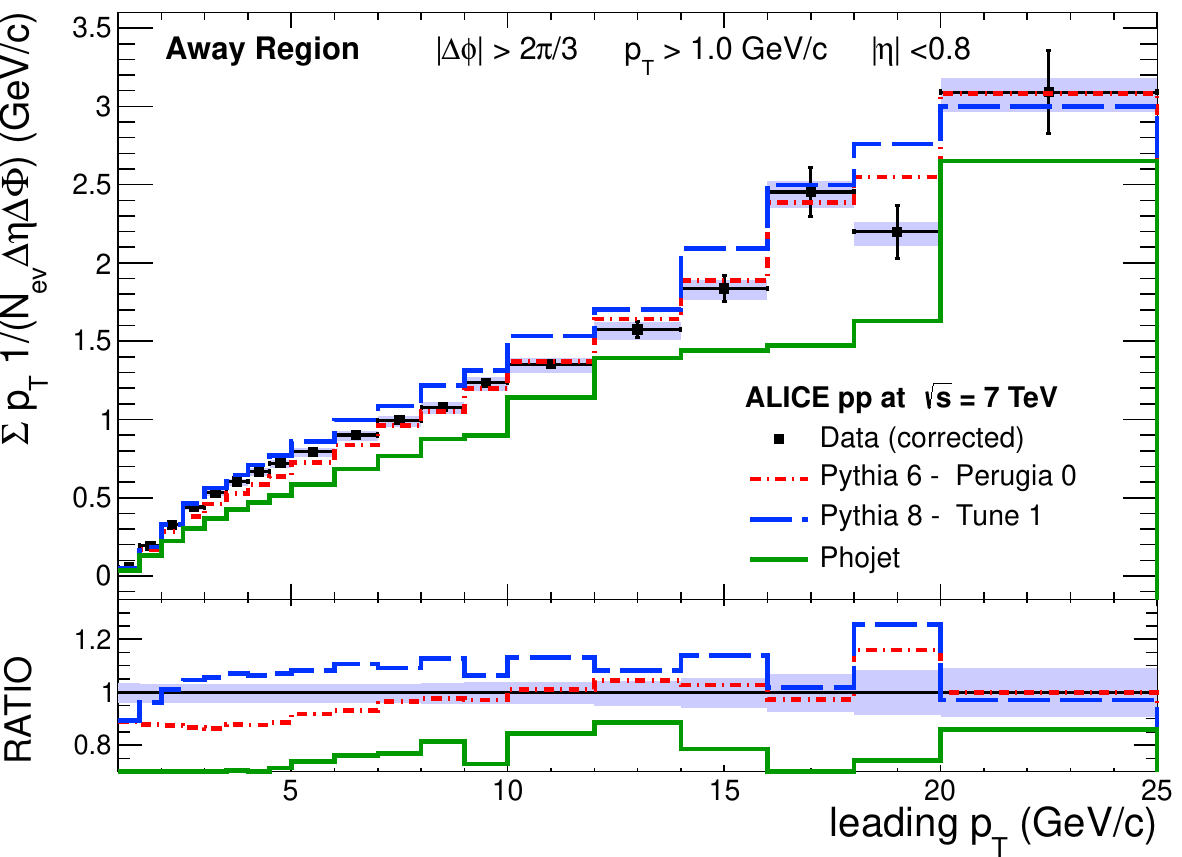}}
{\caption{\textit{Summed  $\pT$ in Toward (top), Transverse (middle) and Away (bottom) regions at $\sqrt{s}=0.9$ TeV (left) and $\sqrt{s}=7$ TeV (right). Right and left vertical scales differ by a factor 4 (3) in the top (middle and bottom) panel. Shaded area in upper plots: systematic uncertainties. Shaded areas in bottom plots: sum in quadrature of statistical and systematic uncertainties. Horizontal error bars: bin width.}}
\label{sumpt_3_away}}

\clearpage

\section*{Azimuthal correlations - track $\pT > 0.15\, \gmom $}
\label{azimuth_015}

\vspace{40pt}
\twoPlotsNoLine[h]{\includegraphics[width=7.6cm]{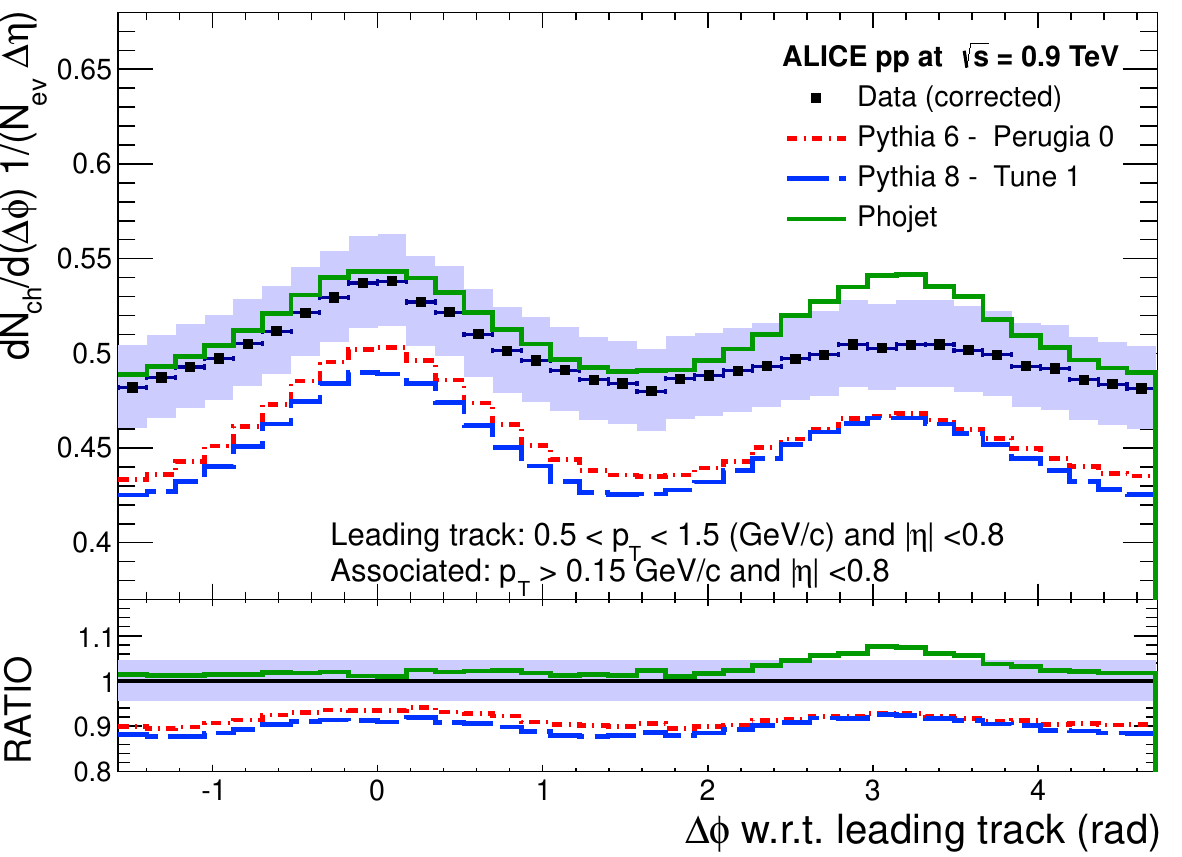}} {\includegraphics[width=7.6cm]{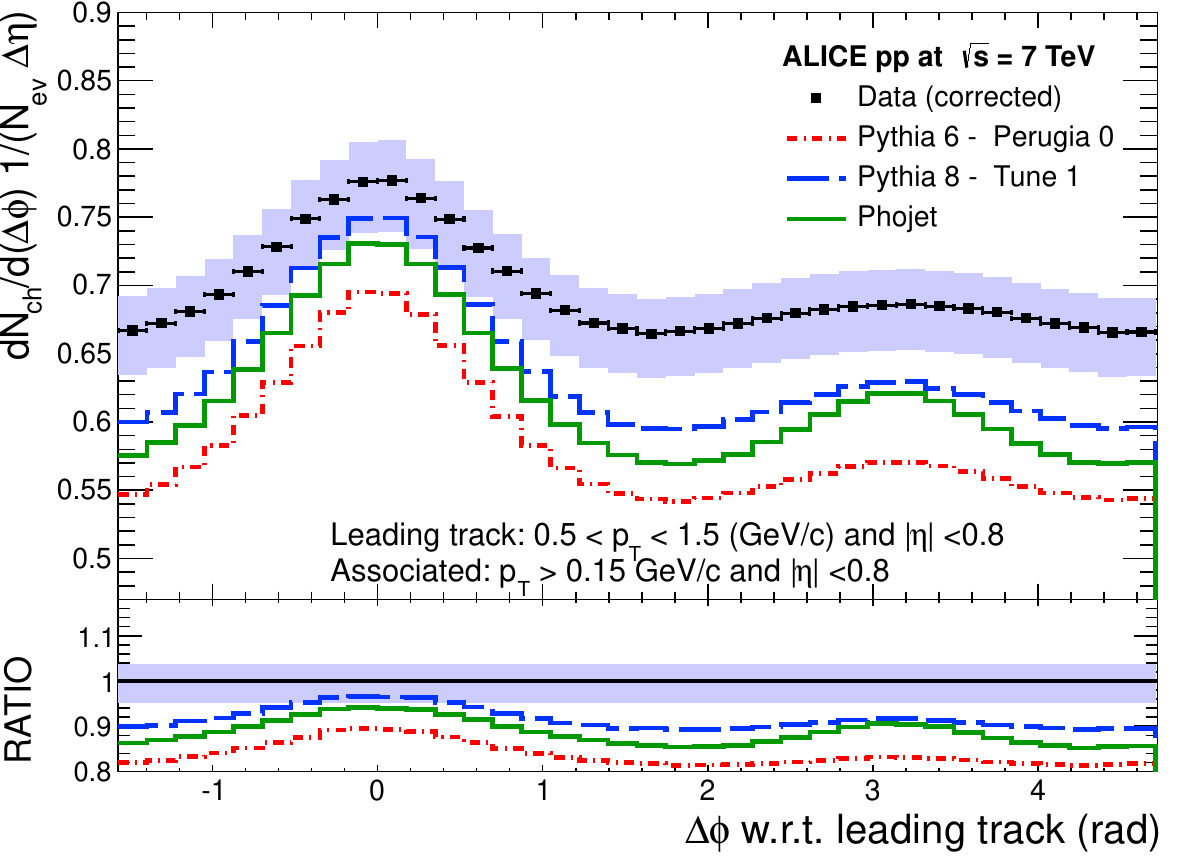}}
{\caption{\textit{Azimuthal correlation at $\sqrt{s}=0.9$ TeV (left) and $\sqrt{s}=7$ TeV (right). Leading-track: 0.5 $<p_{T,LT}<$ \unit[1.5]{GeV/$c$}. For visualization purposes the $\Delta \phi$ axis is not centered around 0. Shaded area in upper plots: systematic uncertainties. Shaded areas in bottom plots: sum in quadrature of statistical and systematic uncertainties. Horizontal error bars: bin width.}}
\label{azimuth_1}}

\vspace{50pt}
\twoPlotsNoLine[h!]{\includegraphics[width=7.6cm]{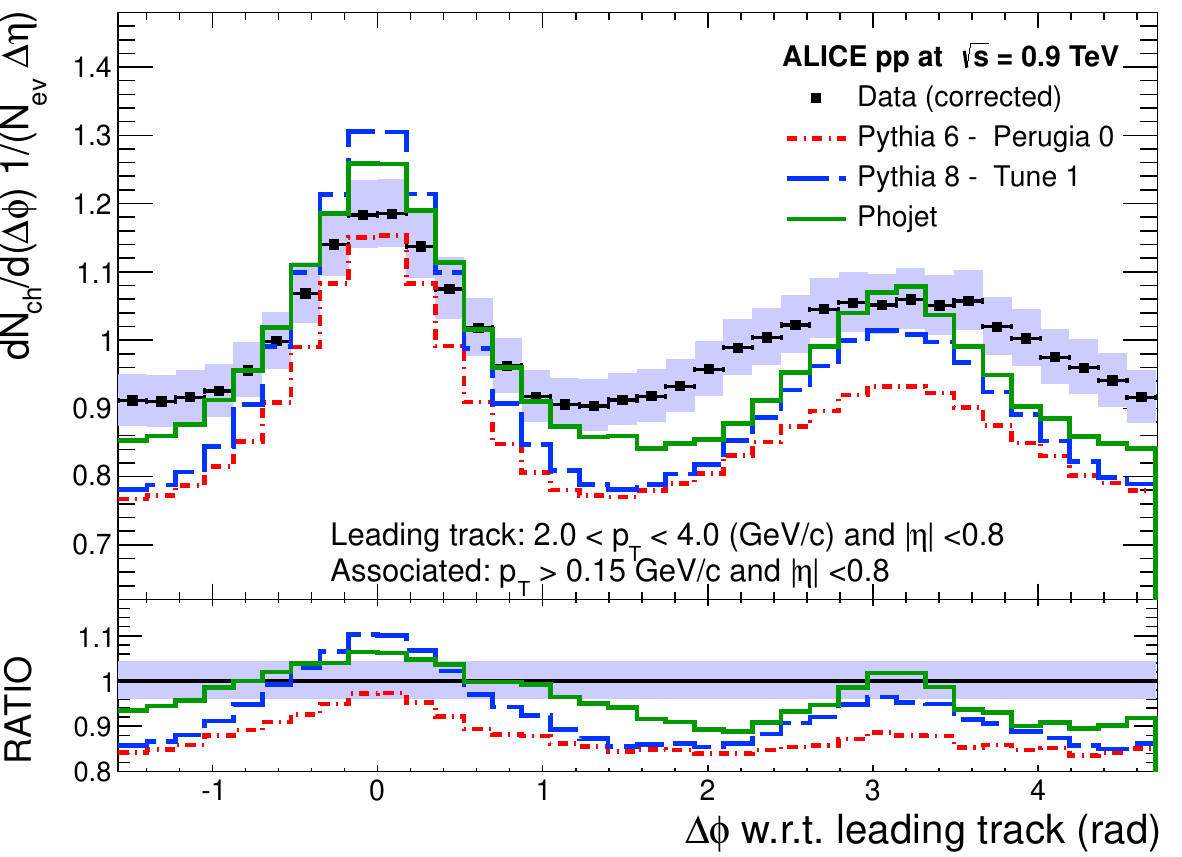}} {\includegraphics[width=7.6cm]{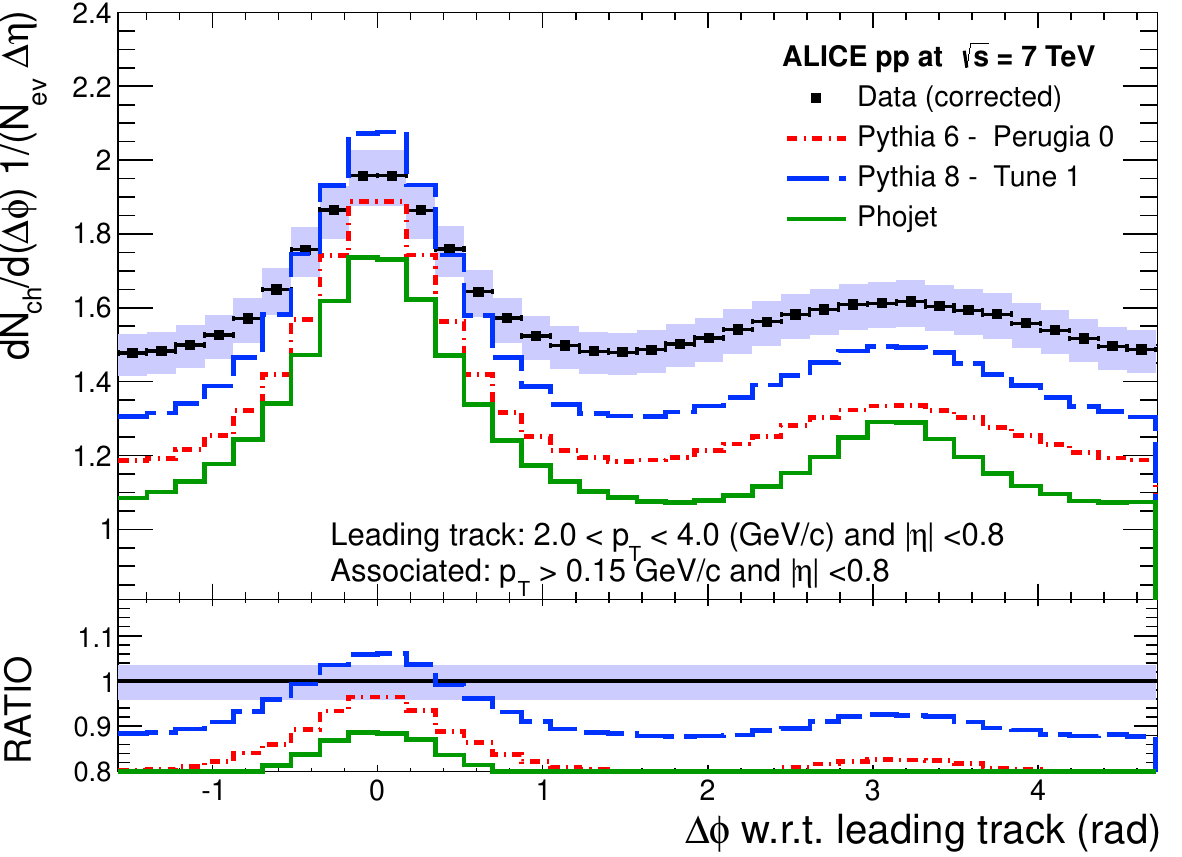}}
{\caption{\textit{Azimuthal correlation at $\sqrt{s}=0.9$ TeV (left) and $\sqrt{s}=7$ TeV (right). Leading-track: 2.0 $<p_{T,LT}<$ \unit[4.0]{GeV/$c$}. For visualization purposes the $\Delta \phi$ axis is not centered around 0. Shaded area in upper plots: systematic uncertainties. Shaded areas in bottom plots: sum in quadrature of statistical and systematic uncertainties. Horizontal error bars: bin width.}}
\label{azimuth_2}}


\twoPlotsNoLine[t!]{\includegraphics[width=7.6cm]{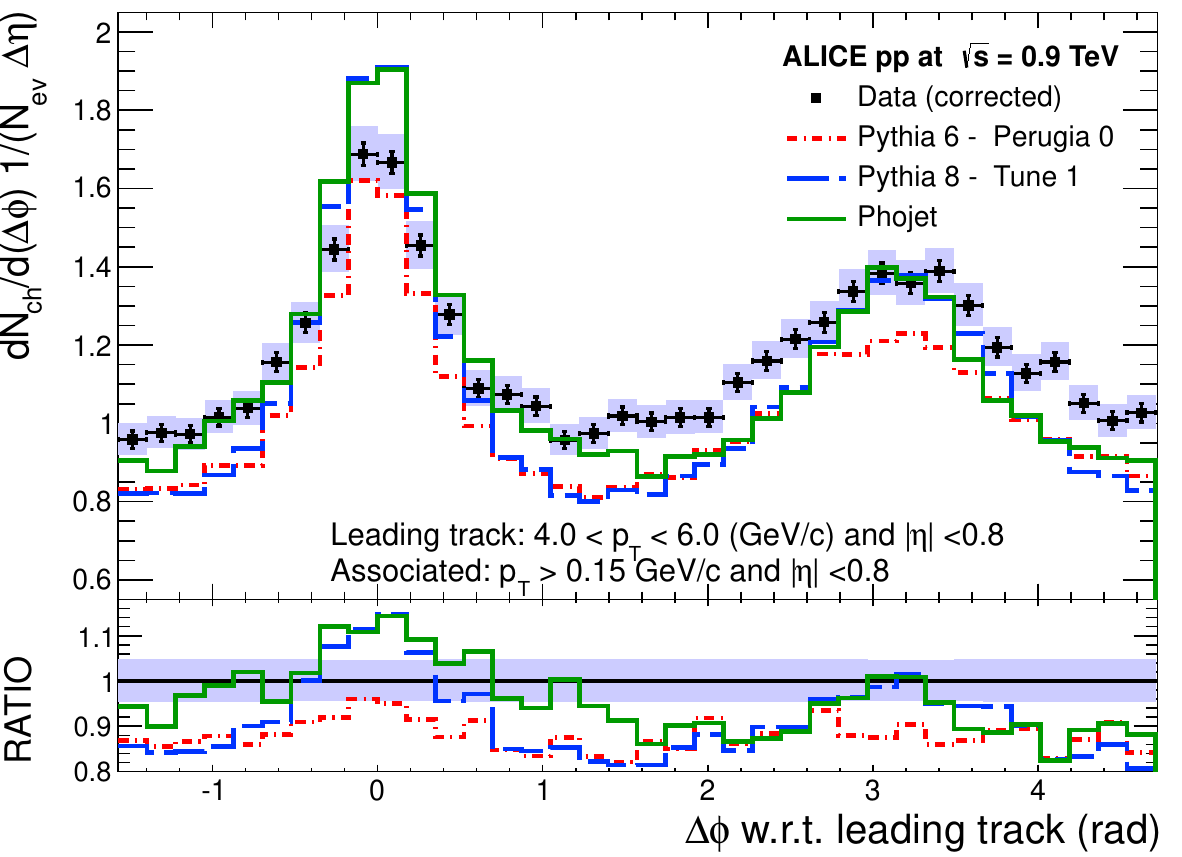}} {\includegraphics[width=7.6cm]{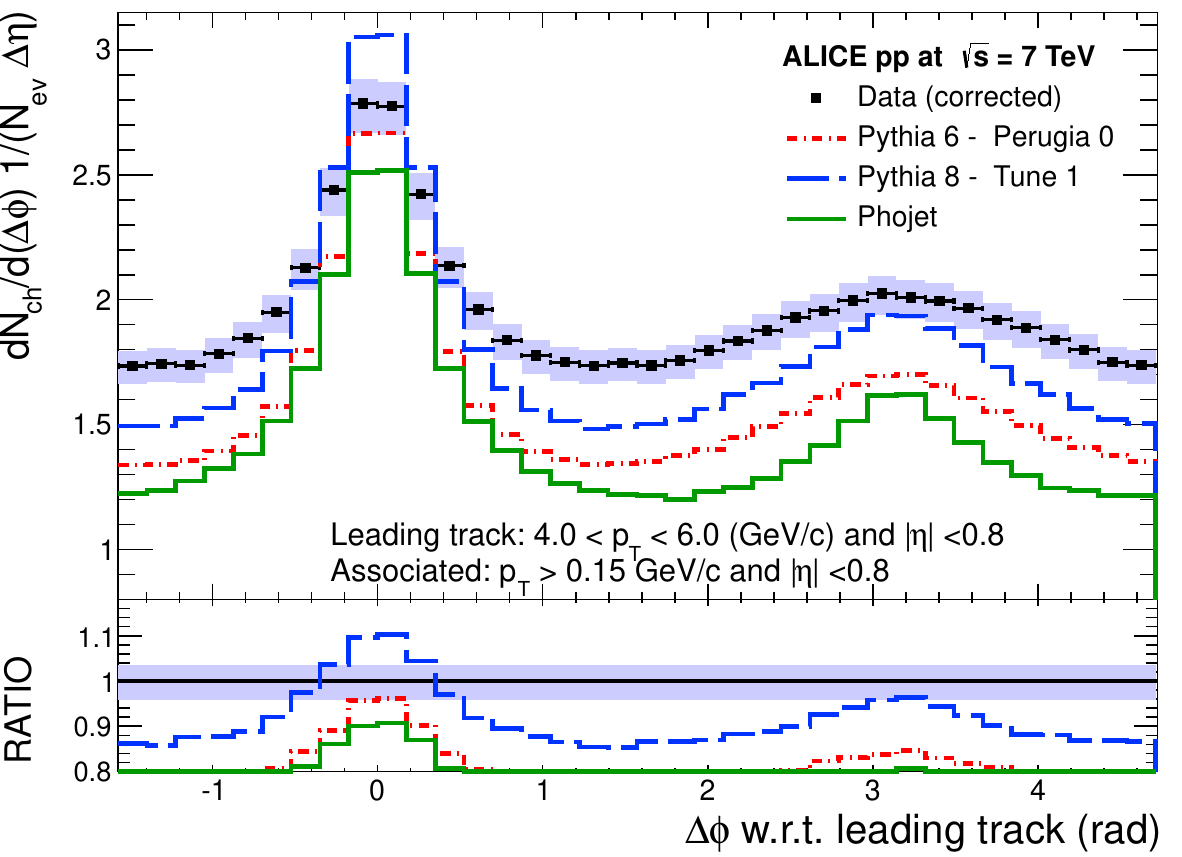}}
{\caption{\textit{Azimuthal correlation at $\sqrt{s}=0.9$ TeV (left) and $\sqrt{s}=7$ TeV (right).  Leading-track: 4.0 $<p_{T,LT}<$ \unit[6.0]{GeV/$c$}. For visualization purposes the $\Delta \phi$ axis is not centered around 0. Shaded area in upper plots: systematic uncertainties. Shaded areas in bottom plots: sum in quadrature of statistical and systematic uncertainties. Horizontal error bars: bin width.}}
\label{azimuth_3}}

\twoPlotsNoLine[b!]{\includegraphics[width=7.6cm]{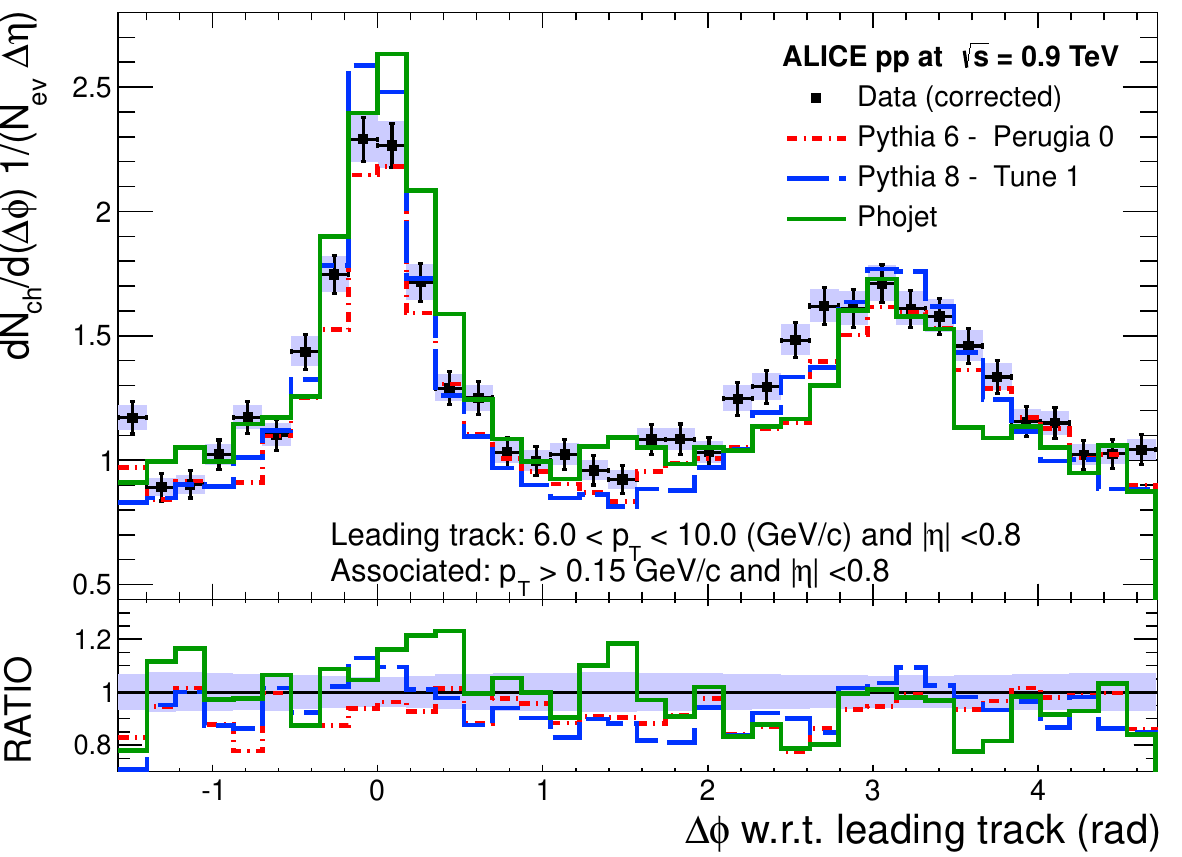}} {\includegraphics[width=7.6cm]{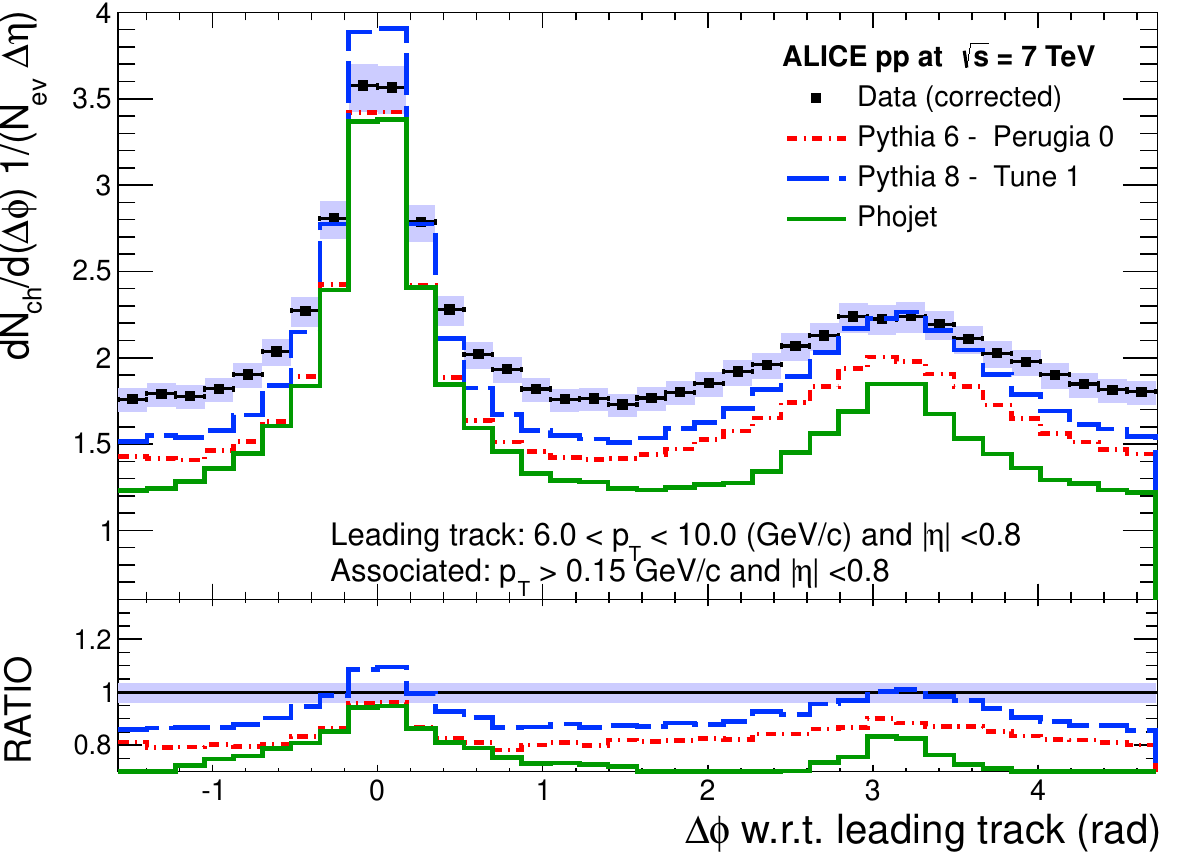}}
{\caption{\textit{Azimuthal correlation at $\sqrt{s}=0.9$ TeV (left) and $\sqrt{s}=7$ TeV (right).  Leading-track: 6.0 $<p_{T,LT}<$ \unit[10.0]{GeV/$c$}. For visualization purposes the $\Delta \phi$ axis is not centered around 0. Shaded area in upper plots: systematic uncertainties. Shaded areas in bottom plots: sum in quadrature of statistical and systematic uncertainties. Horizontal error bars: bin width.}}
\label{azimuth_4}}

\clearpage
\section*{Azimuthal correlations - track $\pT > 0.5 \, \gmom $}
\label{azimuth_05}
\vspace{40pt}
\twoPlotsNoLine[h!]{\includegraphics[width=7.6cm]{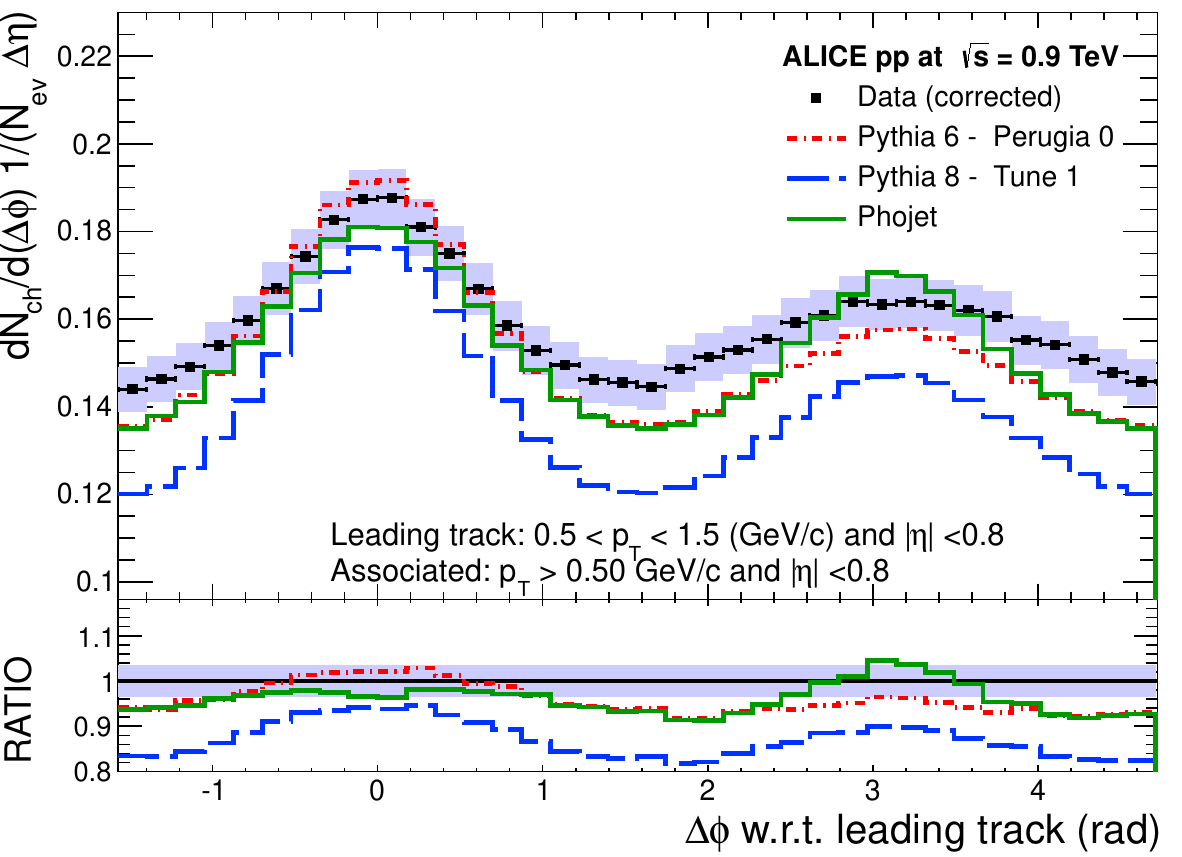}} {\includegraphics[width=7.6cm]{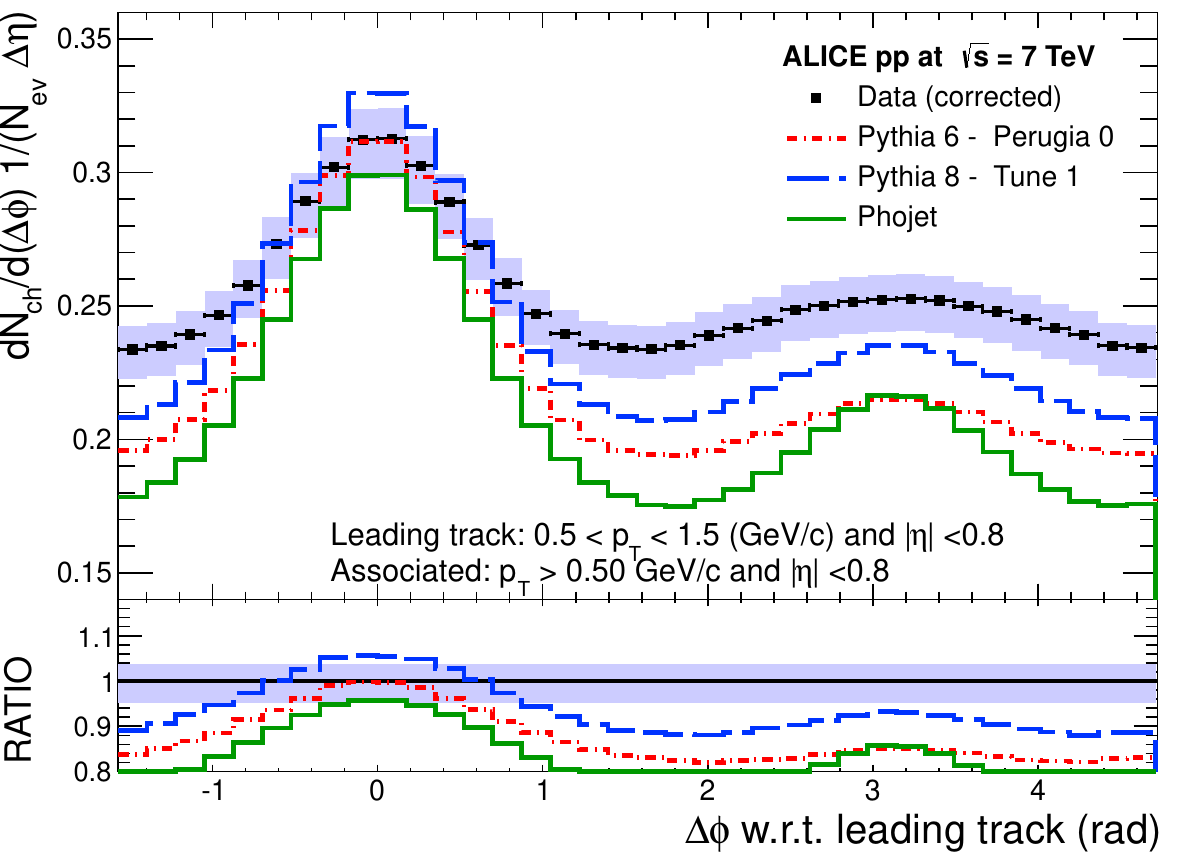}}
{\caption{\textit{Azimuthal correlation at $\sqrt{s}=0.9$ TeV (left) and $\sqrt{s}=7$ TeV (right).  Leading-track: 0.5 $<p_{T,LT}<$ \unit[1.5]{GeV/$c$}. For visualization purposes the $\Delta \phi$ axis is not centered around 0. Shaded area in upper plots: systematic uncertainties. Shaded areas in bottom plots: sum in quadrature of statistical and systematic uncertainties. Horizontal error bars: bin width.}}
\label{azimuth_5}}
\vspace{50pt}
\twoPlotsNoLine[h!]{\includegraphics[width=7.6cm]{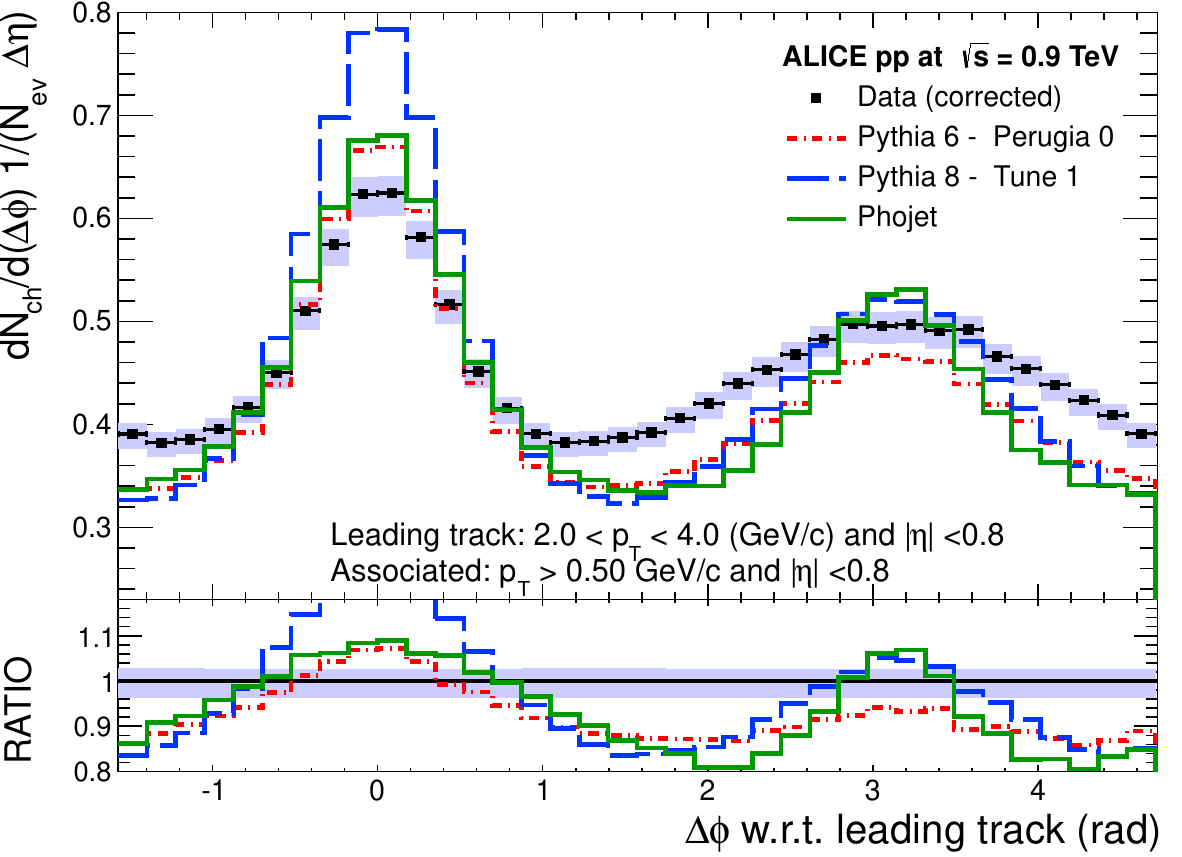}} {\includegraphics[width=7.6cm]{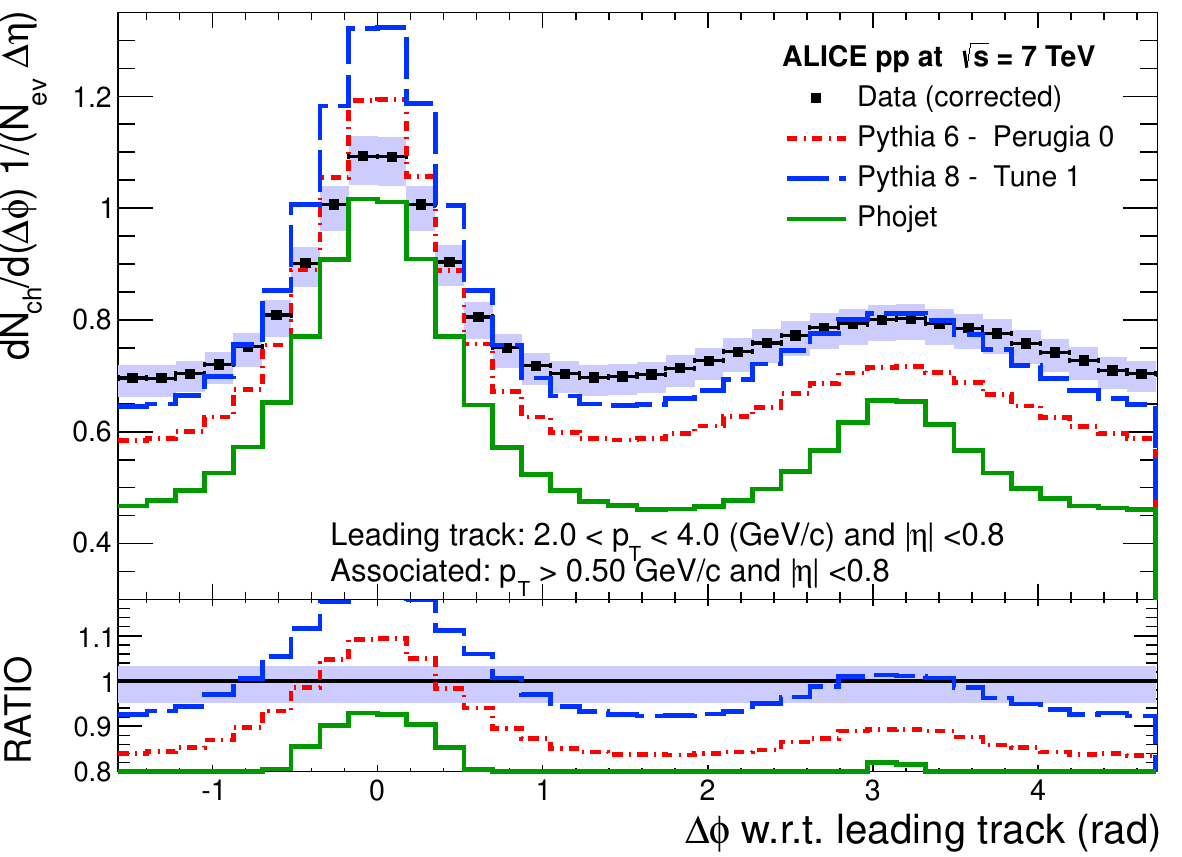}}
{\caption{\textit{Azimuthal correlation at $\sqrt{s}=0.9$ TeV (left) and $\sqrt{s}=7$ TeV (right). Leading-track: 2.0 $<p_{T,LT}<$ \unit[4.0]{GeV/$c$}. For visualization purposes the $\Delta \phi$ axis is not centered around 0. Shaded area in upper plots: systematic uncertainties. Shaded areas in bottom plots: sum in quadrature of statistical and systematic uncertainties. Horizontal error bars: bin width.}}
\label{azimuth_6}}

\twoPlotsNoLine[t!]{\includegraphics[width=7.6cm]{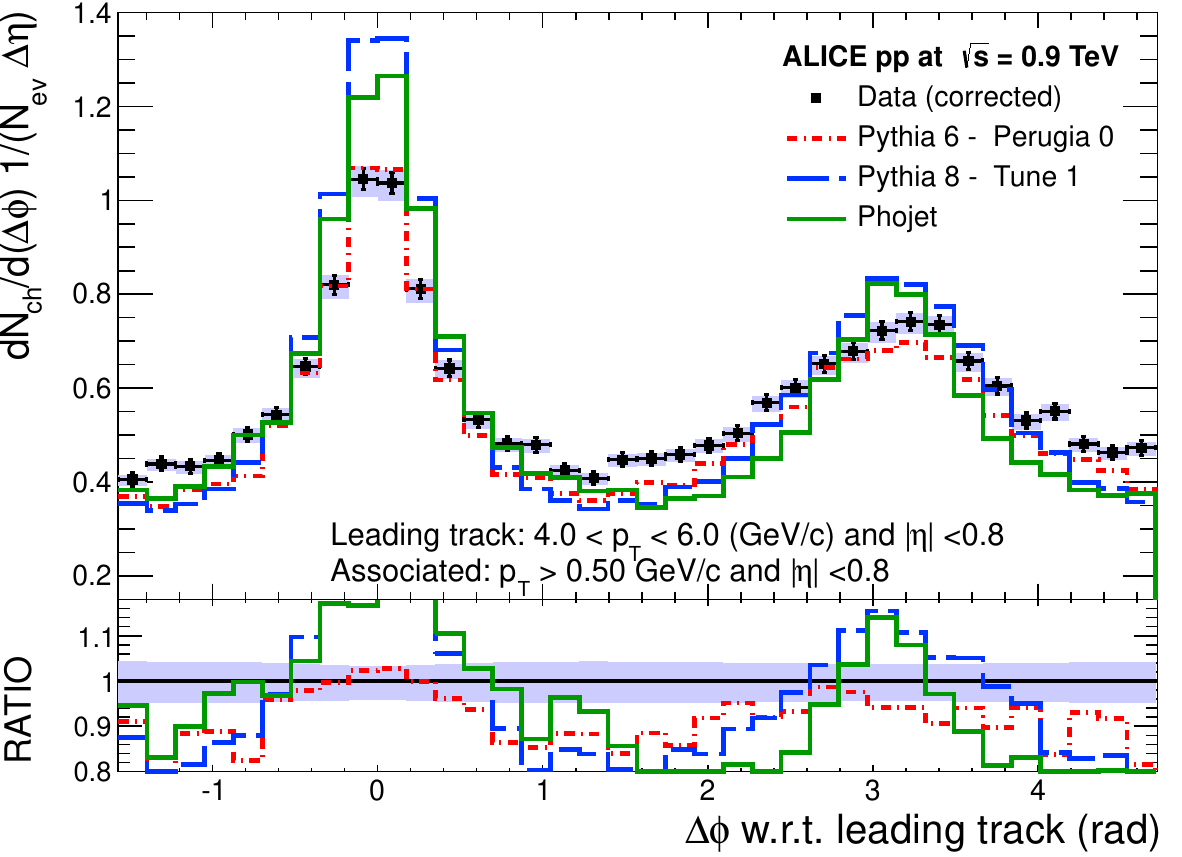}} {\includegraphics[width=7.6cm]{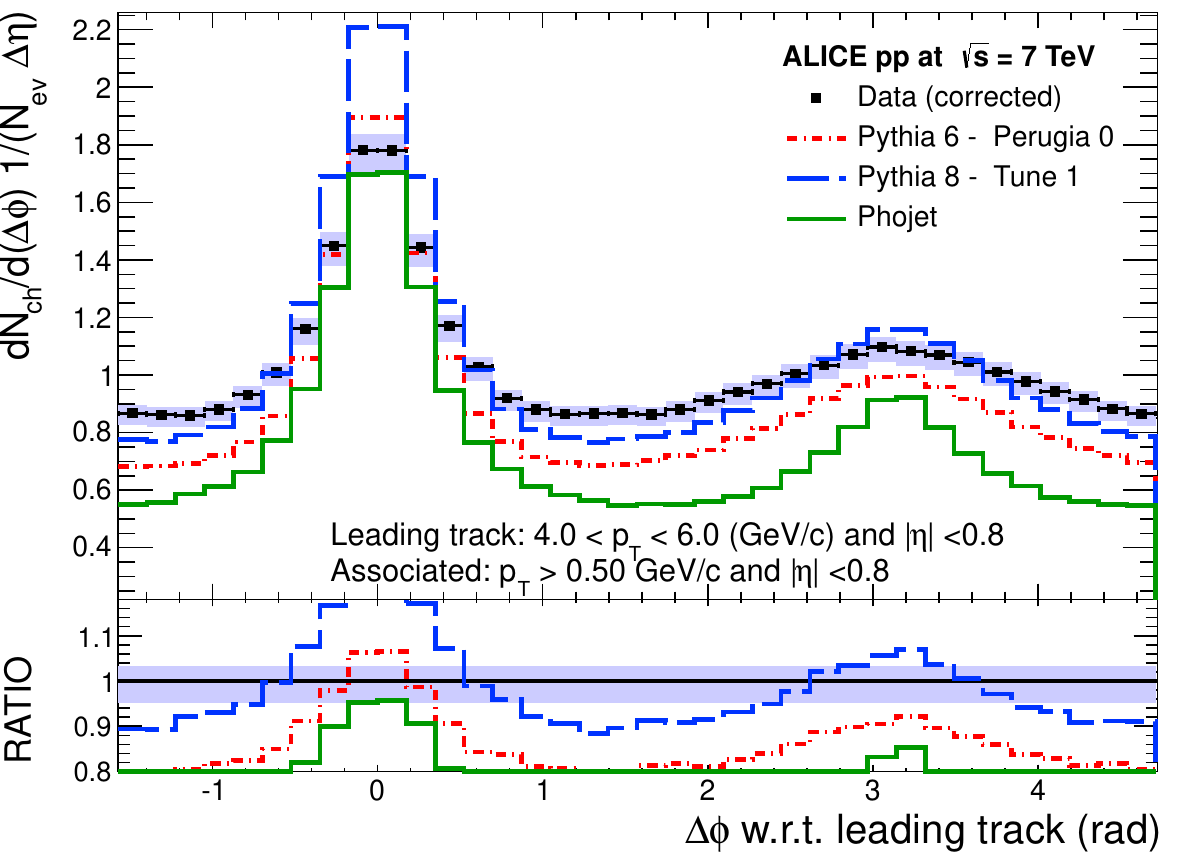}}
{\caption{\textit{Azimuthal correlation at $\sqrt{s}=0.9$ TeV (left) and $\sqrt{s}=7$ TeV (right).  Leading-track: 4.0 $<p_{T,LT}<$ \unit[6.0]{GeV/$c$}. For visualization purposes the $\Delta \phi$ axis is not centered around 0. Shaded area in upper plots: systematic uncertainties. Shaded areas in bottom plots: sum in quadrature of statistical and systematic uncertainties. Horizontal error bars: bin width.}}
\label{azimuth_7}}

\twoPlotsNoLine[b!]{\includegraphics[width=7.6cm]{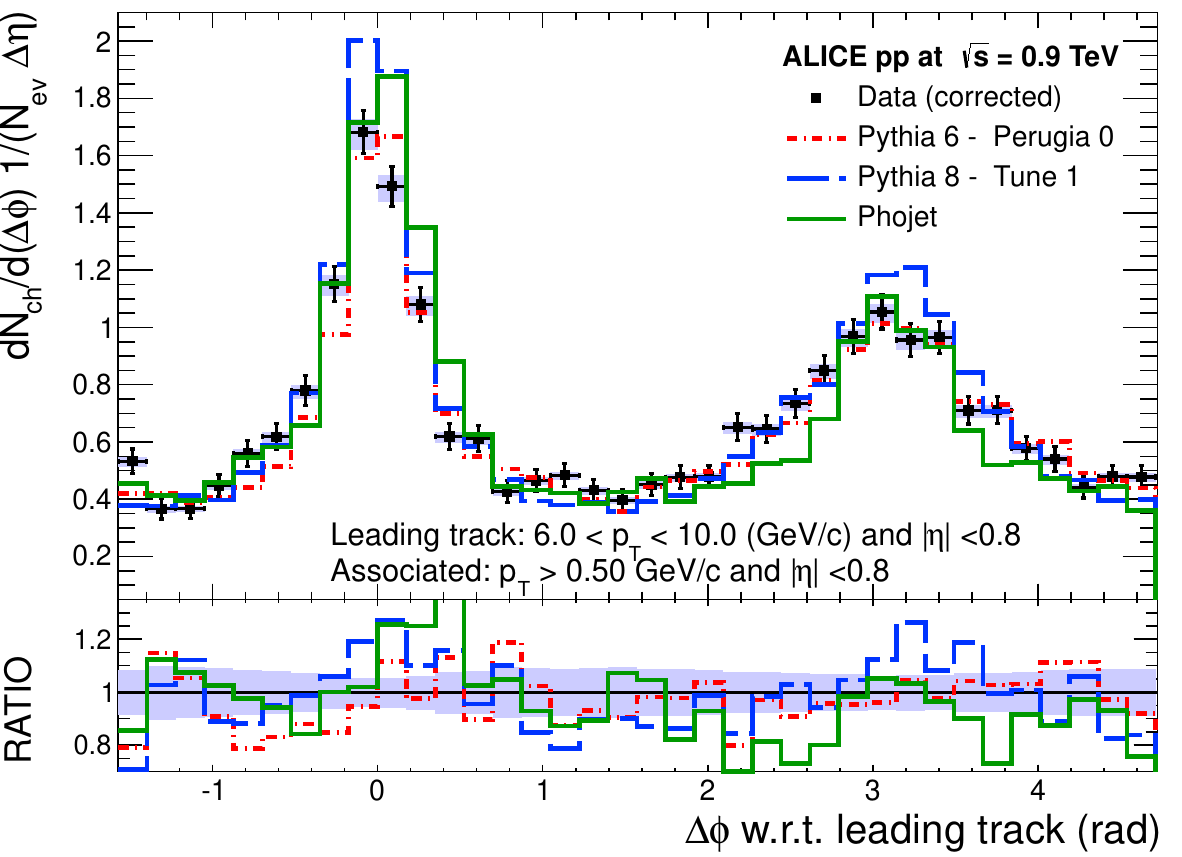}} {\includegraphics[width=7.6cm]{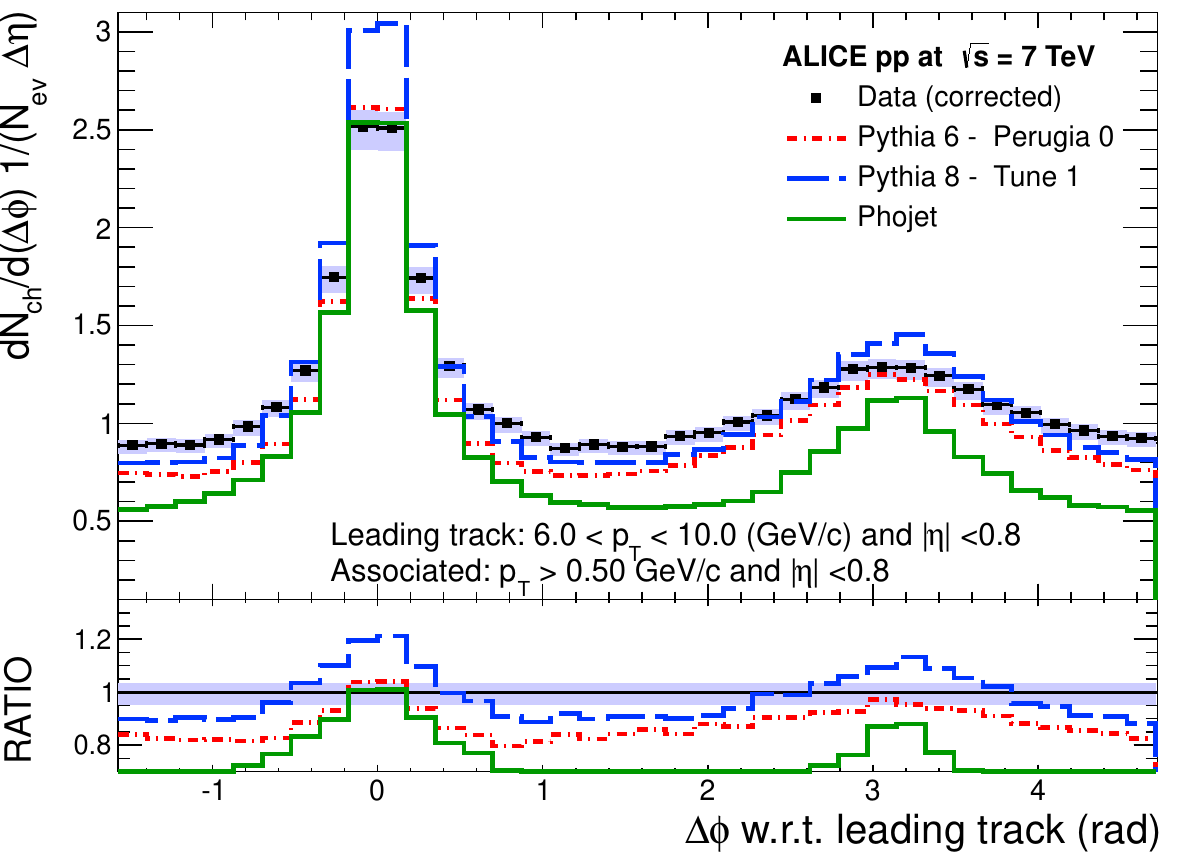}}
{\caption{\textit{Azimuthal correlation at $\sqrt{s}=0.9$ TeV (left) and $\sqrt{s}=7$ TeV (right). Leading-track: 6.0 $<p_{T,LT}<$ \unit[10.0]{GeV/$c$}.  For visualization purposes the $\Delta \phi$ axis is not centered around 0. Shaded area in upper plots: systematic uncertainties. Shaded areas in bottom plots: sum in quadrature of statistical and systematic uncertainties. Horizontal error bars: bin width.}}
\label{azimuth_8}}

\clearpage
\section*{Azimuthal correlations - track $\pT >  1.0 \, \gmom$}
\vspace{40pt}
\twoPlotsNoLine[h!]{\includegraphics[width=7.6cm]{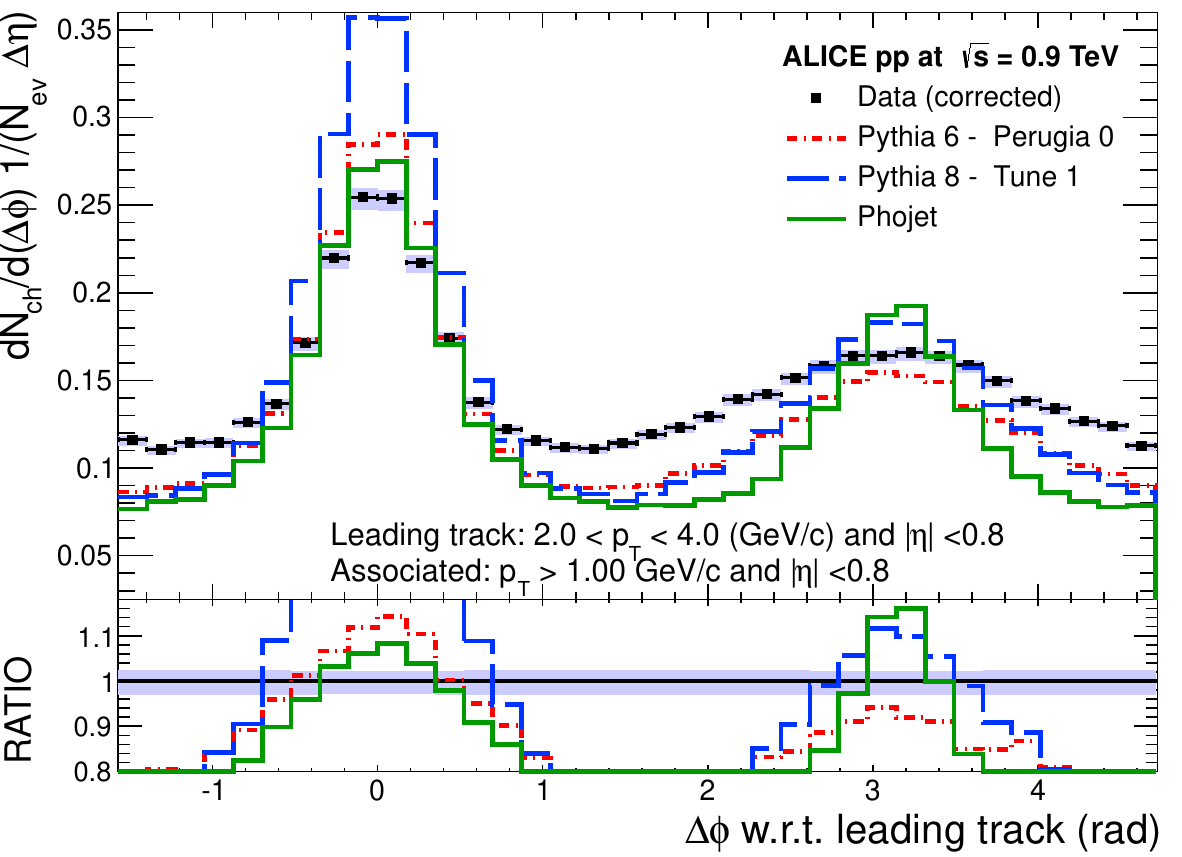}} {\includegraphics[width=7.6cm]{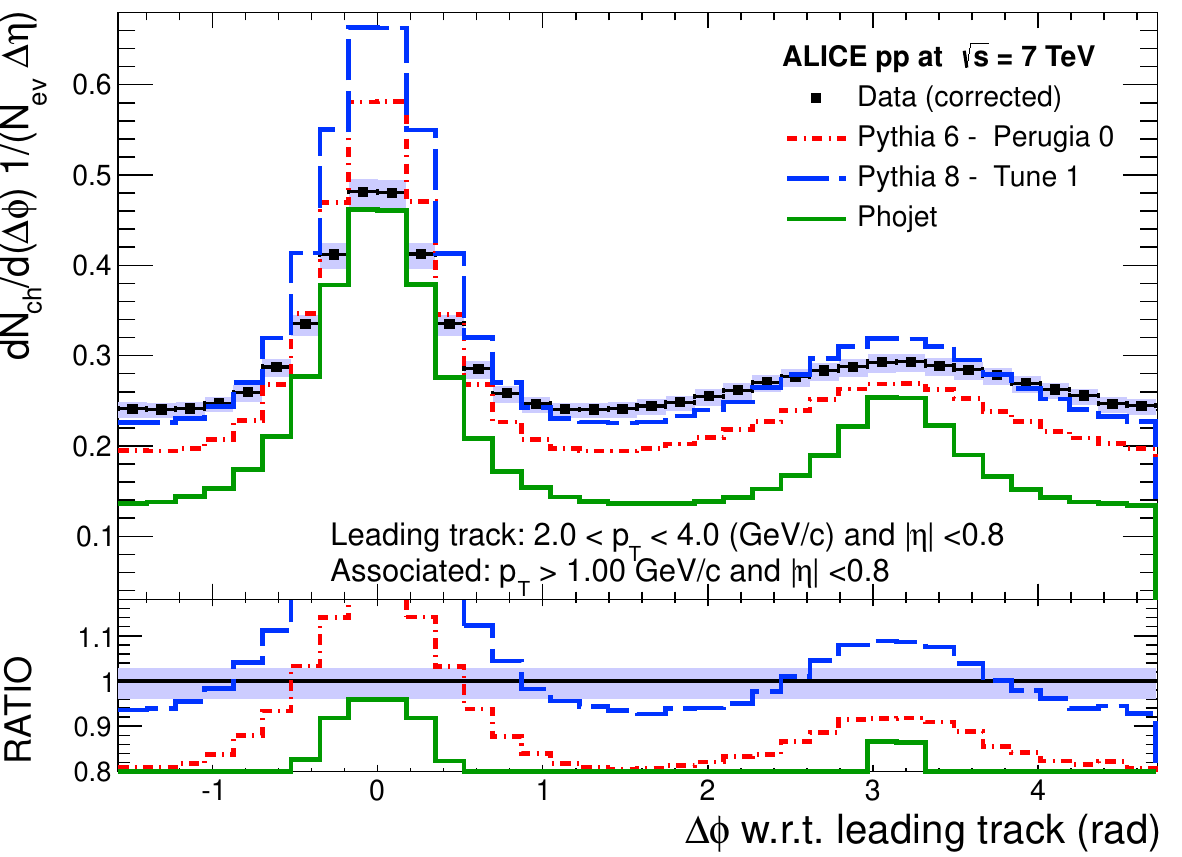}}
{\caption{\textit{Azimuthal correlation at $\sqrt{s}=0.9$ TeV (left) and $\sqrt{s}=7$ TeV (right). Leading-track: 2.0 $<p_{T,LT}<$ \unit[4.0]{GeV/$c$}. For visualization purposes the $\Delta \phi$ axis is not centered around 0. Shaded area in upper plots: systematic uncertainties. Shaded areas in bottom plots: sum in quadrature of statistical and systematic uncertainties. Horizontal error bars: bin width.}}
\label{azimuth_9}}
\vspace{50pt}
\twoPlotsNoLine[h!]{\includegraphics[width=7.6cm]{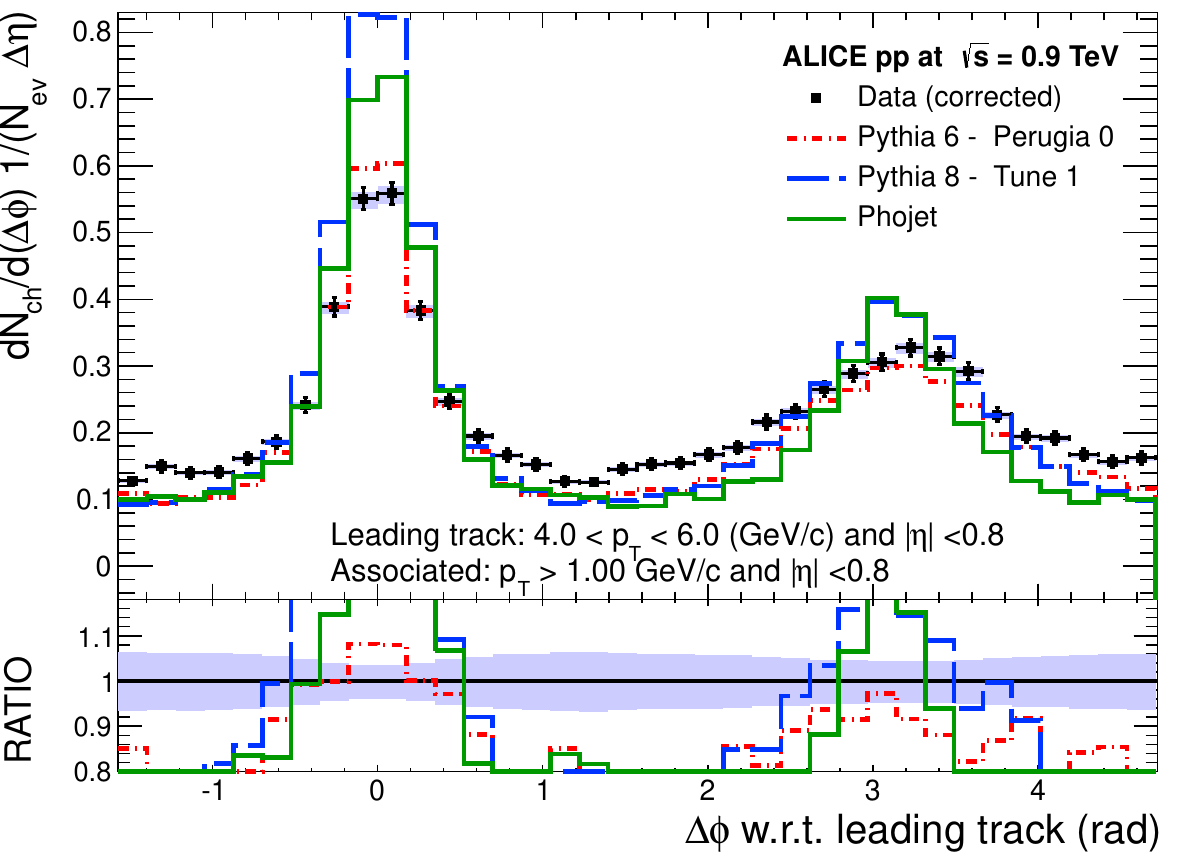}} {\includegraphics[width=7.6cm]{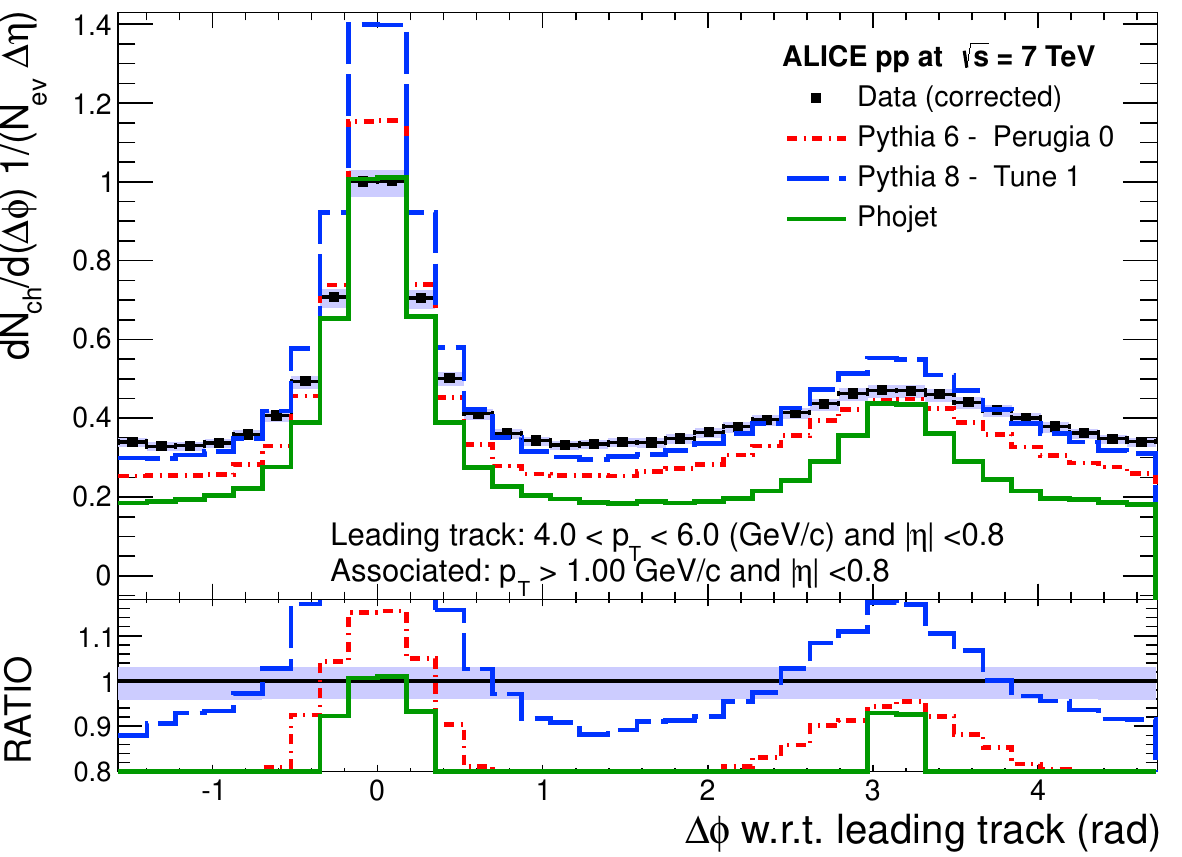}}
{\caption{\textit{Azimuthal correlation at $\sqrt{s}=0.9$ TeV (left) and $\sqrt{s}=7$ TeV (right).  Leading-track: 4.0 $<p_{T,LT}<$ \unit[6.0]{GeV/$c$}. For visualization purposes the $\Delta \phi$ axis is not centered around 0. Shaded area in upper plots: systematic uncertainties. Shaded areas in bottom plots: sum in quadrature of statistical and systematic uncertainties. Horizontal error bars: bin width.}}
\label{azimuth_10}}

\clearpage
\twoPlotsNoLine[h!]{\includegraphics[width=7.6cm]{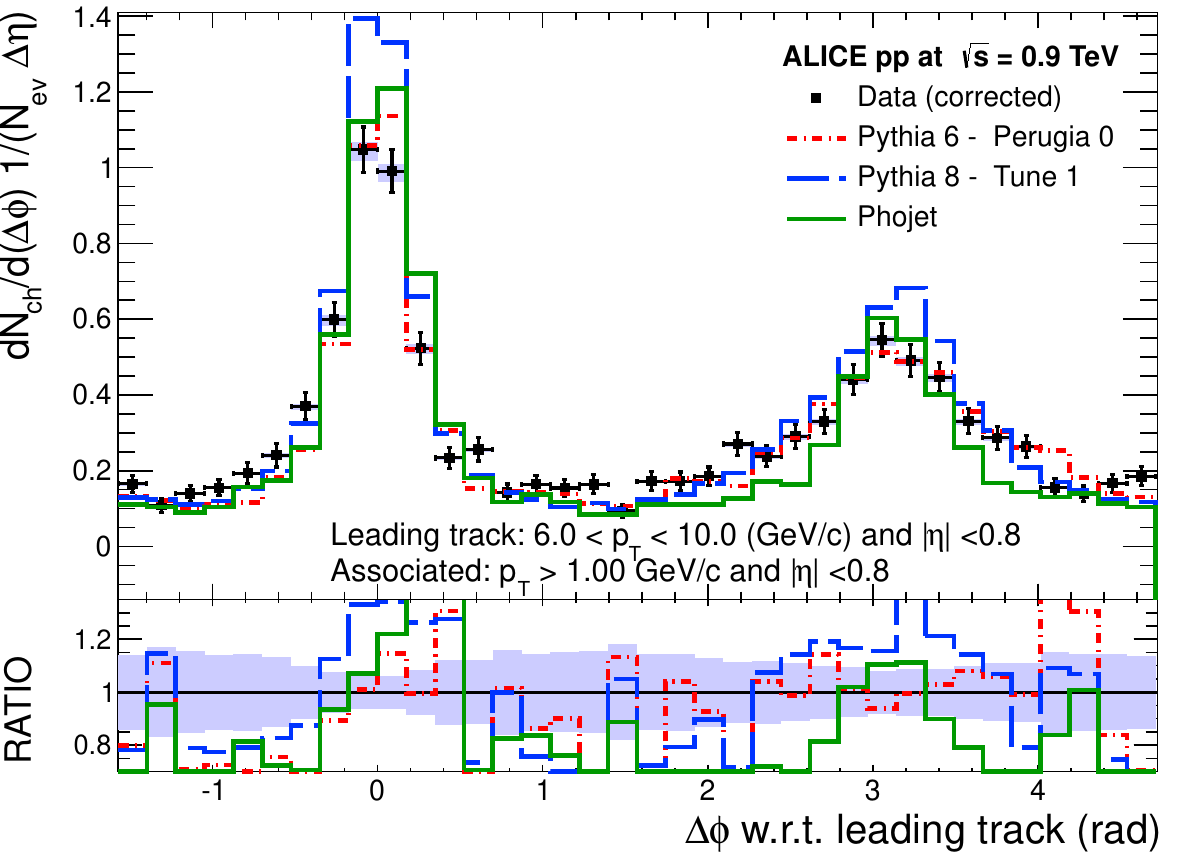}} {\includegraphics[width=7.6cm]{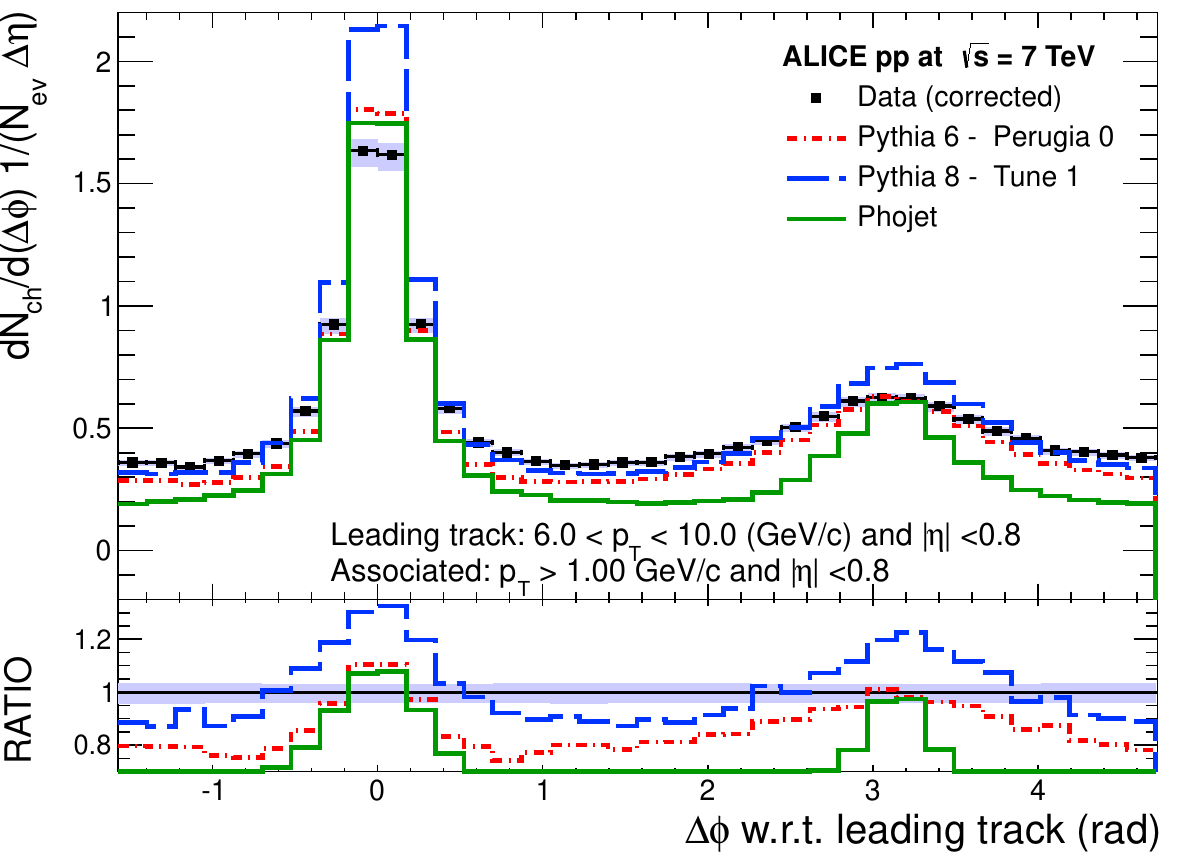}}
{\caption{\textit{Azimuthal correlation at $\sqrt{s}=0.9$ TeV (left) and $\sqrt{s}=7$ TeV (right).  Leading-track: 6.0 $<p_{T,LT}<$ \unit[10.0]{GeV/$c$}. For visualization purposes the $\Delta \phi$ axis is not centered around 0. Shaded area in upper plots: systematic uncertainties. Shaded areas in bottom plots: sum in quadrature of statistical and systematic uncertainties. Horizontal error bars: bin width.}}
\label{azimuth_11}}

%% file: acknowledgements_Nov2011.tex
The ALICE collaboration would like to thank all its engineers and technicians for their invaluable contributions to the construction of the experiment and the CERN accelerator teams for the outstanding performance of the LHC complex.
\\
The ALICE collaboration acknowledges the following funding agencies for their support in building and
running the ALICE detector:
 \\
Calouste Gulbenkian Foundation from Lisbon and Swiss Fonds Kidagan, Armenia;
 \\
Conselho Nacional de Desenvolvimento Cient\'{\i}fico e Tecnol\'{o}gico (CNPq), Financiadora de Estudos e Projetos (FINEP),
Funda\c{c}\~{a}o de Amparo \`{a} Pesquisa do Estado de S\~{a}o Paulo (FAPESP);
 \\
National Natural Science Foundation of China (NSFC), the Chinese Ministry of Education (CMOE)
and the Ministry of Science and Technology of China (MSTC);
 \\
Ministry of Education and Youth of the Czech Republic;
 \\
Danish Natural Science Research Council, the Carlsberg Foundation and the Danish National Research Foundation;
 \\
The European Research Council under the European Community's Seventh Framework Programme;
 \\
Helsinki Institute of Physics and the Academy of Finland;
 \\
French CNRS-IN2P3, the `Region Pays de Loire', `Region Alsace', `Region Auvergne' and CEA, France;
 \\
German BMBF and the Helmholtz Association;
\\
General Secretariat for Research and Technology, Ministry of
Development, Greece;
\\
Hungarian OTKA and National Office for Research and Technology (NKTH);
 \\
Department of Atomic Energy and Department of Science and Technology of the Government of India;
 \\
Istituto Nazionale di Fisica Nucleare (INFN) of Italy;
 \\
MEXT Grant-in-Aid for Specially Promoted Research, Ja\-pan;
 \\
Joint Institute for Nuclear Research, Dubna;
 \\
National Research Foundation of Korea (NRF);
 \\
CONACYT, DGAPA, M\'{e}xico, ALFA-EC and the HELEN Program (High-Energy physics Latin-American--European Network);
 \\
Stichting voor Fundamenteel Onderzoek der Materie (FOM) and the Nederlandse Organisatie voor Wetenschappelijk Onderzoek (NWO), Netherlands;
 \\
Research Council of Norway (NFR);
 \\
Polish Ministry of Science and Higher Education;
 \\
National Authority for Scientific Research - NASR (Autoritatea Na\c{t}ional\u{a} pentru Cercetare \c{S}tiin\c{t}ific\u{a} - ANCS);
 \\
Federal Agency of Science of the Ministry of Education and Science of Russian Federation, International Science and
Technology Center, Russian Academy of Sciences, Russian Federal Agency of Atomic Energy, Russian Federal Agency for Science and Innovations and CERN-INTAS;
 \\
Ministry of Education of Slovakia;
 \\
Department of Science and Technology, South Africa;
 \\
CIEMAT, EELA, Ministerio de Educaci\'{o}n y Ciencia of Spain, Xunta de Galicia (Conseller\'{\i}a de Educaci\'{o}n),
CEA\-DEN, Cubaenerg\'{\i}a, Cuba, and IAEA (International Atomic Energy Agency);
 \\
Swedish Reseach Council (VR) and Knut $\&$ Alice Wallenberg Foundation (KAW);
 \\
Ukraine Ministry of Education and Science;
 \\
United Kingdom Science and Technology Facilities Council (STFC);
 \\
The United States Department of Energy, the United States National
Science Foundation, the State of Texas, and the State of Ohio.

%% file: authorlist1.tex
\begingroup
\small
\begin{flushleft}
B.~Abelev\Irefn{org1234}\And
A.~Abrahantes~Quintana\Irefn{org1197}\And
D.~Adamov\'{a}\Irefn{org1283}\And
A.M.~Adare\Irefn{org1260}\And
M.M.~Aggarwal\Irefn{org1157}\And
G.~Aglieri~Rinella\Irefn{org1192}\And
A.G.~Agocs\Irefn{org1143}\And
A.~Agostinelli\Irefn{org1132}\And
S.~Aguilar~Salazar\Irefn{org1247}\And
Z.~Ahammed\Irefn{org1225}
% ...
\renewcommand\labelenumi{\textsuperscript{\theenumi}~}
\section*{Affiliation notes}
\renewcommand\theenumi{\roman{enumi}}
\begin{Authlist}
\item \Adef{0}Deceased
\item \Adef{Dipartimento di Fisica dell'Universita, Udine, Italy}Also at: Dipartimento di Fisica dell'Universita, Udine, Italy
\item \Adef{M.V.Lomonosov Moscow State University, D.V.Skobeltsyn Institute of Nuclear Physics, Moscow, Russia}Also at: M.V.Lomonosov Moscow State University, D.V.Skobeltsyn Institute of Nuclear Physics, Moscow, Russia
\item \Adef{Institute of Nuclear Sciences, Belgrade, Serbia}Also at: "Vin\v{c}a" Institute of Nuclear Sciences, Belgrade, Serbia
\end{Authlist}
\section*{Collaboration Institutes}
\renewcommand\theenumi{\arabic{enumi}~}
\begin{Authlist}
\item \Idef{org1279}Benem\'{e}rita Universidad Aut\'{o}noma de Puebla, Puebla, Mexico
\item \Idef{org1220}Bogolyubov Institute for Theoretical Physics, Kiev, Ukraine
\item \Idef{org1262}Budker Institute for Nuclear Physics, Novosibirsk, Russia
\item \Idef{org1292}California Polytechnic State University, San Luis Obispo, California, United States
\item \Idef{org14939}Centre de Calcul de l'IN2P3, Villeurbanne, France
\item \Idef{org1197}Centro de Aplicaciones Tecnol\'{o}gicas y Desarrollo Nuclear (CEADEN), Havana, Cuba
\item \Idef{org1242}Centro de Investigaciones Energ\'{e}ticas Medioambientales y Tecnol\'{o}gicas (CIEMAT), Madrid, Spain
\item \Idef{org1244}Centro de Investigaci\'{o}n y de Estudios Avanzados (CINVESTAV), Mexico City and M\'{e}rida, Mexico
\item \Idef{org1335}Centro Fermi -- Centro Studi e Ricerche e Museo Storico della Fisica ``Enrico Fermi'', Rome, Italy
\item \Idef{org17347}Chicago State University, Chicago, United States
\item \Idef{org1118}China Institute of Atomic Energy, Beijing, China
\item \Idef{org1288}Commissariat \`{a} l'Energie Atomique, IRFU, Saclay, France
\item \Idef{org1294}Departamento de F\'{\i}sica de Part\'{\i}culas and IGFAE, Universidad de Santiago de Compostela, Santiago de Compostela, Spain
\item \Idef{org1106}Department of Physics Aligarh Muslim University, Aligarh, India
\item \Idef{org1121}Department of Physics and Technology, University of Bergen, Bergen, Norway
\item \Idef{org1162}Department of Physics, Ohio State University, Columbus, Ohio, United States
\item \Idef{org1300}Department of Physics, Sejong University, Seoul, South Korea
\item \Idef{org1268}Department of Physics, University of Oslo, Oslo, Norway
\item \Idef{org1132}Dipartimento di Fisica dell'Universit\`{a} and Sezione INFN, Bologna, Italy
\item \Idef{org1315}Dipartimento di Fisica dell'Universit\`{a} and Sezione INFN, Trieste, Italy
\item \Idef{org1145}Dipartimento di Fisica dell'Universit\`{a} and Sezione INFN, Cagliari, Italy
\item \Idef{org1270}Dipartimento di Fisica dell'Universit\`{a} and Sezione INFN, Padova, Italy
\item \Idef{org1285}Dipartimento di Fisica dell'Universit\`{a} `La Sapienza' and Sezione INFN, Rome, Italy
\item \Idef{org1154}Dipartimento di Fisica e Astronomia dell'Universit\`{a} and Sezione INFN, Catania, Italy
\item \Idef{org1290}Dipartimento di Fisica `E.R.~Caianiello' dell'Universit\`{a} and Gruppo Collegato INFN, Salerno, Italy
\item \Idef{org1312}Dipartimento di Fisica Sperimentale dell'Universit\`{a} and Sezione INFN, Turin, Italy
\item \Idef{org1103}Dipartimento di Scienze e Tecnologie Avanzate dell'Universit\`{a} del Piemonte Orientale and Gruppo Collegato INFN, Alessandria, Italy
\item \Idef{org1114}Dipartimento Interateneo di Fisica `M.~Merlin' and Sezione INFN, Bari, Italy
\item \Idef{org1237}Division of Experimental High Energy Physics, University of Lund, Lund, Sweden
\item \Idef{org1192}European Organization for Nuclear Research (CERN), Geneva, Switzerland
\item \Idef{org1227}Fachhochschule K\"{o}ln, K\"{o}ln, Germany
\item \Idef{org1122}Faculty of Engineering, Bergen University College, Bergen, Norway
\item \Idef{org1136}Faculty of Mathematics, Physics and Informatics, Comenius University, Bratislava, Slovakia
\item \Idef{org1274}Faculty of Nuclear Sciences and Physical Engineering, Czech Technical University in Prague, Prague, Czech Republic
\item \Idef{org1229}Faculty of Science, P.J.~\v{S}af\'{a}rik University, Ko\v{s}ice, Slovakia
\item \Idef{org1184}Frankfurt Institute for Advanced Studies, Johann Wolfgang Goethe-Universit\"{a}t Frankfurt, Frankfurt, Germany
\item \Idef{org1215}Gangneung-Wonju National University, Gangneung, South Korea
\item \Idef{org1212}Helsinki Institute of Physics (HIP) and University of Jyv\"{a}skyl\"{a}, Jyv\"{a}skyl\"{a}, Finland
\item \Idef{org1203}Hiroshima University, Hiroshima, Japan
\item \Idef{org1329}Hua-Zhong Normal University, Wuhan, China
\item \Idef{org1254}Indian Institute of Technology, Mumbai, India
\item \Idef{org1266}Institut de Physique Nucl\'{e}aire d'Orsay (IPNO), Universit\'{e} Paris-Sud, CNRS-IN2P3, Orsay, France
\item \Idef{org1277}Institute for High Energy Physics, Protvino, Russia
\item \Idef{org1249}Institute for Nuclear Research, Academy of Sciences, Moscow, Russia
\item \Idef{org1320}Nikhef, National Institute for Subatomic Physics and Institute for Subatomic Physics of Utrecht University, Utrecht, Netherlands
\item \Idef{org1250}Institute for Theoretical and Experimental Physics, Moscow, Russia
\item \Idef{org1230}Institute of Experimental Physics, Slovak Academy of Sciences, Ko\v{s}ice, Slovakia
\item \Idef{org1127}Institute of Physics, Bhubaneswar, India
\item \Idef{org1275}Institute of Physics, Academy of Sciences of the Czech Republic, Prague, Czech Republic
\item \Idef{org1139}Institute of Space Sciences (ISS), Bucharest, Romania
\item \Idef{org27399}Institut f\"{u}r Informatik, Johann Wolfgang Goethe-Universit\"{a}t Frankfurt, Frankfurt, Germany
\item \Idef{org1185}Institut f\"{u}r Kernphysik, Johann Wolfgang Goethe-Universit\"{a}t Frankfurt, Frankfurt, Germany
\item \Idef{org1177}Institut f\"{u}r Kernphysik, Technische Universit\"{a}t Darmstadt, Darmstadt, Germany
\item \Idef{org1256}Institut f\"{u}r Kernphysik, Westf\"{a}lische Wilhelms-Universit\"{a}t M\"{u}nster, M\"{u}nster, Germany
\item \Idef{org1246}Instituto de Ciencias Nucleares, Universidad Nacional Aut\'{o}noma de M\'{e}xico, Mexico City, Mexico
\item \Idef{org1247}Instituto de F\'{\i}sica, Universidad Nacional Aut\'{o}noma de M\'{e}xico, Mexico City, Mexico
\item \Idef{org23333}Institut of Theoretical Physics, University of Wroclaw
\item \Idef{org1308}Institut Pluridisciplinaire Hubert Curien (IPHC), Universit\'{e} de Strasbourg, CNRS-IN2P3, Strasbourg, France
\item \Idef{org1182}Joint Institute for Nuclear Research (JINR), Dubna, Russia
\item \Idef{org1143}KFKI Research Institute for Particle and Nuclear Physics, Hungarian Academy of Sciences, Budapest, Hungary
\item \Idef{org18995}Kharkiv Institute of Physics and Technology (KIPT), National Academy of Sciences of Ukraine (NASU), Kharkov, Ukraine
\item \Idef{org1199}Kirchhoff-Institut f\"{u}r Physik, Ruprecht-Karls-Universit\"{a}t Heidelberg, Heidelberg, Germany
\item \Idef{org20954}Korea Institute of Science and Technology Information
\item \Idef{org1160}Laboratoire de Physique Corpusculaire (LPC), Clermont Universit\'{e}, Universit\'{e} Blaise Pascal, CNRS--IN2P3, Clermont-Ferrand, France
\item \Idef{org1194}Laboratoire de Physique Subatomique et de Cosmologie (LPSC), Universit\'{e} Joseph Fourier, CNRS-IN2P3, Institut Polytechnique de Grenoble, Grenoble, France
\item \Idef{org1187}Laboratori Nazionali di Frascati, INFN, Frascati, Italy
\item \Idef{org1232}Laboratori Nazionali di Legnaro, INFN, Legnaro, Italy
\item \Idef{org1125}Lawrence Berkeley National Laboratory, Berkeley, California, United States
\item \Idef{org1234}Lawrence Livermore National Laboratory, Livermore, California, United States
\item \Idef{org1251}Moscow Engineering Physics Institute, Moscow, Russia
\item \Idef{org1140}National Institute for Physics and Nuclear Engineering, Bucharest, Romania
\item \Idef{org1165}Niels Bohr Institute, University of Copenhagen, Copenhagen, Denmark
\item \Idef{org1109}Nikhef, National Institute for Subatomic Physics, Amsterdam, Netherlands
\item \Idef{org1283}Nuclear Physics Institute, Academy of Sciences of the Czech Republic, \v{R}e\v{z} u Prahy, Czech Republic
\item \Idef{org1264}Oak Ridge National Laboratory, Oak Ridge, Tennessee, United States
\item \Idef{org1189}Petersburg Nuclear Physics Institute, Gatchina, Russia
\item \Idef{org1170}Physics Department, Creighton University, Omaha, Nebraska, United States
\item \Idef{org1157}Physics Department, Panjab University, Chandigarh, India
\item \Idef{org1112}Physics Department, University of Athens, Athens, Greece
\item \Idef{org1152}Physics Department, University of Cape Town, iThemba LABS, Cape Town, South Africa
\item \Idef{org1209}Physics Department, University of Jammu, Jammu, India
\item \Idef{org1207}Physics Department, University of Rajasthan, Jaipur, India
\item \Idef{org1200}Physikalisches Institut, Ruprecht-Karls-Universit\"{a}t Heidelberg, Heidelberg, Germany
\item \Idef{org1325}Purdue University, West Lafayette, Indiana, United States
\item \Idef{org1281}Pusan National University, Pusan, South Korea
\item \Idef{org1176}Research Division and ExtreMe Matter Institute EMMI, GSI Helmholtzzentrum f\"ur Schwerionenforschung, Darmstadt, Germany
\item \Idef{org1334}Rudjer Bo\v{s}kovi\'{c} Institute, Zagreb, Croatia
\item \Idef{org1298}Russian Federal Nuclear Center (VNIIEF), Sarov, Russia
\item \Idef{org1252}Russian Research Centre Kurchatov Institute, Moscow, Russia
\item \Idef{org1224}Saha Institute of Nuclear Physics, Kolkata, India
\item \Idef{org1130}School of Physics and Astronomy, University of Birmingham, Birmingham, United Kingdom
\item \Idef{org1338}Secci\'{o}n F\'{\i}sica, Departamento de Ciencias, Pontificia Universidad Cat\'{o}lica del Per\'{u}, Lima, Peru
\item \Idef{org1146}Sezione INFN, Cagliari, Italy
\item \Idef{org1115}Sezione INFN, Bari, Italy
\item \Idef{org1313}Sezione INFN, Turin, Italy
\item \Idef{org1133}Sezione INFN, Bologna, Italy
\item \Idef{org1155}Sezione INFN, Catania, Italy
\item \Idef{org1316}Sezione INFN, Trieste, Italy
\item \Idef{org1286}Sezione INFN, Rome, Italy
\item \Idef{org1271}Sezione INFN, Padova, Italy
\item \Idef{org1322}Soltan Institute for Nuclear Studies, Warsaw, Poland
\item \Idef{org1258}SUBATECH, Ecole des Mines de Nantes, Universit\'{e} de Nantes, CNRS-IN2P3, Nantes, France
\item \Idef{org1304}Technical University of Split FESB, Split, Croatia
\item \Idef{org1168}The Henryk Niewodniczanski Institute of Nuclear Physics, Polish Academy of Sciences, Cracow, Poland
\item \Idef{org17361}The University of Texas at Austin, Physics Department, Austin, TX, United States
\item \Idef{org1173}Universidad Aut\'{o}noma de Sinaloa, Culiac\'{a}n, Mexico
\item \Idef{org1296}Universidade de S\~{a}o Paulo (USP), S\~{a}o Paulo, Brazil
\item \Idef{org1149}Universidade Estadual de Campinas (UNICAMP), Campinas, Brazil
\item \Idef{org1239}Universit\'{e} de Lyon, Universit\'{e} Lyon 1, CNRS/IN2P3, IPN-Lyon, Villeurbanne, France
\item \Idef{org1205}University of Houston, Houston, Texas, United States
\item \Idef{org20371}University of Technology and Austrian Academy of Sciences, Vienna, Austria
\item \Idef{org1222}University of Tennessee, Knoxville, Tennessee, United States
\item \Idef{org1310}University of Tokyo, Tokyo, Japan
\item \Idef{org1318}University of Tsukuba, Tsukuba, Japan
\item \Idef{org21360}Eberhard Karls Universit\"{a}t T\"{u}bingen, T\"{u}bingen, Germany
\item \Idef{org1225}Variable Energy Cyclotron Centre, Kolkata, India
\item \Idef{org1306}V.~Fock Institute for Physics, St. Petersburg State University, St. Petersburg, Russia
\item \Idef{org1323}Warsaw University of Technology, Warsaw, Poland
\item \Idef{org1179}Wayne State University, Detroit, Michigan, United States
\item \Idef{org1260}Yale University, New Haven, Connecticut, United States
\item \Idef{org1332}Yerevan Physics Institute, Yerevan, Armenia
\item \Idef{org15649}Yildiz Technical University, Istanbul, Turkey
\item \Idef{org1301}Yonsei University, Seoul, South Korea
\item \Idef{org1327}Zentrum f\"{u}r Technologietransfer und Telekommunikation (ZTT), Fachhochschule Worms, Worms, Germany
\end{Authlist}
\endgroup